\documentclass[12pt]{article}
\usepackage{latexsym}
\usepackage{comment}
\usepackage[english]{babel}
\usepackage[font={small}]{caption}
\usepackage{epsfig,amssymb,euscript,slashed}
\usepackage[linktocpage]{hyperref}
\usepackage{amsmath}
\usepackage[noadjust]{cite}
\usepackage{mathtools}
\usepackage{physics}
\usepackage{extarrows}
\usepackage{xcolor}
\usepackage{float}
\usepackage{graphicx}
\usepackage{caption}
\usepackage{subcaption}
\usepackage{tcolorbox}
\usepackage{MnSymbol}
\usepackage{tikz}
\usetikzlibrary{positioning,shapes.multipart,arrows.meta}
\usepackage{adjustbox}
\usepackage{array,calc}
\usepackage[]{graphicx}
\usepackage{color}
\usepackage{bbm}
\usepackage{fancybox}
\usepackage{ytableau}
\usepackage{enumerate}
\usepackage[shortlabels]{enumitem}
\numberwithin{equation}{section}

\def\Ad{\dot{A}}
\def\Bd{\dot{B}}
\def\Cd{\dot{C}}
\def\Dd{\dot{D}}

\def\zt{\tilde{z}}

\def\ep{\epsilon}
\def\vep{\varepsilon}

\def\ald{\dot{\alpha}}
\def\bed{\dot{\beta}}
\def\yb{\bar{y}}
\def\ty{\tilde{y}}
\def\tj{\tilde{j}}
\def\tq{\tilde{q}}
\def\ty{\tilde{y}}
\def\qb{\bar{q}}

\def\lam{\lambda}

\def\pd{\partial}

\def\lto{\longrightarrow}

\makeatletter
\g@addto@macro\bfseries{\boldmath}
\makeatother

\def\cA{{\cal A}}

\def\cE{{\cal E}}

\def\cG{{\cal G}}
\def\cH{{\cal H}}

\def\cM{{\cal M}}
\def\cN{{\cal N}}
\def\cO{{\cal O}}

\def\cZ{{\cal Z}}

\def\bbR{{\mathbb{R}}}

\def\bbZ{{\mathbb{Z}}}

\begin{document}
\font\cmss=cmss10 \font\cmsss=cmss10 at 7pt

\begin{flushright}{  
\scriptsize YITP-25-74}
\end{flushright}
\hfill
\vspace{18pt}
\begin{center}
{\Large 
\textbf{Fortuity and Supergravity}}

\end{center}

\vspace{8pt}
\begin{center}
{\textsl{Marcel R. R. Hughes$^{\,a}$ and Masaki Shigemori$^{\,a,b}$}}

\vspace{1cm}

\textit{\small ${}^a$ Department of Physics, Nagoya University\\
Furo-cho, Chikusa-ku, Nagoya 464-8602, Japan} \\ \vspace{6pt}

\textit{\small ${}^b$ 
Center for Gravitational Physics,\\
Yukawa Institute for Theoretical Physics, Kyoto University\\
Kitashirakawa Oiwakecho, Sakyo-ku, Kyoto 606-8502, Japan
}\\
\vspace{6pt}

\end{center}

\vspace{12pt}

\begin{center}
\textbf{Abstract}
\end{center}

\vspace{4pt} {\small
\noindent 
BPS states in holographic CFTs naturally split into those describing black holes in the bulk and those that do not, with black hole states only existing above a certain energy threshold. In the context of the AdS$_3$/CFT$_2$ duality this can be seen from the agreement of the CFT and supergraviton supersymmetric indices up to a certain central charge-scaling conformal dimension. However, there also exist additional smooth horizonless bulk configurations called singletons that have not been previously accounted for when distinguishing black hole states from non-black hole states. These singletons describe bulk degrees of freedom that are non-trivial on the AdS$_3$ boundary. From a detailed analysis of BPS states in the D1-D5 system we identify singleton states in $\mathrm{Sym}^N(T^4)$ and $\mathrm{Sym}^N(K3)$ and explicitly incorporate them into a generalised supergraviton index for low levels for the case of $N=2$, leading to an enhanced matching with the CFT index in the case of $T^4$. Singleton states are monotone and, under the assumption that together with supergraviton states they span the monotone Hilbert space, the generalised supergraviton index represents the full monotone index. This allows us to define fortuitous indices for these theories and for $\mathrm{Sym}^2(K3)$ we construct the first explicit fortuitous states.
}

\vspace{1cm}

\thispagestyle{empty}

\vfill
\vskip 5.mm
\hrule width 5.cm
\vskip 2.mm
{
\noindent  {\scriptsize e-mails:  {\texttt{hughes.mrr@eken.phys.nagoya-u.ac.jp, masaki.shigemori@nagoya-u.jp}} }
}

\setcounter{footnote}{0}
\setcounter{page}{0}

\newpage

\tableofcontents
%\newpage

%%%%%%%%%%%%%%%%%%%%%%%%%%%%%%%%%%%%%%%%%

\section{Introduction and summary}
\label{sec:intro}

Matching states between the two sides of the AdS/CFT correspondence is
a longstanding and continually evolving problem of central importance in string theory.  In these efforts, supersymmetric or BPS states have been the primary focus and supersymmetry indices have been the principal tools.
It is natural to classify the states being matched into two categories:  ``supergraviton states'' and ``black hole states.''   In the bulk, supergraviton states mean light, BPS states of particles that appear as perturbations around the empty AdS vacuum. (They do not literally mean gravitons but BPS combinations of excitations of the metric and other fields that exist in the theory.)
On the other hand, black hole states are heavy BPS states that appear above an energy threshold, described by a black hole with a horizon.  
In reality, the distinction between the two categories cannot be made based on energy. Black hole states appear above an energy threshold of order $N$ above the AdS vacuum, where $N$ is proportional to the central charge of the boundary CFT\@.  If we excite a few supergravitons around empty AdS, the energy is $\cO(1)$ but, if we excite an $\cO(N)$ number of them, we can have a coherent gas of supergravitons which backreacts on the AdS background and becomes a smooth horizonless geometry.  Such smooth geometric states can exist even above the black-hole energy threshold.

To discuss the CFT side, let us be concrete and consider the ${\rm AdS_3/CFT_2}$ correspondence in the D1-D5 system, where the bulk is type IIB string theory in ${\rm AdS}_3\times S^3\times \cM_4$ with $\cM_4=T^4$ or $K3$, and the boundary theory is a $d=2$, $\cN=(4,4)$ CFT called the D1-D5 CFT\@.  This system is obtained by wrapping $n_1$ D1-branes and $n_5$ D5-branes on $S^1\times \cM_4$ and taking a decoupling limit.
This is the setup in which the first successful matching of the number of black-hole states on the bulk and boundary sides was found \cite{Strominger:1996sh}.
In this setup, single-particle supergraviton states correspond in CFT to short-multiplet states excited on the vacuum.  On the other hand, black-hole states are long-multiplet states that appear above the energy threshold $h\approx\frac{N}{4}$, where $h=L_0$ is the left-moving conformal weight in the Neveu-Schwarz (NS) sector\footnote{Our convention is such that the NS and Ramond (R) vacua both have $h=0$.} and $N=n_1 n_5$ ($N=n_1 n_5+1$) for $\cM_4=T^4$ ($\cM_4=K3$).  As in the bulk, we cannot distinguish black-hole states from supergraviton states just by the energy; if we excite multiple supergravitons, they become long-multiplet states whose energy can be $\cO(N)$ if their number is $\cO(N)$.

Comparison of supersymmetry indices between bulk and boundary can be performed much more precisely than in \cite{Strominger:1996sh} and, in particular, it was shown in \cite{deBoer:1998us} that the CFT index (elliptic genus) for $\cM_4=K3$ precisely agrees with the supergraviton elliptic genus up to the energy bound $h\le \frac{N+1}{4}$ for any value of $j=J^3_0$, the left-moving angular momentum. Here the supergraviton elliptic genus is defined by enumerating all supergraviton states in the bulk, with an exclusion principle \cite{Maldacena:1998bw} that limits the number of supergravitons.
This is an interesting result which indicates precisely at what value of~$h$ black hole states start to exist, sharpening the question of what distinguishes black hole states from others.  This result was later extended to the $T^4$ case~\cite{Maldacena:1999bp}, where the
modified elliptic genus was shown to agree between the two sides within the bound $h<\frac{N}{4}$ for any $j$.  We will refer to these bounds as the de Boer bounds.

More recently, a new proposal for classifying CFT states into supergraviton states and black hole states based on ``fortuity'' was put forward \cite{Chang:2022mjp, Chang:2024zqi}.
In this proposal, one classifies BPS states in CFT into ``monotone'' states and ``fortuitous'' states based on their behaviour as one changes $N$. Roughly speaking (we will make this more precise below), monotone states are BPS states for all $N$,\footnote{The discussion here is based on $d=4$, $\cN=4$ super-Yang-Mills theory.  The situation is subtly different for the case of the D1-D5 system, which we will discuss below.} while fortuitous states are BPS for particular finite ranges of $N$ due to the specific structure of the Hilbert space at finite~$N$. 
Fortuity was originally proposed in the setting of AdS$_5$/CFT$_4$, in which the first black-hole states were searched for in $d=4$, $\cN=4$ super-Yang-Mills theory using $Q$-cohomology \cite{Kinney:2005ej, Grant:2008sk, Chang:2013fba}, first constructed in~\cite{Chang:2022mjp}, and investigated further in~\cite{Choi:2022caq, Choi:2023vdm, Chang:2023zqk, Choi:2023znd, Budzik:2023vtr, deMelloKoch:2024pcs}. It was conjectured \cite{Chang:2024zqi} that monotone states are dual to perturbative excitations and smooth horizonless geometries (see the reviews~\cite{Bena:2022ldq,Bena:2022rna,Shigemori:2020yuo} and references therein) while fortuitous states are dual to typical black hole microstates, based on the fact that the monotone partition function agrees with the supergraviton partition function in this example~\cite{Chang:2024zqi}. More recently the same ideas have been applied to the SYK model \cite{Chang:2024lxt} and the D1-D5 system~\cite{Chang:2025rqy}.

\bigskip

In the current paper, we study in detail the spectrum of BPS states in the D1-D5 system using supersymmetry indices, and re-examine the matching of the bulk and boundary indices observed in \cite{deBoer:1998us, Maldacena:1999bp} and the idea of using fortuity to distinguish between black hole states and non-black hole ones.  We find that there are degrees of freedom (so-called singletons, or diffeomorphisms that do not vanish at the AdS boundary \cite{Brown:1986nw}) that have not been properly taken into account, and that incorporating them leads to matching beyond the $T^4$ de Boer bound. 

Furthermore, while these singleton degrees of freedom are na\"{i}vely fortuitous, they are in fact monotone due to a subtlety in the definition of monotonicity in the D1-D5 system.\footnote{In an earlier version of~\cite{Chang:2025rqy} the subtleties of the definition of monotonicity for the D1-D5 system had not been fully understood. This led to the labelling of singletons as fortuitous states in the first version of the present paper. We thank Chi-Ming Chang and Haoyu Zhang for detailed discussions on this point.}

\bigskip

In the rest of the introduction, we describe our results in more detail, after briefly introducing background material needed for developing relevant notions.

Let us quickly recall the structure of the D1-D5 CFT; a more complete presentation will be given in section \ref{sec:background}. At the orbifold point in its moduli space, this CFT is a symmetric orbifold theory ${\rm Sym}^N(\cM_4)$, in which a general state $\ket{\Psi}$ can be written as a product of ``strands'' as  
\begin{align}
   \ket{\Psi}=\ket{\psi_1}_{a_1}\ket{\psi_2}_{a_2}\ket{\psi_3}_{a_3}\,\cdots~,
   \label{gen_CFT_state_intro_1}
\end{align}
where $a_i\in\bbZ_{>0}$ are the length of strands, satisfying the ``stringy exclusion principle'' $\sum_i a_i=N$; {\it i.e.}, the total strand length must be equal to $N$.  A strand of length $a$ is made of $a$ copies of the $\cM_4$ CFT, twisted together by a twist operator.  The D1-D5 CFT is an $\cN=(4,4)$ SCFT and has the symmetry group $SU(1,1\,|\,2)_L\times SU(1,1\,|\,2)_R$. The theory being a 2D CFT, the generators of the symmetry include \emph{global} generators, {\it i.e.}, generators of the anomaly-free subalgebra, such as $L_0,L_{-1},J^3_0$, and \emph{affine} generators, such as $L_{-n}$ with $n>1$.\footnote{In the standard usage, affine generators include global generators, but we often mean by affine generators non-global generators.  The meaning must be clear from the context but, when confusion may arise, we explicitly use words such as non-global generators.}
We will focus only on left-moving part of the algebra because we are considering BPS states in which only the left-moving sector is excited.\footnote{The right-moving part of the states is not trivial and plays a role in state counting, but we ignore it for the rest of the introduction to simplify presentation.}
Let us denote (left-moving) global generators collectively by $\cG$ and affine generators by $\cA$.  In the NS sector, the state of each strand, $\ket{\psi_i}_{a_i}$, can be written as a $\tfrac12$-BPS (chiral-primary) state $\ket{\phi_i}_{a_i}$ acted upon by some operator, as $\ket{\psi_i}_{a_i}=\cO_i\!\ket{\phi_i}_{a_i}$.
Here $\cO_i$ can contain both global and affine generators. So, more explicitly, the general CFT state \eqref{gen_CFT_state_intro_1} can be written as
\begin{align}
   \ket{\Psi}=\cO_1\!\ket{\phi_1}_{a_1}\cO_2\!\ket{\phi_2}_{a_2}\,\cdots~.
   \label{gen_CFT_state_intro_2}
\end{align}
We will refer to generators acting on a single strand, such as $\cO_1$ in $\cO_1\!\ket{\phi_1}_{a_1}$ above, as \emph{individual} generators.  We will also consider a \emph{total} generator
$\cO^{\rm (T)}=\sum_i \cO_i$ where $\cO_i$ acts only on the $i$th strand.
Among chiral primaries $\ket{\phi}_a$ is the special state $\ket{0}_1$, the NS vacuum strand, which is annihilated by all global generators: $\cG \ket{0}_1=0$.
Even if we move away from the orbifold point of the CFT, the total generators $\cG^{\rm (T)},\cA^{\rm (T)}$ continue to be the generators of superconformal transformations.  In particular, the total global generators $\cG^{\rm (T)}$ are dual to the super-isometries of the bulk ${\rm AdS}_3\times S^3$ background.

\begin{figure}[tb]
\begin{center}
\begin{tabular}{c@{~~~~~~}c}
\includegraphics[width=0.45\textwidth]{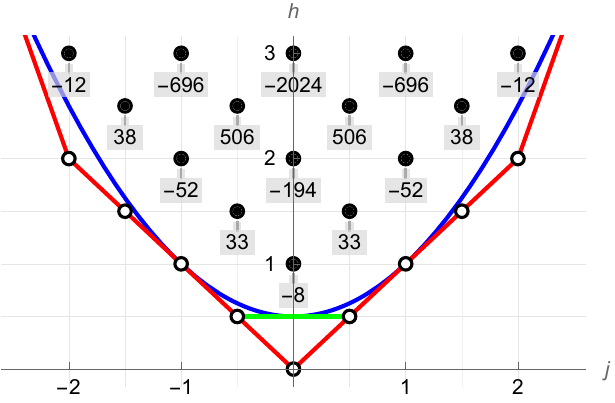}&
\includegraphics[width=0.45\textwidth]{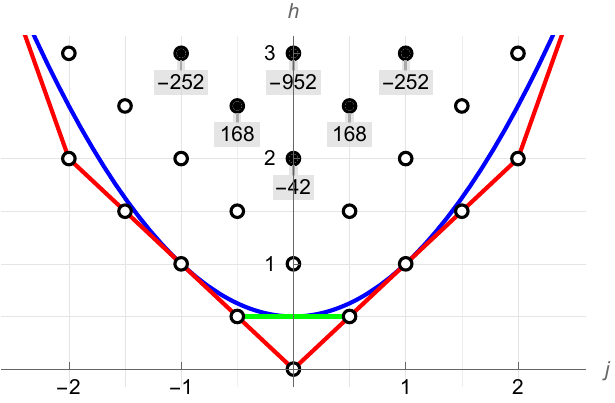}\\
$(a)$~~&$(b)~~$
\end{tabular}
\caption{\it The difference between the boundary and bulk indices (modified elliptic genera) shown on the $(j,h)=(J^3_0,L_0)$ plane in the NS sector for the $T^4$ theory for $N=2$. 
(a): The CFT index minus the supergraviton index.  (b): The CFT index minus the generalized supergraviton index.  The black dots indicate points $(j,h)$ where the difference is non-vanishing, with the difference shown in a call-out.  The white dots indicate points where the indices agree.  Above the blue parabola $h=j^2/N+N/4$, the black hole exists in the bulk \cite{Breckenridge:1996is}.  The red polygon is the unitarity bound below which there exists no state.  The green horizontal line $h=N/4$ is the de Boer bound for $T^4$ \cite{Maldacena:1999bp} up to which the CFT and supergraviton indices agree. Note that the contribution to the generalised index from additional affine characters, if such characters indeed arise, as speculated at the end of \ref{ssec:subtract}, has not been included.
\label{fig:MEGdiff}}
\end{center}
\end{figure}

In this language, a single-particle supergraviton is the single-strand state $\cG\! \ket{\phi}_a$, where $\cG$ is a product of global generators.  The full state is
\begin{align}
    \cG\!\ket{\phi}_a \underbrace{\ket{0}_1 \cdots \ket{0}_1}_{N-a}~,
\end{align}
where we inserted copies of the vacuum strand $\ket{0}_1$ so that the total strand length is~$N$. A multi-supergraviton state is obtained by multiplying single-supergraviton states together:
\begin{align}
    \cG_1\!\ket{\phi_1}_{a_1} \cG_2\!\ket{\phi_2}_{a_2} \cdots\,\,
    \ket{0}_1\cdots\ket{0}_1~,
    \label{sgton_state_intro}
\end{align}
where again the vacuum strands were inserted to make the total strand length equal to~$N$. 
These states \eqref{sgton_state_intro} describe smooth horizonless geometries.\footnote{To be precise we have to take a coherent sum of such states with different numbers of supergravitons \cite{Kanitscheider:2007wq,Bena:2017xbt,Giusto:2015dfa}, but we ignore this point.} In \cite{deBoer:1998us,Maldacena:1999bp}, the supergraviton states~\eqref{sgton_state_intro} were counted using supersymmetry indices ((modified) elliptic genera) and it was found that these supergraviton indices are equal to the full CFT indices that count states of the general form \eqref{gen_CFT_state_intro_2}, up to the de Boer bounds. 
In such counting, the stringy exclusion principle that puts an upper bound on the number of supergravitons plays a crucial role \cite{deBoer:1998ip, deBoer:1998us, Maldacena:1998bw}.
See figure \ref{fig:MEGdiff}(a) for comparison between CFT and supergraviton indices for the $T^4$ case for $N=2$.

However, states of the form \eqref{sgton_state_intro} are not the most general states whose bulk dual is a smooth horizonless geometry.  The total affine modes $\cA^{\rm (T)}$ are known to be dual to ``singletons,'' {\it i.e.}, diffeomorphisms that do not vanish at the AdS infinity \cite{Brown:1986nw}, and its superpartners.  The state obtained by adding a singleton to the multi-supergraviton state~\eqref{sgton_state_intro} is 
\begin{align}
&\cA^{\rm (T)} \Bigl(
    \cG_1\!\ket{\phi_1}_{a_1} \cG_2\!\ket{\phi_2}_{a_2} \cdots\,\,\ket{0}_1\cdots\ket{0}_1
    \Bigr)
\notag\\
&=  
    \Bigl(\cA\cG_1\!\ket{\phi_1}_{a_1} \cG_2\!\ket{\phi_2}_{a_2} \cdots\,\,\ket{0}_1\cdots\ket{0}_1\Bigr)
    +\Bigl(\cG_1\!\ket{\phi_1}_{a_1} \cA\cG_2\!\ket{\phi_2}_{a_2} \cdots\,\,\ket{0}_1\cdots\ket{0}_1\Bigr)
    +\cdots
    \notag\\
    &~
    +\Bigl(\cG_1\!\ket{\phi_1}_{a_1} \cG_2\!\ket{\phi_2}_{a_2} \cdots\,\,\cA\!\ket{0}_1\ket{0}_1\cdots\ket{0}_1\Bigr)
    +\Bigl(\cG_1\!\ket{\phi_1}_{a_1} \cG_2\!\ket{\phi_2}_{a_2} \cdots\,\,\ket{0}_1\cA\!\ket{0}_1\cdots\ket{0}_1\Bigr)
    +\cdots~,
    \label{gen_sgton_state_intro}
\end{align}
which is not in general a supergraviton state \eqref{sgton_state_intro}.  We can keep adding more singletons by repeatedly acting with total affine modes.   Such states are expected to represent a smooth horizonless geometry, because we are only applying a diffeomorphism on a smooth horizonless geometry \eqref{sgton_state_intro}.  For explicit constructions of smooth geometries involving singletons and their superpartners, see \cite{Mathur:2011gz, Mathur:2012tj, Lunin:2012gp, Giusto:2013bda}.

In this paper, we develop a systematic procedure to compute \emph{generalised supergraviton indices} which incorporate the contributions from singletons and their superpartners.  This is quite non-trivial, because singletons mix with supergravitons and we need to carefully keep only new contributions without over-counting.
We explicitly carry out the programme for $N=2$ for both $T^4$ and $K3$, and find that, for $T^4$, the generalised supergraviton index (modified elliptic genus) agrees with the CFT index up to $h={3\over 2}$, even though the original supergraviton index agrees with the CFT index only up to $h={1\over 2}$; see figure \ref{fig:MEGdiff}(b).
We write down the explicit form of the extra singleton states responsible for the improved matching.  
For $K3$, we do not see a similar improved matching; both the original and generalised supergraviton elliptic genera agree with the CFT one up to $h={1\over 2}$. We also list CFT states that are responsible for this disagreement at $h=1$ -- the first fortuitous states -- in \eqref{eq.BHstate322}.

Let us turn to fortuity. Our claim is that the state \eqref{gen_sgton_state_intro} is monotone, in line with it representing a smooth horizonless geometry, and that incorporating these states into an index along with supergraviton states captures the \emph{full} monotone index. In doing so, this allows us to properly define a fortuitous index.
As mentioned above, fortuity is defined by the behaviour of BPS states
as we change~$N$. In the D1-D5 system, the statement is as follows \cite{Chang:2024zqi,Chang:2025rqy,Chang:2025,HaoyuTalk}.

Given the BPS Hilbert space $\cH_N$ in a theory with some value of $N$, the physical states $|\Psi\rangle_N\in\cH_N$ can have any strand structure of total strand length $N$. The task is to uplift the state $|\Psi\rangle_N\in\cH_N$ to $|\Psi\rangle_{N+1}\in\cH_{N+1}$ such that quantum numbers are kept fixed. At the level of Hilbert spaces the embedding should be such that the projection operator maps $\pi_N:\cH_{N+1} \to \cH_{N}$ by killing any states not of the form $|\Psi\rangle_N|0\rangle_1$ and projects $|\Psi\rangle_N|0\rangle_1 \mapsto |\Psi\rangle_N$. It is non-trivial that such an embedding exists since we demand that the uplifted state is also BPS\@. Note that both $|\Psi\rangle_N|0\rangle_1$ and $|\Psi\rangle_N|0\rangle_1 + |\Psi'\rangle_N|\phi\rangle_1$, where $|\phi\rangle_1\neq |0\rangle_1$ and $|\Psi'\rangle_N$ is arbitrary, are potentially valid uplifts of the state $|\Psi\rangle_N$.\footnote{We can even consider a state that does not have the structure of a state in $\cH_N$ times a strand of length one as a potential uplift.} If this procedure can be iterated and a state in $\cH_N$ is uplifted to a state in $\cH_{N'}$ for any $N'>N$ then the state is said to be monotone, otherwise it is fortuitous.

For supergraviton states \eqref{sgton_state_intro} the uplift is done trivially by appending an additional vacuum strand $\ket{0}_1$ and so these are examples of monotone states. Doing the same thing to singleton states \eqref{gen_sgton_state_intro} results in non-BPS states since each term on the right-hand side contains a state of the form $\cA \cG\ket{\phi}$, which was shown in \cite{Gava:2002xb} to become non-BPS upon turning on interaction due to mixing with states with different strand lengths.  Therefore, the reason why \eqref{gen_sgton_state_intro} is BPS is due to subtle cancellations among all the terms, and due to the stringy exclusion principle that truncates states that go beyond length $N$. Therefore, to preserve these cancellations we have to find a particular way of uplifting. This is done by
\begin{align}
    \cA^{(\mathrm{T})} |\Phi\rangle_N \longmapsto \big(\cA^{(\mathrm{T})} |\Phi\rangle_N\big)|0\rangle_1 + |\Phi\rangle_N\cA|0\rangle_1 = \cA^{(\mathrm{T})} \big(|\Phi\rangle_N|0\rangle_1\big) \ ,
\end{align}
where $|\Phi\rangle_N$ is a supergraviton state of the form \eqref{sgton_state_intro}. The last expression shows that the resulting state is a singleton state in $\cH_{N+1}$. These singleton states are therefore also examples of monotone states since an uplift of this form can be found for any $N$.

We expect that these are the only classes of monotone states and therefore by capturing both supergraviton and singleton states in an index we can find an index for fortuitous states, thought to be dual to typical black hole microstates.

\bigskip

The organization of the rest of the paper is as follows.
In section~\ref{sec:background}, we give a brief review of the key features of the D1-D5 CFT, focusing on the symmetric orbifold description, the BPS spectrum and the supersymmetric indices which can them. The concept of singleton states as total modes in the CFT is then discussed in section~\ref{ssec:totalaffine} and a method of incorporating them into an index is introduced in section~\ref{ssec:VirEx} for a toy theory with only Virasoro symmetry.
In section~\ref{sec:T4}, we provide a detailed subtraction and promotion procedure in the case of $\mathrm{Sym}^2(T^4)$ for the generalisation of the supergraviton index to include these singleton states without over-counting. The explicit form of low-level singleton states and the enhanced de Boer bound of the generalised supergraviton index is analysed in section~\ref{ssec:Fortuity}.
In section~\ref{sec:K3}, this process is repeated for $\mathrm{Sym}^2(K3)$ and the explicit form of the lowest-dimension fortuitous states are presented in section~\ref{sec:App2}.
We conclude and discuss future directions in section~\ref{sec:conc}. In appendix~\ref{sec:App1}, we give our conventions for the D1-D5 CFT, its symmetry algebra and character formulae.

\section{Background}
\label{sec:background}

\subsection{The D1-D5 CFT}
\label{ssec:D1D5cft}

We consider type IIB string theory on $\mathbb{R}^{1,4}\times S^1_{\sigma}\times \cM_4$, where $\cM_4$ can be either $K3$ or $T^4$, with $n_1$ D1-branes wrapping the $S^1_{\sigma}$ and $n_5$ D5-branes wrapping $S^1_{\sigma}\times\cM_4$. The size of the compact $\cM_4$ is taken to be string scale, whereas the $S^1_{\sigma}$ is kept macroscopic. In the decoupling limit of this brane system the background has ${\rm AdS}_3\times S^3\times\cM_4$ asymptotics. This IR physics has an alternative description in terms of a holographic 2-dimensional CFT \cite{Maldacena:1997re}. This so-called D1-D5 CFT\footnote{For more detail about the D1-D5 CFT, see \cite{David:2002wn,Avery:2009tu}. Our conventions follow \cite{Avery:2009tu}.} is strongly believed to be described by the moduli space of the symmetric orbifold theory $\mathrm{Sym}^N(\cM_4)$ where $N=n_1n_5$ ($N=n_1n_5+1$) for $\cM_4=T^4$ ($\cM_4=K3$)~\cite{David:2002wn,Seiberg:1999xz}.

In a particular region of the moduli space of the D1-D5 CFT, dubbed the free locus, the CFT is described by a free symmetric orbifold theory $\mathrm{Sym}^N(\cM_4)$ (where $\cM_4$ can be either $K3$ or $T^4$) with a seed theory of the sigma model with $\cM_4$ target space and $\cN=(4,4)$ supersymmetry. In~\cite{Gaberdiel:2018rqv,Eberhardt:2018ouy,Eberhardt:2019ywk,Eberhardt:2020akk,Eberhardt:2019qcl,Dei:2019osr} it was shown that the D1-D5 CFT on this free locus is dual at large $N$ to a tensionless string theory in AdS${}_3\times S^3\times T^4$ with one unit of NS-NS flux. We first describe this free symmetric orbifold CFT in detail for the case of $\cM_4=T^4$ and then describe just the differences for the case of $\cM_4=K3$ that are relevant to the present paper.

\subsubsection*{Seed theory field content}

The free seed CFT$_2$ used to construct the symmetric orbifold theory defined on the Euclidean cylinder%
\footnote{We will also often consider the theory on the Riemann sphere with coordinates $(z,\tilde{z})$, related to the Euclidean cylinder coordinates via the conformal map $z = e^{(t-i\sigma)/R_{\sigma}}$, $\tilde{z}=e^{(t+i\sigma)/R_{\sigma}}$.} %
$\mathbb{R}_t \times S^1_{\sigma}$ base space contains four bosons from the coordinates on the target space $T^4$, along with four left-moving and four right-moving fermionic superpartners
\begin{align} \label{eq.XpsiFields}
    X^{\Ad A}(t,\sigma) \ ,\ \ \psi^{\alpha \Ad}(t,\sigma) \ ,\ \ \tilde{\psi}^{\ald \Ad}(t,\sigma) \ ,
\end{align}
with the various $SU(2)$ doublet indices taking the values
\begin{equation} \label{eq.doubletdefs}
    \alpha\in\{+,-\}\ ,\ \ald\in\{\dot{+},\dot{-}\}\ ,\ A\in\{1,2\}\ ,\ \Ad\in\{\dot{1},\dot{2}\} \ ,
\end{equation}
where $\alpha$ and $\ald$ are doublet indices for the left and right $SU(2)$ R-symmetry groups from the ``external"
\begin{equation}
    SO(4)_E\cong SU(2)_L\times SU(2)_R \ ,
\end{equation}
coming from the isometry group of $S^3$. The four free bosons transform in the vector representation of the ``internal" $SO(4)_I$ coming from the isometry group of the $T^4$; however, it is generally a broken symmetry here due to the compactification. Nevertheless, it is useful for organising the spectrum and in particular we will make use of the isomorphism
\begin{equation}
    SO(4)_I\cong SU(2)_1\times SU(2)_2 \ ,
\end{equation}
with $A$ and $\Ad$ in \eqref{eq.doubletdefs} being doublet representation indices for $SU(2)_1$ and $SU(2)_2$ respectively.

These free bosons are taken to be periodic around the $S^1_{\sigma}$ and the free fermions to be anti-periodic\footnote{This choice of anti-periodic fermions on the cylinder is a choice of the Neveu-Schwarz sector of the theory. We will further discuss this choice later.}, \textit{i.e.}\footnote{The periodicity/anti-periodicity of fermions reverses when mapping between the Euclidean cylinder and the Riemann sphere.}
\begin{equation}
    \pd X^{\Ad A}(t,\sigma+2\pi R_{\sigma}) = \pd X^{\Ad A}(t,\sigma) \ \ ,\ \ \psi^{\alpha\Ad}(t,\sigma+2\pi R_{\sigma}) = -\psi^{\alpha\Ad}(t,\sigma) \ .
\end{equation}

The symmetry algebra of this theory is the 2d contracted large $\cN=(4,4)$ superconformal algebra with central charge $c_{\mathrm{seed}}=6$. This algebra contains as a subalgebra the small $\cN=(4,4)$, generated by the left-moving stress tensor $T$, $SU(2)_L$ R-symmetry currents $J^{\pm,3}$ and supersymmetry generators $G^{\alpha A}$, along with the analogous right-moving currents (which we denote with a tilde). These currents are formed from combinations of the free bosons and fermions in the usual way, as given in \eqref{eq.currentsdef}. In total there are 16 real supersymmetry generators in this theory: the four left-moving $G^{\alpha A}$, four right-moving $\tilde{G}^{\ald A}$, as well as the corresponding superconformal generators $(G^{\alpha A})^{\dagger}$ and $(\tilde{G}^{\ald A})^{\dagger}$. Along with these generators the contracted large $\cN=(4,4)$ algebra also contains the free bosons and fermions as generators.

Decomposing the various symmetry currents into modes on the complex plane, \textit{e.g.} for the free bosons and fermions via
\begin{align} \label{eq.Xpsimodes}
     \pd X^{\dot{A}A}(z) = \sum_{n\in\mathbb{Z}} \alpha^{\dot{A}A}_{-n} \,z^{n-1} \ ,\ \  \psi^{\alpha\dot{A}}(z) = \sum_{r\in \mathbb{Z}+\frac12} \psi^{\alpha\dot{A}}_{-r}\, z^{r-\frac12} \ ,\ \ \tilde{\psi}^{\ald\dot{A}}(\tilde{z}) = \sum_{r\in \mathbb{Z}+\frac12} \tilde{\psi}^{\ald\dot{A}}_{-r}\, \tilde{z}^{r-\frac12}\ ,
\end{align}
and similarly for the small $\cN=(4,4)$ generators using the mode expansion \eqref{eq.modes} to get
\begin{equation} \label{eq.currentModes}
    \Big\{ L_{n}\ ,\ J^{\pm,3}_{n}\ ,\ G^{\alpha A}_{r}\Big\}\ \ ,\ \ \Big\{ \tilde{L}_{n}\ ,\ \tilde{J}^{\dot{\pm},3}_{n}\ ,\ \tilde{G}^{\ald A}_{r}\Big\} \ ,
\end{equation}
where $n\in\mathbb{Z}$, $r\in\mathbb{Z}+\tfrac12$ here. All together these current modes generate the contracted large $\cN=(4,4)$ mode algebra given in \eqref{eq.commcurrents2} and \eqref{eq.commcurrents3}. In full, the R-symmetry group is $SU(2)_L\times SU(2)_R$ and the 16 supercharges are $G^{\alpha A}_{\pm\frac12}$, $\tilde{G}^{\ald A}_{\pm\frac12}$.

\subsubsection*{$\mathrm{Sym}^N(\cM_4)$ Hilbert space}

The seed theory Hilbert space $\cH_{\mathrm{seed}}$ is the Fock space generated from the free boson and fermion modes $\alpha^{\Ad A}_{-n}$, $\psi^{\alpha\Ad}_{-r}$ and $\tilde{\psi}^{\ald\Ad}_{-r}$ defined in \eqref{eq.Xpsimodes}. This can then be used to construct the Hilbert space of the symmetric orbifold theory from the $S_N$-invariant subspace of $N$ tensor copies of this seed theory Hilbert space
\begin{equation} \label{eq.SymOrbH}
    \cH\big(\mathrm{Sym}^N(\cM_4)\big) \equiv \cH_{\mathrm{seed}}^{\otimes N}\big/ S_N = \bigoplus_{[g]} \cH_{[g]}\ ,
\end{equation}
where the symmetric group $S_N$ acts by permuting the tensor factors of $\cH_{\mathrm{seed}}$.
This orbifold Hilbert space admits a useful decomposition into so-called twist sectors $\cH_{[g]}$, as above, labelled by the conjugacy classes of $S_N$: equivalence classes of elements of $S_N$ under the conjugacy map $\varphi_h:\ g \longmapsto hgh^{-1}$ where $h\in S_N$, \textit{i.e.},
\begin{equation}
    [g] \equiv \Big\{ \tilde{g}\in S_N \ \Big|\ \exists\ h\in S_N\ \mathit{s.t.}\  h\tilde{g}h^{-1} = g \Big\} \ ,
\end{equation}
where $g$ is called the representative element of $[g]$. Conjugation of permutations preserves cycle structure and so conjugacy classes of $S_N$ consist of permutations with a given cycle structure $g=(1)^{n_1}(2)^{n_2}\cdots(N)^{n_N}$, where $(a)$ represents an $a$-cycle, subject to the constraint
\begin{equation} \label{eq.strandBudget}
    \sum_{a=1}^{N} a\,n_a = N \ .
\end{equation}
For a general twisted sector $\mathcal{H}_{[g]}$ labelled by the numbers $n_a$ of cycles of length $a$, the Hilbert space decomposes further into single-cycle Hilbert spaces $\mathcal{H}_a$ as
\begin{equation} \label{eq.HtwistSector}
    \mathcal{H}_{\{n_a\}} = \bigotimes_{a=1}^{N} \bigg(\underbrace{\mathcal{H}_{a}\otimes\cdots\otimes\mathcal{H}_a}_{n_a}\bigg)^{S_{n_a}} \ ,
\end{equation}
where $(\mathcal{H}\otimes\cdots\otimes\mathcal{H})^{S_n}$ represents the $S_n$-invariant subspace of the tensor product space $\mathcal{H}\otimes\cdots\otimes\mathcal{H}$. The single $a$-cycle Hilbert space -- often referred to as strand of length $a$ -- is the Hilbert space of the theory defined on a spatial $S^1_{\sigma}$ of length $2a\pi R_{\sigma}$. The untwisted sector is defined by $n_1=N$ and $n_{a}=0$ for $a>1$ and comes from the conjugacy class of the identity, thus it is present for any $N$. The states in the untwisted sector Hilbert space are simply the $S_N$ invariant combinations of tensor products of free seed theory states. Other twist sectors contain new states that were not present in $\cH_{\mathrm{seed}}$.

A twisted sector labelled by the numbers $n_a$ can also be thought of as a partition of~$N$ which we often denote by 
\begin{align}
    N=\{\underbrace{1,\dots,1}_{n_1},\underbrace{2,\dots,2}_{n_2},\dots\}.
\end{align}
For example, the untwisted sector is denoted by $N=\{1,1,\dots,1\}$ and the maximally twisted sector with $n_1=\cdots=n_{N-1}=0$, $n_N=1$ by $N=\{N\}$.

\subsubsection*{$\mathrm{Sym}^N(T^4)$ field content}

Due to the $N$ copies of the seed theory involved in defining the symmetric orbifold Hilbert space \eqref{eq.SymOrbH}, fields now have a `copy label' $(i)$ to indicate which tensor factor it originated from: \textit{i.e.}\ we now have the fields
\begin{equation} \label{eq.Nfields}
    \Big\{ \pd X^{\Ad A(i)} \ ,\ \psi^{\alpha\Ad(i)} \ ,\ \tilde{\psi}^{\ald\Ad(i)} \Big\} \ ,
\end{equation}
where $i=1,\dots,N$. Likewise the current modes \eqref{eq.currentModes} also now gain a copy label with the surviving symmetry algebra being the diagonal $\cN=4$ with central charge $c=6N$, generators of which we refer to as ``total modes". These are defined, using the example of the Virasoro modes, as
\begin{equation} \label{eq.totalModes}
    L^{(\mathrm{T})}_{n} \equiv \sum_{i=1}^{N} L^{(i)}_n \ ,
\end{equation}
with the remaining total symmetry algebra modes defined similarly.

On a strand of length $a$ the bosonic fields satisfy a periodicity condition: \textit{e.g.}
\begin{align} \label{eq.aCycleFields}
    \pd X^{\Ad A(i)}(t,\sigma+2\pi R_{\sigma}) &= \pd X^{\Ad A(i+1)}(t,\sigma) \quad\text{for } \ i=1,\dots, a-1 \ ,\nonumber\\
    \pd X^{\Ad A(a)}(t,\sigma+2\pi R_{\sigma}) &= \pd X^{\Ad A(1)}(t,\sigma) \ ,
\end{align}
with the analogous anti-periodicity condition satisfied by the fermions. This periodicity on a multiply-wound spatial circle leads to $\tfrac1a$ fractional moding.  Within a twist sector labelled by $\{n_a\}$ the individual states are not $S_N$ invariant, but instead invariant under the centraliser subgroup
\begin{equation} \label{eq.Centraliser}
    \mathcal{C}_{\{n_a\}} = \prod_{a=1}^{N} \mathcal{C}_{n_a} = \prod_{a=1}^{N} \Big(S_{n_a}\! \ltimes \mathbb{Z}_a^{\,n_a}\Big) \ ,
\end{equation}
where each $\mathbb{Z}_a$ acts to cyclically permute copies within $a$-wound strands and $S_{n_a}$ permutes identical strands of length $a$. Invariance of the spectrum under the full $S_N$ group can then be broken down into invariance under the centraliser and invariance under the conjugation orbit.

\subsubsection*{Spectral flow}

Up until now we have taken the fermionic fields (both the $\psi$ and the $G$) to be anti-periodic around the $S^1_{\sigma}$ (and thus periodic on the complex plane) which led to associated half-integer fermion modes in \eqref{eq.Xpsimodes} and \eqref{eq.currentModes}. This resulting theory is said to be in the Neveu-Schwarz (NS) sector. However, we could equally have chosen our fermions to be periodic around the $S^1_{\sigma}$ (anti-periodic on the complex plane) and thus integer moded. This choice would have resulted in the Ramond (R) sector.

In fact the $\cN=4$ symmetry algebra \eqref{eq.commcurrents2} is equally valid for the fermion mode numbers $r,s$ being integer or half integer. This observation is a consequence of the $\mathcal{N}=4$ superconformal algebra in two dimensions having an $SO(4)$ automorphism group, with an $SU(2)$ subgroup being the inner automorphism group and $SO(4)/SU(2)\cong SU(2)$ being the outer automorphism group \cite{Schwimmer:1986mf}. Algebras in different conjugacy classes of the outer automorphism group are inequivalent, whereas algebras that differ only by the conjugacy class of the inner automorphism group are equivalent.

The set of equivalent $\cN=4$ algebras is thus labelled by an angle $2\pi \eta$, which changes the boundary conditions of the fermions or equivalently the moding of their Fourier expansions. This family of equivalent generators can be written in terms of a deformation of the $\eta=0$ generators as
\begin{equation} \label{eq.symmModeDeform}
\begin{aligned}
    L_n(\eta) &= L_n + \eta J^3_n + \eta^2\frac{c}{24} \delta_{n,0} \ , \\
    J^3_n(\eta) &= J^3_n + \eta \frac{c}{12} \delta_{n,0} \ , \\
    J^{\pm}_{n}(\eta) &= J^{\pm}_{n\mp\eta} \ , \\
    G^{\pm A}_{r}(\eta) &= G^{\pm A}_{r\mp \frac{\eta}{2}} \ .
\end{aligned}
\end{equation}
In particular, $\eta\in\mathbb{Z}$ interpolates between the NS and R sectors.

\subsubsection*{The BPS spectrum}

The spectrum of the D1-D5 CFT can be organised by eigenvalues of the Cartan subalgebra generators of the contracted large $\cN=(4,4)$; namely the eigenvalues of $L_0$ and $J^3_0$ which we denote $h$ and $j$ respectively (and likewise $\tilde{h}$, $\tj$ for the right-moving generators). In the NS sector there is a unique vacuum state with $h=j=0$ which is annihilated by the generators of the anomaly-free subalgebra of \eqref{eq.commcurrents2} (related to the generators that are globally defined on the Riemann sphere, hence we also refer to it as the global subalgebra)
\begin{equation} \label{eq.globalSubAlg}
    \Big\{ L_{\pm1}\,,\ L_0\,,\ G^{\alpha A}_{\pm\frac12}\,,\ J^a_0 \Big\} \ ,
\end{equation}
which generate the supergroup $SU(1,1\,|\,2)_L$. The analogous right-moving global subalgebra forms the group $SU(1,1\,|\,2)_R$.

Primary states $|\phi\rangle$ are those states annihilated by the positive modes of the $\cN=4$ algebra, \textit{i.e.}\ in the NS sector
\begin{equation}
\begin{aligned} \label{eq.primaryDef}
    L_n|\phi\rangle &= J^a_n|\phi\rangle = G^{\alpha A}_{r}|\phi\rangle = 0 \quad\text{for } n>0,\ r\geq\frac12 \ , \\
    L_0|\phi\rangle &= h|\phi\rangle \quad,\ \ J^3_0|\phi\rangle = j|\phi\rangle \ ,
\end{aligned}
\end{equation}
and so are the bottom members of representations of the symmetry algebra. In addition to this, unitarity imposes the following bounds on the spectrum
\begin{equation}
    h\geq |j| \ , \ \ h\geq J \ ,
\end{equation}
where $J$ is the eigenvalue of the quadratic Casimir of $SU(2)_L$, with equivalent bounds for the right-moving states.

A primary that saturates this unitarity bound with $h=j$ is called a chiral primary\footnote{More precisely $h=j$ defines a left-chiral state. Right-chiral states have $\tilde{h}=\tj$ and are annihilated by $\tilde{G}^{\dot{+}A}_{-\frac12}$.}; these are states which satisfy \eqref{eq.primaryDef} along with the additional condition
\begin{equation} \label{eq.chiraldef}
    G^{+A}_{-\frac12}|\phi\rangle = 0 \ .
\end{equation}
From unitarity constraints on a strand of length $a$, \textit{i.e.}\ on the Hilbert space $\mathcal{H}_a$, one finds that chiral primaries exist only at dimensions
\begin{equation} \label{eq.CPdims}
    h = j = \frac{a-1}{2},\ \frac{a}{2},\ \frac{a+1}{2} \ .
\end{equation}
In the case of the $T^4$ symmetric orbifold theory we denote the left-moving chiral primary states by
\begin{equation} \label{eq.T4LCP}
    \begin{aligned}
        |-\rangle_a\ \ &:\ \ h=j=\frac{a-1}{2}\ ,\\
        |\Ad\rangle_a\ \ &:\ \ h=j=\frac{a}{2}\ ,\\
        |+\rangle_a \ \ &:\ \ h=j=\frac{a+1}{2}\ ,
    \end{aligned}
\end{equation}
with the analogous right-moving chiral primaries being 
\begin{equation} \label{eq.T4RCP}
    \begin{aligned}
        |\dot{-}\rangle_a\ \ &:\ \ \tilde{h}=\tj=\frac{a-1}{2}\ ,\\
        |\Ad\rangle_a\ \ &:\ \ \tilde{h}=\tj=\frac{a}{2}\ ,\\
        |\dot{+}\rangle_a \ \ &:\ \ \tilde{h}=\tj=\frac{a+1}{2}\ .
    \end{aligned}
\end{equation}
The full set of chiral-chiral primaries is then any combination of \eqref{eq.T4LCP} and \eqref{eq.T4RCP}.

BPS states are given in the NS-NS sector of the symmetric orbifold theory by tensoring left- and right-moving parts as
\begin{align}
    \text{$\tfrac12$-BPS} &: \ \ |\mathrm{chiral\ primary}\rangle \; |\overline{\mathrm{chiral\ primary}}\rangle \ , \label{eq.1/2BPSNSNS}\\[.5ex]
    \text{$\tfrac14$-BPS} &: \ \ |\mathrm{anything}\rangle \; |\overline{\mathrm{chiral\ primary}}\rangle \ ,\label{eq.1/4BPSNSNS}
\end{align}
where in the D1-D5 CFT $\frac12$-BPS states preserve 8 real supersymmetries, whereas $\frac14$-BPS states preserve 4 (due to \eqref{eq.chiraldef} and its right-moving equivalent).\footnote{Note that, in some literature, $\frac12$-BPS and $\frac14$-BPS states here are referred to as $\frac14$-BPS and $\frac18$-BPS states, respectively, counting the number of supersymmetries relative to the 32 supersymmetries of type IIB theory in 10 dimensions.} On a general primary state $|\phi\rangle$ one can act with the supersymmetry modes $G^{\alpha A}_{-\frac{1}{2}}$ to fill out a long SUSY multiplet, schematically of the form
\begin{equation}
    (1)\ \ |\phi\rangle\,,\quad\ 
    (4)\ \ G|\phi\rangle\,,\quad\  
    (6)\ \ G^2|\phi\rangle\,,\quad\ 
    (4) \ \ G^3|\phi\rangle\,,\quad\ 
    (1)\ \  G^4|\phi\rangle\ ,
\end{equation}
with multiplicities of these states in parentheses. Due to \eqref{eq.chiraldef}, left-chiral primary states are bottom members of short supersymmetry multiplets filled out by $G^{-A}_{-\frac12}$ and of the schematic form
\begin{equation} \label{eq.shortrep}
    (1)\ \ |\phi\rangle\,,\quad\
    (2)\ \ G|\phi\rangle\,,\quad\ 
    (1)\ \ G^2|\phi\rangle \ .
\end{equation}
Right-moving states also form analogous short and long multiplets generated by the right-moving supercharges.

Under the spectral flow transformation of the $\mathcal{N}=4$ algebra generators \eqref{eq.symmModeDeform}, the quantum numbers $h,j$ of states in the Hilbert space will also transform. Namely by
\begin{equation} \label{eq.hjSF}
    h(\eta) = h +\eta j + \frac{c}{24}\eta^2 \ , \qquad 
    j(\eta) = j + \frac{c}{12}\eta \ .
\end{equation}
Using $\eta=-1$ to flow to the Ramond sector, the chiral primary states in the NS sector (with $h=j$) map to the highly degenerate Ramond ground states with $h_R=\frac{c}{24}$. BPS states in the R-R sector of the symmetric orbifold theory are given by states of the form (simply the spectral flow of the NS-NS sector BPS states \eqref{eq.1/2BPSNSNS} and \eqref{eq.1/4BPSNSNS})
\begin{align}
    \text{$\tfrac12$-BPS} &: \ \ |\mathrm{R\ ground\ state}\rangle\; |\overline{\mathrm{R\ ground\  state}}\rangle \ ,\label{eq.1/2BPSRR}\\[.5ex ]
    \text{$\tfrac14$-BPS} &: \ \ |\mathrm{anything}\rangle\; |\overline{\mathrm{R\ ground\ state}}\rangle \ .\label{eq.1/4BPSRR}
\end{align}

\subsubsection*{$\mathrm{Sym}^N(K3)$}

While there are differences in the above exposition of the symmetric orbifold theory for the case of $\cM_4=K3$, for the requirements of the present paper only the changes to the symmetry algebra and the $\frac12$-BPS spectrum are needed. 

The symmetry algebra of $\mathrm{Sym}^N(K3)$ is just the small $\cN=(4,4)$ superconformal algebra with central charge $c=6N$, involving the current modes \eqref{eq.currentModes} with the mode algebra \eqref{eq.commcurrents2}.

For the BPS spectrum we can use the description of $K3$ as $T^4/\mathbb{Z}_2$. In this description $\mathbb{Z}_2$ acts on the free bosons and fermions of the seed sigma model in \eqref{eq.XpsiFields} as
\begin{equation}
     X^{\Ad A} \to -X^{\Ad A}  \ ,\qquad \psi^{\alpha \Ad}\to -\psi^{\alpha \Ad}\ ,\qquad \tilde{\psi}^{\ald \Ad}\to-\tilde{\psi}^{\ald \Ad} \ ,
\end{equation}
with the Hilbert space of the seed theory decomposes into a $\mathbb{Z}_2$-invariant sector and a twisted sector under the $\mathbb{Z}_2$ action\footnote{This Hilbert space decomposition is much like the symmetric orbifold Hilbert space given in \eqref{eq.SymOrbH}, though with a $\mathbb{Z}_2$ orbifold theory instead of $S_N$.}. With this description of the seed theory on $T^4/\mathbb{Z}_2$, the symmetric orbifold $\mathrm{Sym}^N(T^4/\mathbb{Z}_2)$ is defined in the same way as for $\mathrm{Sym}^N(T^4)$. To understand the single-strand $\frac12$-BPS states of $\mathrm{Sym}^N(K3)$ it is sufficient to look at the BPS spectrum of $\mathrm{Sym}^N(T^4/\mathbb{Z}_2)$: the set of chiral-chiral primaries is given by
\begin{equation} \label{eq.K3LCP}
    \begin{aligned}
        |{-}\dot{\pm}\rangle_a\ \ &:\ \ h=j=\frac{a-1}{2}\ ,\ \tilde{h}=\tj=\frac{a\pm1}{2}\ ,\\
        |I\rangle_a\ \ &:\ \ h=j=\tilde{h}=\tj=\frac{a}{2}\ ,\\
        |{+}\dot{\pm}\rangle_a \ \ &:\ \ h=j=\frac{a+1}{2}\ ,\ \tilde{h}=\tj=\frac{a\pm1}{2}\ ,
    \end{aligned}
\end{equation}
where the $|I\rangle_a$ contain the 4 states $|\Ad\Bd\rangle_a$, as well as 16 states $|r\rangle_a$ coming from the $\mathbb{Z}_2$ twisted sector.

\subsubsection*{$\mathrm{Sym}^N(\cM_4)$ moduli space}

Due to the existence of 20 exactly marginal operators which preserve the superconformal algebra, the D1-D5 CFT has a 20-dimensional moduli space. One such modulus in the doubly-wound sector is equivalent to turning on R flux in the bulk language. Under flow in this conformal manifold the spectrum changes; in fact even the $\frac14$-BPS spectrum is not invariant under such deformations of the symmetric orbifold theory. This occurs due to the possibility for short multiplets of the free symmetric orbifold theory (\textit{i.e.}\ of the right-chiral states in \eqref{eq.1/4BPSNSNS}) to combine into long multiplets in the deformed theory. The associated BPS states thus gain anomalous dimensions and ``lift".

In other words, if the coupling associated to a particular modulus is $\lam$, despite the right-moving supercharges $\tilde{G}^{\dot{+}A}_{-\frac12}$ annihilating states of the form \eqref{eq.1/4BPSNSNS} at $\lambda=0$, in the theory that has been deformed such states are not necessarily annihilated by the deformed supercharges $\tilde{G}^{\dot{+}A}_{-\frac12}(\lambda)$. This phenomenon of lifting has been studied for the D1-D5 CFT variously in \cite{Gava:2002xb,Gaberdiel:2015uca,Hampton:2018ygz,Benjamin:2021zkn,Guo:2019ady,Guo:2020gxm,Guo:2022ifr,Hughes:2023apl,Hughes:2023fot,Fabri:2025rok}. A partial lifting of the BPS spectrum of the symmetric orbifold theory is also expected from the point of view of holography; it is the mechanism that allows the Hagedorn growth of BPS spectrum of $\mathrm{Sym}^N(\cM_4)$ to be compatible with the sub-exponential growth of the KK spectrum of supergravity on AdS$_3\times S^3$ (see \cite{Benjamin:2016pil} for a more detailed discussion on this).

\subsubsection*{Supergravity}

The D1-D5 CFT in a strongly coupled region of its moduli space and at large $N$ has a description in terms of supergravity on AdS$_3\times S^3$, the field content of which comes from the massless modes of the internal $\cM_4$ compactification. For $\cM_4=T^4$ and $K3$ the relevant AdS$_3\times S^3$ supergravities are the maximal $\cN=(2,2)$ and half-maximal $\cN=(2,0)$ theories with $n_T=5$ and $n_T=21$ tensor multiplets respectively. The KK modes obtained from the $S^3$ correspond to supergraviton excitations in $\mathrm{AdS}_3$, the spectra of which decompose into short representations of the global subalgebra $SU(1,1\,|\,2)_L\times SU(1,1\,|\,2)_R$, where the generators of $SU(1,1\,|\,2)_L$ are given in \eqref{eq.globalSubAlg}. The most general multi-particle supergraviton Hilbert space is then generated by arbitrary products of these KK modes.

In \cite{Maldacena:1998bw,deBoer:1998ip,Deger:1998nm,Larsen:1998xm} the KK spectrum on $S^3$ of these supergravity theories was determined via representation theory and matched to the chiral primary spectrum of the symmetric orbifold $\mathrm{Sym}^N(\cM_4)$. Matching at finite $N$ requires the constraint on the multi-particle supergraviton Hilbert space coming from the stringy exclusion principle \cite{Maldacena:1998bw}.

The multi-particle $\frac12$-BPS supergraviton spectrum is obtained from arbitrary tensor products of single-particle $\frac12$-BPS supergravitons, \textit{i.e.}\ the lowest-weight states of the short $SU(1,1\,|\,2)_L\times SU(1,1\,|\,2)_R$ multiplets in the KK spectrum, or when realised in the CFT just single-strand chiral-chiral primaries of the form \eqref{eq.1/2BPSNSNS}. Given the set of such $\frac12$-BPS single-particle supergraviton states $\{|\psi\rangle\}$, the multi-particle $\frac12$-BPS supergraviton states are of the form
\begin{align} \label{eq.1/2BPSsupergrav}
    \prod_{\psi} \Big(|\psi\rangle\Big)^{N_\psi} \ .
\end{align}
The $\frac14$-BPS multi-supergravitons are obtained from products of the full short $SU(1,1\,|\,2)_L$ multiplets making up the KK spectrum. These are realised in the CFT by states of the form
\begin{equation} \label{eq.1/4BPSsupergrav}
\begin{aligned}
    &\prod_{\psi,n,m} \Big[\big(L_{-1}\big)^n\big(J^-_{0}\big)^m|\psi\rangle\Big]^{N_{\psi,n,m}} \ , \\
    &\prod_{\psi,n,m} \Big[\big(L_{-1}\big)^n\big(J^-_{0}\big)^m G^{-A}_{-\frac12}|\psi\rangle\Big]^{N^{'}_{\psi,n,m}} \ ,\\
    &\prod_{\psi,n,m} \Big[\big(L_{-1}\big)^n\big(J^-_{0}\big)^m \big( G^{-1}_{-\frac12}G^{-2}_{-\frac12} - \tfrac1{2h}L_{-1}J^{-}_0\big)|\psi\rangle\Big]^{N^{''}_{\psi,n,m}} \ ,
\end{aligned}
\end{equation}
where $h$ is the conformal weight of $\ket{\psi}$. 

The full multi-particle Hilbert space at finite $N$ takes the form
\begin{align} \label{eq.sgravH}
    \cH^{\mathrm{graviton}}_{N} = \bigoplus_{\{j_,\tj_i;d_i\}}{}^{\!\!\!\!\!\!\!\!'} \ \,\bigotimes_{i}\, (j_i,\tj_i;d_i)_s \ ,
\end{align}
where $(j_i,\tj_i;d_i)_s$ are the $SU(1,1\,|\,2)_L\times SU(1,1\,|\,2)_R$ short multiplets of the single-particle $S^3$ KK spectrum. Each multiplet is assigned the integer $d_i$ (historically referred to as the degree) which is nothing but the strand length of the corresponding state in the CFT description. The tensor sum $\bigoplus{}^{\!'}$ in \eqref{eq.sgravH} is restricted to products of single-particle KK modes with a total strand length $d=\sum_i d_i\leq N$ in line with \cite{deBoer:1998us,Maldacena:1998bw}. This is often referred to as the stringy exclusion principle.

\subsection{Elliptic genera}
\label{ssec:index}

The partition function of a CFT$_2$ with at least $\cN=(2,2)$ superconformal symmetry and central charge $c$ can be written using the Ramond-Ramond sector Hilbert space as
\begin{equation} \label{eq.ZRCFT}
    Z^{\mathrm{CFT}}_{R}(q,y,\tq,\ty) \equiv \Tr_{\cH^{\mathrm{CFT}}_R} [(-1)^{F} q^{L_0-\frac{c}{24}} y^{2J^3_0} \tq^{\tilde{L}_0-\frac{c}{24}} \ty^{2\tilde{J}^3_0} ]\ ,
\end{equation}
where the fermion number operator here is $F=2J^3_0-2\tilde{J}^3_0$ and in our conventions the various fugacities are
\begin{align} \label{eq.qydefs}
    q=e^{2\pi i\tau}\ ,\ \ y=e^{2\pi i\nu}\ ,\ \ \tq=e^{2\pi i\tilde{\tau}}\ ,\ \ \ty=e^{2\pi i\tilde{\nu}} \ .
\end{align}
We have included a factor of $(-1)^F$ in the partition function (sometimes referred to as ``signed" or ``twisted" partition function) in order to connect with the elliptic genus which this paper is focussed on. However, setting $y=\ty=-1$ in \eqref{eq.ZRCFT} still recovers a true count of states.

From this twisted partition function the elliptic genus, an index for $\frac14$-BPS states can be defined as
\begin{align} \label{eq.EGRCFT}
    \cE^{\mathrm{CFT}}_{R}(q,y) \equiv Z^{\mathrm{CFT}}_{R}(q,y,1,1) = \Tr_{\cH^{\mathrm{CFT}}_R} [(-1)^{F} q^{L_0-\frac{c}{24}} y^{2J^3_0}] \ .
\end{align}
Non-BPS states form long supersymmetry multiplets on both the left and the right, the contribution of which vanishes in the elliptic genus. Analogous formulae can be defined for the NS-NS sector of the theory, obtained from spectral flowing \eqref{eq.ZRCFT} and \eqref{eq.EGRCFT}: using \eqref{eq.symmModeDeform} with $\eta=1$ (along with the analogous right-moving flow) the resulting NS twisted partition function is
\begin{equation} \label{eq.ZNSCFT}
    Z^{\mathrm{CFT}}_{\mathrm{NS}}(q,y,\tq,\ty) \equiv (y\ty)^{\frac{c}{6}}(q\tq)^{\frac{c}{12}}Z^{\mathrm{CFT}}_{\mathrm{R}}(q,yq^{\frac12},\tq,\ty\tq^{\frac12}) = \Tr_{\cH^{\mathrm{CFT}}_{\mathrm{NS}}} [(-1)^{F} q^{L_0}y^{2J^3_0} \tq^{\tilde{L}_0} \ty^{2\tilde{J}^3_0}] \ ,
\end{equation}
where the generators on the right-hand side are in the NS sector and we have taken the unique NS-NS ground state to have $h=\tilde{h}=j=\tj=0$. The NS sector elliptic genus is then given by
\begin{align} \label{eq.EGNSCFT}
    \cE^{\mathrm{CFT}}_{\mathrm{NS}}(q,y) \equiv Z^{\mathrm{CFT}}_{\mathrm{NS}}(q,y,1,1) = \Tr_{\cH^{\mathrm{CFT}}_{\mathrm{NS}}} [(-1)^{F} q^{L_0} y^{2J^3_0}] \ ,
\end{align}
which receives contributions from states of the form \eqref{eq.1/2BPSNSNS} and \eqref{eq.1/4BPSNSNS}. In this paper we will be primarily focussed on the BPS spectrum in the NS sector since that allows for direct comparison with the supergraviton spectrum.

On the other hand, for the seed sigma model with $T^4$ target space the elliptic genus vanishes due to every left-moving state coming with a quartet generated in the NS sector by the right-moving fermion modes $\tilde{\psi}^{+\Ad}_{-\frac12}$ (in the R sector by the zero modes $\tilde{\psi}^{+\Ad}_{0}$). The contribution to the elliptic genus of each such quartet vanishes identically. The correct choice of $\frac14$-BPS index for the $T^4$ theory is instead the modified elliptic genus \cite{Maldacena:1999bp}
\begin{align} \label{eq.EMEGNS}
    \hat{\cE}^{\mathrm{CFT}}_{\mathrm{NS}}(q,y) \equiv \frac12 \big(\ty\pd_{\ty}\big)^2 Z^{\mathrm{CFT}}_{\mathrm{NS}}(q,y,\tq,\ty)\Big|_{\tq=\ty=1} = \frac12\Tr_{\cH_{\mathrm{CFT}}^{\mathrm{NS}}} \Big[(-1)^{F} q^{L_0} y^{2J^3_0} \big(2\tilde{J}^3_0\big)^2\Big] \ ,
\end{align}
with the NS partition function $Z^{\mathrm{CFT}}_{\mathrm{NS}}$ defined in \eqref{eq.ZNSCFT}. Terms inside the trace of \eqref{eq.EMEGNS} that are linear in $\tilde{J}^3_0$ coming from the action of the $\ty$ derivative dropped out since such contributions from a right-moving fermion quartet also vanish. To this modified elliptic genus, a given right-moving quartet generated by $\tilde{\psi}^{+\Ad}_{-\frac12}$ on a state with quantum numbers $h,j,\tj$ contributes
\begin{align}
    \hat{\cE}^{\mathrm{CFT}}_{\mathrm{NS}}(q,y) &\supset \frac12(-1)^{2j}q^{h} y^{2j} \Big[(-1)^{2\tj}(2\tj)^2 + 2(-1)^{2\tj+1}(2\tj+1)^2 + (-1)^{\tj+2}(2\tj+2)^2\Big] \nonumber\\
    &= (-1)^{2j-2\tj}q^{h} y^{2j} \ .
\end{align}
This modification to the elliptic genus thus effectively absorbs the right-moving degeneracy coming from right-moving fermion quartets present in the $T^4$ theory.

\subsubsection{$K3$} \label{sssec:K3index}

The elliptic genus of the symmetric orbifold theory $\mathrm{Sym}^N(K3)$ can be computed from the seed sigma model elliptic genus: in the R sector this is given by
\begin{align} \label{eq.EGRseed}
    \vep^{\mathrm{CFT}}_{R,K3}(q,y) = 8\sum_{i=2}^4 \frac{\vartheta_i(\nu,\tau)}{\vartheta_i(0,\tau)} = \sum_{m,\ell} c_{R}^{K3}(m,\ell)\, q^m y^{\ell} \ ,
\end{align}
where $\nu$ and $\tau$ are related to $q$ and $y$ via \eqref{eq.qydefs}. In \eqref{eq.Jtheta} the $\vartheta_i$ are defined; these are the four Jacobi theta functions with characteristics and definite parity under $\nu\to-\nu$~\cite{DHoker:2022dxx}. The expansion coefficients $c_R^{K3}(m,\ell)$ for this index are in fact only dependant on one combination of the parameters $m$ and $\ell$, namely\footnote{This property of the expansion coefficients of the elliptic genus is a consequence of it being a weak Jacobi form of weight $0$ and index $\frac{c}{6}$. For the seed theory $c_{\mathrm{seed}}/6=1$.}
\begin{equation} \label{eq.cR}
    c_R^{K3}(m,\ell) = c(4m-\ell^2) \ ,
\end{equation}
with the coefficients vanishing for $4m-\ell^2<-1$.

From this seed theory elliptic genus, the generating function of elliptic genera of $\mathrm{Sym}^N(K3)$ is then given by \cite{Dijkgraaf:1996xw}
\begin{equation} \label{eq.cftEGgenR}
    \cZ_{R,K3}^{\mathrm{CFT}}(p,q,y) \equiv \sum_{N=1}^{\infty} p^N \cE^{\mathrm{CFT}}_{R}(q,y;N) = \prod_{n,m,\ell} \big(1-p^nq^my^{\ell}\big)^{-c_{R}^{K3}(nm,\ell)} \ ,
\end{equation}
where the product ranges over $n\in\mathbb{Z}_{\geq1}$, $m\in\mathbb{Z}_{\geq0}$, $\ell\in\mathbb{Z}$. The generating function of NS-sector elliptic genera can then be defined in a similar way from
\begin{equation} \label{eq.cftEGgenNS}
    \cZ_{\mathrm{NS},K3}^{\mathrm{CFT}}(p,q,y) \equiv \sum_{N=1}^{\infty} p^N \cE^{\mathrm{CFT}}_{\mathrm{NS}}(q,y;N) = \prod_{n,m,\ell} \big(1-p^nq^my^{\ell}\big)^{-c_{\mathrm{NS}}^{K3}(n,m,\ell)} \ ,
\end{equation}
where the NS coefficients $c_{\mathrm{NS}}(n,m,\ell)$ are given in terms of those in \eqref{eq.cR} by
\begin{equation} \label{eq.cNS}
    c_{\mathrm{NS}}^{K3}(n,m,\ell) = c(4nm-n^2-\ell^2) \ .
\end{equation}
The product in \eqref{eq.cftEGgenNS} is over the values $n\in\mathbb{Z}_{\geq1}$, $2m\in\mathbb{Z}_{\geq0}$, $m-\frac{\ell}{2}\in\mathbb{Z}_{\geq0}$, $m\geq\frac{|\ell|}{2}$.

By using the supergraviton Hilbert space \eqref{eq.sgravH}, and in particular the CFT realisation of the states in \eqref{eq.1/4BPSsupergrav}, one can also define a supergraviton elliptic genus. For the purposes of identifying contributions to the supergraviton elliptic genus from individual states in section~\ref{sec:K3} we choose to refine\footnote{This type of refined elliptic genus for the CFT is also sometimes referred to as the Hodge elliptic genus~\cite{Kachru:2016igs}. Some of its properties were studied in \cite{Benjamin:2016pil,Benjamin:2017rnd}.} by the right-moving R-charge $\tj$. This right-moving quantum number is counted by the fugacity $\ty$ and we decompose states into representations of $SU(2)_R$.

This refined index can be computed starting from the single-particle index
\begin{align} \label{eq.K31ptEG}
    z_{K3}^{\mathrm{graviton}}(p,q,y,\ty) &= \sum _{k=1}^{\infty} p^k \bigg[\phi^{(s)}_{\frac{k-1}{2}}\!(q,y)\,\tilde{\chi}_{\frac12;k}(\ty) + 20\phi^{(s)}_{\frac{k}{2}}\!(q,y)\,\tilde{\chi}_{0;k}(\ty) + \phi^{(s)}_{\frac{k+1}{2}}\!(q,y)\,\tilde{\chi}_{\frac12;k}(\ty) \bigg] \nonumber\\
    &= \frac{p}{(1-q)(1-y^2)} \left[\frac{\big(1-\sqrt{q} y\big)^2 \Big(\tilde{\chi}_{\frac12;k}(\ty) + 20\sqrt{q}y^{-1}\tilde{\chi}_{0;k}(\ty) + qy^{-2}\tilde{\chi}_{\frac12;k}(\ty)\Big)}{1-p \sqrt{q}y^{-1}}\right. \nonumber\\
    &\hspace{2cm}- \left.\frac{y^2\big(1-\sqrt{q} y^{-1}\big)^2 \Big(\tilde{\chi}_{\frac12;k}(\ty) + 20\sqrt{q}y\tilde{\chi}_{0;k}(\ty) + qy^{2}\tilde{\chi}_{\frac12;k}(\ty) \Big)}{1-p \sqrt{q}y} \right] \nonumber\\
    &= \sum_{n,m,\ell,\tilde{\ell}} c^{K3}_{\mathrm{graviton}}(n,m,\ell,\tilde{\ell})\, p^nq^my^{\ell}\ty^{\tilde{\ell}} \ ,
\end{align}
where $\phi^{(s)}_{j}(q,y)$ is a short $SU(1,1\,|\,2)_L$ character given in \eqref{eq.globalN4s}, built on a $h=j$ left-chiral primary, and $\tilde{\chi}_{\tj}(\ty)$ is an $SU(2)_R$ character, defined in \eqref{eq.SU2Char}. Note we define characters \emph{without} the $(-1)^{2j}$  factor in front; namely, it starts with a positive sign, even when the highest-weight state is fermionic.
Defining
\begin{equation} \label{eq.XK3def}
    X_{k}^{K3}(q,y,\ty) \equiv \phi^{(s)}_{\frac{k-1}{2}}\!(q,y)\,\tilde{\chi}_{\frac12;k}(\ty) + 20\phi^{(s)}_{\frac{k}{2}}\!(q,y)\,\tilde{\chi}_{0;k}(\ty) + \phi^{(s)}_{\frac{k+1}{2}}\!(q,y)\,\tilde{\chi}_{\frac12;k}(\ty) \ ,
\end{equation}
the generating function of multi-particle (refined) elliptic genera is then the plethystic exponential of the single-particle index
\begin{align} \label{eq.K3NptEG}
    \cZ_{K3}^{\mathrm{graviton}}(p,q,y,\ty) &\equiv \sum_{N=1}^{\infty} p^N Z_{K3}^{\mathrm{graviton}}(q,y,\ty;N) \nonumber\\
    &= \mathrm{PE}\big[z_{K3}^{\mathrm{graviton}}\big](p,q,y,\ty) \nonumber\\
    &\equiv \exp\bigg(\sum_{n=1}^{\infty}\frac1{n}\sum_{k=1}^{\infty} p^{nk}X^{K3}_{k}(q^n,y^n,\ty^n)  \bigg) \nonumber\\
    &= \prod_{n,m,\ell,\tilde{\ell}} \Big(1-p^nq^my^{\ell}\ty^{\tilde{\ell}}\Big)^{-c^{K3}_{\mathrm{graviton}}(n,m,\ell,\tilde{\ell})} \ .
\end{align}
To recover the standard elliptic genus one can simply set $\ty=1$ in the refined index
\begin{equation} \label{eq.K3gravEGdef}
    \cE^{\mathrm{graviton}}_{K3}(q,y;N) = Z_{K3}^{\mathrm{graviton}}(q,y,1;N) \ .
\end{equation}

\subsubsection{$T^4$} \label{sssec:T4index}

The partition function of $\mathrm{Sym}^N(T^4)$ can be computed from the seed sigma model partition function: in the Ramond sector this is given by\footnote{In this paper we do not consider the contribution of states with target space momentum or winding. See, for example, \cite{ArabiArdehali:2024lyz} for a recent discussion of these topologically charged sectors.} \cite{Maldacena:1999bp}
\begin{align} \label{eq.ZseedT4}
    z^{\mathrm{CFT}}_{R,T^4}(q,y,\tq,\ty) = \bigg(\frac{\vartheta_1(\nu,\tau)}{\eta(q)^3}\bigg)^{2} \bigg(\frac{\vartheta_1(\tilde{\nu},\tilde{\tau})}{\eta(\tq)^3}\bigg)^{2} = \sum_{m,\ell,\tilde{m},\tilde{\ell}} c_{R}^{T^4}(m,\ell,\tilde{m},\tilde{\ell})\, q^my^{\ell}\tq^{\tilde{m}}\ty^{\tilde{\ell}} \ ,
\end{align}
where $\nu$, $\tau$, $\tilde{\nu}$, $\tilde{\tau}$ are related to the fugacities $q$, $y$, $\tq$, $\ty$ via \eqref{eq.qydefs}. This can then be used to construct the generating function for partition functions of $\mathrm{Sym}^N(T^4)$ with a small modification of \eqref{eq.cftEGgenR}
\begin{align} \label{eq.ZNT4}
    \cZ^{\mathrm{CFT}}_{R,T^4}(p,q,y,\tq,\ty) \equiv \sum_{N=1}^{\infty} p^N Z^{\mathrm{CFT}}_{R,T^4}(q,y,\tq,\ty;N) = \prod_{n,m,\ell,\tilde{m},\tilde{\ell}} \big(1-p^nq^my^{\ell}\tq^{\tilde{m}}\ty^{\tilde{\ell}}\big)^{-c_{R}^{T^4}(nm,\ell,n\tilde{m},\tilde{\ell})} \ .
\end{align}
The equivalent NS sector partition function generating function is then given by spectral flow
\begin{align} \label{eq.ZNT4NS}
    Z^{\mathrm{CFT}}_{\mathrm{NS},T^4}(q,y,\tq,\ty;N) = (y\ty)^{\frac{c}{6}}(q\tq)^{\frac{c}{12}} Z^{\mathrm{CFT}}_{R,T^4}(q,y q^{\frac12},\tq,\ty \tq^{\frac12};N) \ ,
\end{align}
with $c=6N$. From this partition function the modified elliptic genus in the NS sector $\hat{\cE}^{\mathrm{CFT}}_{\mathrm{NS}}$ can then be found by the use of \eqref{eq.EMEGNS}.

From the supergraviton Hilbert space \eqref{eq.sgravH} we can define a partition function, starting from the single-particle partition function
\begin{align} \label{eq.T41ptPF}
    z_{T^4}^{\mathrm{graviton}}(p,q,y,\tq,\ty) &= \sum_{k=1}^{\infty} p^k \bigg[\phi^{(s)}_{\frac{k-1}{2}}(q,y) - 2\phi^{(s)}_{\frac{k}{2}}(q,y) + \phi^{(s)}_{\frac{k+1}{2}}(q,y) \bigg] \,\omega_k(\tq,\ty) \nonumber\\
    &= \sum_{n,m,\ell,\tilde{m},\tilde{\ell}} c_{T^4}^{\mathrm{graviton}}(n,m,\ell,\tilde{m},\tilde{\ell})\, p^nq^my^{\ell} \tq^{\tilde{m}}\ty^{\tilde{\ell}} \ ,
\end{align}
where we define the contribution from the choice of right-moving chiral primaries \eqref{eq.T4RCP} as
\begin{equation} \label{eq.omegakn}
    \omega_k(\tq^n,\ty^n) \equiv \tq^{n\frac{k-1}{2}}\ty^{n(k-1)}\big( 1 - \tq^{\frac{n}{2}}\ty^n \big)^2 \ .
\end{equation}
Defining
\begin{equation} \label{eq.XT4def}
    X_k^{T^4}(q,y) \equiv \phi^{(s)}_{\frac{k-1}{2}}(q,y) - 2\phi^{(s)}_{\frac{k}{2}}(q,y) + \phi^{(s)}_{\frac{k+1}{2}}(q,y) \ ,
\end{equation}
the generating function for the multi-supergraviton partition function is the plethystic exponential of \eqref{eq.T41ptPF}
\begin{align} \label{eq.T4NptPF}
    \cZ ^{\mathrm{graviton}}_{T^4}(p,q,y,\tq,\ty) &\equiv \sum_{N=1}^{\infty} p^N Z_{T^4}^{\mathrm{graviton}}(q,y,\tq,\ty;N) \nonumber\\
    &= \mathrm{PE}\big[z_{T^4}^{\mathrm{graviton}}\big](p,q,y,\tq,\ty) \nonumber\\
    &= \exp\bigg(\sum_{n=1}^{\infty}\frac{1}{n}\sum_{k=1}^{\infty} p^{nk}X^{T^4}_{k}\!(q^n,y^n) \,\omega_k(\tq^n,\ty^n) \bigg) \nonumber\\
    &= \prod_{n,m,\ell,\tilde{m},\tilde{\ell}} \Big(1-p^nq^my^{\ell} \tq^{\tilde{m}} \ty^{\tilde{\ell}}\Big)^{-c_{T^4}^{\mathrm{graviton}}(n,m,\ell,\tilde{m},\tilde{\ell})} \ .
\end{align}
A modified elliptic genus $\hat{\cE}^{\mathrm{graviton}}_{T^4}$ can then be defined from the supergraviton partition function $Z^{\mathrm{graviton}}_{T^4}$  analogously to \eqref{eq.EMEGNS}.

\subsubsection{The de Boer bound}
Whilst the $\frac12$-BPS spectrum in both the $K3$ and $T^4$ theories is protected under deformations of the symmetric orbifold theories on their conformal manifolds, the $\frac14$-BPS spectrum is not invariant. In other words the CFT partition function \eqref{eq.ZRCFT} restricted to BPS states varies with the coupling $\lambda$ of the symmetric orbifold theory to an exactly marginal operator. The elliptic genus \eqref{eq.EGRCFT}, however, is invariant under such flows in the moduli space. This is because the contribution of free theory ($\lambda=0$) short multiplets which lift by combining into a long multiplet in the deformed theory ($\lambda\neq0$) identically cancel.

In \cite{deBoer:1998us}, it was shown that the $K3$ NS CFT elliptic genus \eqref{eq.cftEGgenNS} is reproduced by the supergraviton elliptic genus (using the Hilbert space \eqref{eq.sgravH}) for states with conformal dimension
\begin{align} \label{deBoer_bound_K3}
    h\leq\frac{N+1}{4} \qquad\qquad (K3) \ .
\end{align}
We will call this the \emph{de Boer bound} for $\mathrm{Sym}^N(K3)$. In other words, the first non-supergraviton states in the CFT appear at $h>\frac{N+1}{4}$. The equivalent bound for $\mathrm{Sym}^N(T^4)$ when comparing the NS CFT modified elliptic genus \eqref{eq.EMEGNS} with that of the relevant supergraviton spectrum is given by \cite{Maldacena:1999bp}
\begin{align} \label{deBoer_bound_T4}
    h<\frac{N}{4}\qquad\qquad (T^4)\ .
\end{align}

\subsection{Total affine modes}
\label{ssec:totalaffine}

In the above, we saw how BPS supergravitons are realised in the CFT Hilbert space, and introduced partition functions that count them.  However, those supergraviton states are not the only BPS states realised within supergravity.
There are ``boundary diffeomorphism'' modes that do not vanish at the infinity of $\rm AdS_3$ \cite{Brown:1986nw}, sometimes called singletons, and their $\cN=4$ extensions.
These modes correspond to total modes in CFT, including affine generators, and exist everywhere in the CFT moduli space
\cite{Maldacena:1999bp}.\footnote{One could consider quotienting the theory by the total algebra and define partition functions in the quotient theory \cite{Chang:2025rqy}, but we do not take that approach.}
For precise counting of states realised in $\rm AdS_3$ supergravity, we must include them and define \emph{generalised supergraviton partition function and index}.

These total affine modes are not at all exotic as they might seem -- in fact, some of them are already in the supergraviton index \cite{deBoer:1998ip,deBoer:1998us}.  For example, using the commutation relation~\eqref{LJcomm2}, one can show that
\begin{align}
    J_{-(n+1)}^+\ket{-}_1={1\over n!}L_{-1}^n\ket{+}_1\ ,\qquad n\ge 0\ ,
\end{align}
where we only wrote the left-moving part of the state.  
Therefore, the states obtained by acting with the affine modes $J_{-(n+1)}^+$ on the vacuum strand $\ket{-}_1$ are part of the supergraviton spectrum.  
More generally, the states $J^a_{-(n+1)}\ket{-}_1$, $G^{\alpha A}_{-(n+3/2)}\ket{-}_1$, and $L_{-(n+2)}\ket{-}_1$ give the entire global multiplet built on the chiral primary $\ket{+}_1$. For $T^4$, the states $\psi^{\alpha\dot{A}}_{-(n+1/2)}\ket{-}_1$ give the entire global (short) multiplet built on $\ket{\dot{A}}_1$.

However, states obtained by acting with total affine generators on \emph{general} multi-supergraviton states are not part of the supergraviton spectrum.
We need to include them for a more complete counting of states describable within $\rm AdS_3$ supergravity.

On the bulk side, a general multi-supergraviton state corresponds to multiple particles sitting inside ${\rm AdS}_3$.
Acting with the total affine modes on it corresponds to, while keeping those particles inside $\rm AdS_3$, changing the way the ${\rm AdS}_3$ spacetime (more precisely the full ${\rm AdS}_3\times S^3\times \mathcal{M}_4$ spacetime) is connected to the ``UV'' spacetime.%
\footnote{Supergravity solutions including total affine modes (singleton degrees of freedom) were studied in~\cite{Mathur:2011gz,Mathur:2012tj,Lunin:2012gp,Giusto:2013bda}.}
This is analogous to the Schwarzian modes that appear in nearly ${\rm AdS}_2$  geometries and describe how the $\rm AdS_2$ spacetime is connected to the UV spacetime \cite{Maldacena:2016upp}. Note that
the total modes are not the centre-of-mass degrees of freedom of the D1-D5 branes; the centre-of-mass modes are not part of the ${\rm Sym}^N(\mathcal{M}_4)$ CFT, which describes the Higgs branch of the D1-D5 brane system coming from D1-D5 strings \cite{vafa:1995bm,Maldacena:1999bp}.\footnote{The centre-of-mass degrees of freedom are described by a decoupled $\bbR^4\times T^4$  CFT \cite{vafa:1995bm,Maldacena:1999bp}.
This is similar to the free $U(1)$ part of the 4D $\cN=4$ super $U(N)$ Yang-Mills theory.}

The rest of the paper will be concerned with how to incorporate the contribution from total affine modes.
Before delving into the discussion of $\cN=4$ algebras, let us first discuss some simpler example.

\subsection{Virasoro example}
\label{ssec:VirEx}

In order to demonstrate the idea of generalising a supergraviton partition function, let us consider the $\cN=0$ case as a toy example; namely, we focus on the Virasoro generators $\{L_{-n}\}$.

Let $\ket{h}$ represent a non-vacuum Virasoro primary state on a strand of length one, with conformal weight $h$.  The $SL(2,\bbR)$ (``global Virasoro'') multiplet built on $\ket{h}$ consists of the states
\begin{align}
  \bigl\{\, \ket{h}, L_{-1}\ket{h}, L_{-1}^2\ket{h},\dots\bigr\}
    =\{L_{-1}^n\ket{h}\}_{n=0}^{\infty}\ .
    \label{SL(2,R)_multiplet_on_h}
\end{align}
These are single-graviton states.  The associated ``single-graviton partition function'' is
\begin{align}
    z(q)=q^h(1+q+q^2+\cdots)={q^h\over 1-q}
    =\chi_h^{\rm global}(q)
    =\sum_m c(m)q^m\ ,
\end{align}
where $\chi_h^{\rm global}(q)$ is nothing but the $SL(2,\bbR)$ (``global Virasoro'') character.  The last expression defines $c(m)$ which counts the degeneracy of states with conformal weight~$m$, although in this case it is simply $c(h+n)=1$, $n=0,1,2,\dots$.  If there is only one species of graviton given by \eqref{SL(2,R)_multiplet_on_h}, the multi-graviton states are obtained by tensoring the single-graviton states as
\begin{align}
    \prod_{n\ge 0}\, [(L_{-1})^n \ket{h}]^{N_n}\ .\label{mpos11Mar24}
\end{align}
The $N$-graviton partition function $Z_N(q)$ is computed by introducing the grand partition function
\begin{align}
    Z(q,p) = \sum_N Z_N(q) p^N\ ,
\end{align}
where the fugacity $p$ counts the strand length, which is the same as the graviton number in the present example.  Because we have $c(m)$ different bosons with weight $m$, this is computed as
\begin{align}
 Z(q,p)
=\prod_{m}{1\over  (1-q^m p)^{c(m)}}
=\prod_{n\ge 0}{1\over  1-q^{h+n} p}\ .
\label{multi_Z_N=0}
\end{align}

Thus, we have constructed multi-graviton states \eqref{mpos11Mar24}, which are made from the action of the individual global generator $L_{-1}$. Now we would like to generalise these states by including the action of the total Virasoro generators $L_{-n}^{(\mathrm{T})}=\sum_{i=1}^N L_{-n}^{(i)}$, where $n\ge 1$ and~$i$ is the copy index. One might think that we can simply multiply the multi-graviton partition function~\eqref{multi_Z_N=0} by $\prod_{n\ge 1}(1-q^n)^{-1}$, where $(1-q^n)^{-1}$ comes from $L_{-n}$,  $n\ge 1$, but that would be over-counting because the total Virasoro generators $\{L_{-n}^{(\mathrm{T})}\}$ include $L_{-1}$ which the state~\eqref{mpos11Mar24} already contains.
The proper way to include the action of the total Virasoro generators without over-counting is to first identify total Virasoro primary states  from the states \eqref{mpos11Mar24},\footnote{This is somewhat similar to finding the highest-weight states for the total spin ${\bf s}^{\rm (T)}$ when we add $N$ spins ${\bf s}_i$, $i=1,\dots,N$.} remove all other states, and finally add back in the Virasoro descendants of all these Virasoro primaries.

Let us see how this can go explicitly in the case of $N=2$, namely for 2-graviton states.  We list states according to the total conformal weight $h^{\rm (T)}$,
\begin{align}
\begin{aligned}
 h^{\rm (T)}&=2h  &&:~  (\ket{h})^2\\
 h^{\rm (T)}&=2h+1&&:~  (L_{-1}\ket{h})\ket{h}\\
 h^{\rm (T)}&=2h+2&&:~  (L_{-1}^2\ket{h})\ket{h},~(L_{-1}\ket{h})^2\\
 h^{\rm (T)}&=2h+3&&:~  (L_{-1}^3\ket{h})\ket{h},~(L_{-1}^2\ket{h})(L_{-1}\ket{h})\\
 ...
\end{aligned}\label{mwau28Mar24}
\end{align}
Let us look for Virasoro primaries in this list.
\begin{itemize}
 \item 
 The $h^{\rm (T)}=2h$ state $(\ket{h})^2$ is a total Virasoro primary.
 \item 
 The $h^{\rm (T)}=2h+1$ state
 $(L_{-1}\ket{h})\ket{h}={1\over 2}L_{-1}^{(\mathrm{T})}[(\ket{h})^2]$ is a total descendant.  

 \item At $h^{\rm (T)}=2h+2$, there are two states, but one linear combination is a total descendant of the $h^{\rm (T)}=2h+1$ state by  $L_{-1}^{(\mathrm{T})}$.  We expect that there is a different linear combination that is a total Virasoro primary.  Actually, for that, we need to also include the total descendant of the $h^{\rm (T)}=2h$ state $(\ket{h})^2$ by $L_{-2}^{(\mathrm{T})}$, namely,  $L_{-2}^{(\mathrm{T})}[(\ket{h})^2]=2(L_{-2}\ket{h})\ket{h}$, which is not among the graviton states in \eqref{mpos11Mar24}. In order for the linear combination of these three states,
\begin{align}
\ket{\phi}=a(L_{-1}^2\ket{h})\ket{h}+2b(L_1L_{-1}\ket{h})(L_{-1}\ket{h})+d(L_{-2}\ket{h})\ket{h}\ ,
\end{align}
to be a Virasoro primary, we must require
\begin{align}
\begin{split} L_1^{(\mathrm{T})}\ket{\phi}
 &=((4h+1)a+4hb+3d)(L_{-1}\ket{h})\ket{h}=0\ ,\\
 L_2^{(\mathrm{T})}\ket{\phi}
 &=(6ha+(\tfrac{c}{2}+4h)d)(\ket{h})^2=0\ ,
 \label{lqrv6Apr25}
 \end{split}
 \end{align}
where $c$ is the central charge.  Solving these, we find the linear combination  that is a total Virasoro primary annihilated by both $L_{1}^{(\mathrm{T})}$ and $L_2^{(\mathrm{T})}$.  We can see from~\eqref{lqrv6Apr25} that the  coefficient $d$ is smaller than the other coefficients $a$ and $b$ by $1/c$ at large~$c$, if $h=\cO(1)$.  Namely, mixing-in of the higher mode $L_{-2}$ is a loop effect in gravity.
\end{itemize}
This suggests that, if ignore the $1/c$ effects, we can identify Virasoro primaries with global primaries.  Or, more precisely,
we expect a one-to-one correspondence between Virasoro primaries and global primaries.  Therefore, as far as counting of states is concerned,
we can incorporate the effect of total Virasoro modes by finding total \emph{global} primaries, removing all other states, and replacing the total global primaries with total Virasoro modules.  In terms of partition function, the procedure is:
\begin{enumerate}
    \item [(i)]
Decompose the multi-graviton partition function into a sum of total \emph{global} characters, and 
\item[(ii)]
``Promote'' each total global character to the corresponding total \emph{affine} (Virasoro) character.
\end{enumerate}

Let us follow this procedure for the present example.  First, we decompose the states~\eqref{mpos11Mar24} into the representations of the total global algebra $\{L_n^{(\mathrm{T})}\}_{n=0,\pm 1}$ by expressing $Z_N(q)$ as a sum of total global characters $\chi_{h'}^{\rm global}(q)$ with different values of $h'$. This is quite easy to do because of the simple form of $\chi_{h'}^{\rm global}(q)=q^{h'}/(1-q)$:
\begin{align}
    Z_N(q)={1\over 1-q}\cdot (1-q)Z_N(q) = \sum_{h'} c_{h'} \chi_{h'}^{\rm global} (q) \ ,
\end{align}
where 
\begin{align}
    (1-q)Z_N(q)\equiv \sum_{h'} c_{h'} q^{h'}\ .
\end{align}
Namely, we can read off the global-algebra representation contents simply by expanding $(1-q)Z_N(q)$ in~$q$.  Next, we promote the total global character $\chi_{h'}^{\rm global}(q)$ to the total Virasoro character:
\begin{align}
    Z_N(q)=\sum_{h'} c_{h'} \chi_{h'}^{\rm global} (q) \to \sum_{h'} c_{h'} \chi^{\rm Vir}_{h'} (q)\ ,\qquad \chi^{\rm Vir}_{h'} (q)={q^{h'}\over \prod_{n\ge 1} (1-q^n)}\ .
\end{align}
This simply amounts to
\begin{align}
    Z_N(q)\to {Z_N(q)\over \prod_{n\ge 2} (1-q^n)}\ ,
\end{align}
namely, we drop the contribution of the generator $L_{-1}^{(\mathrm{T})}$.

The above procedure of replacing global characters by affine characters fails if two global multiplets $\chi_{h'}^{\rm global}$ and $\chi_{h''}^{\rm global}$ that appear in the global-character decomposition are related to each other by the action of some affine generator.  
If this is the case, promoting
$\chi^{\rm global}_{h'} \to \chi^{\rm affine}_{h'}$ would add states that have already appeared in $\chi_{h''}^{\rm global}$ and lead to over-counting.  However, this complication does not happen in the present example, because the added states all include higher modes $L_{-n}$ with $n\ge 2$, which are not part of the graviton states \eqref{mpos11Mar24}.  However, we will see that this complication does happen for $\cN=4$ algebras and we have to work harder for a proper counting of states.

\section{$T^4$ generalised supergraviton index}
\label{sec:T4}

As motivated by the toy example in section \ref{ssec:VirEx}, in order to generalise the supergraviton partition function to one that also counts total affine descendants -- multi-supergraviton states of the form \eqref{eq.1/4BPSsupergrav} acted on by the non-global modes of the left-moving $\cN=4$ algebra -- it would be natural to proceed by the following steps:
\begin{enumerate}[start=1,
    labelindent=\parindent,
    leftmargin =\parindent,
    label=\roman*)]
    \item \label{list:i}
    Decompose the multi-supergraviton partition function into characters of the total global $SU(1,1\,|\,2)_L$ algebra. This character expansion identifies the set of total global primaries.
    \item \label{list:ii}
    With the assumption that this set of total global primaries is in one-to-one correspondence with the set of total affine primaries, we can replace each global character with an affine character built on a highest-weight state of the same quantum numbers. We refer to this replacement step as \emph{promotion}.
\end{enumerate}
However, as mentioned in section \ref{ssec:VirEx} this procedure leads to an \emph{over-counting} of states in the resulting generalised supergraviton partition function when there exist total global primary supergraviton states that are related to each other by total affine modes. When this is the case, the above step \ref{list:ii} should be replaced by:
\begin{enumerate}[start=1,
    labelindent=\parindent,
    leftmargin =\parindent,
    label=ii\alph*)]
    \item \label{list:iia}
    Identify global primaries in the decomposition whose multiplets are related by the action of total affine modes. In this set of related characters, discard all but the character whose multiplet has the lowest dimension highest-weight state. We refer to this step as \emph{subtraction}.
    \item \label{list:iib}
    The remaining total global characters in the partition function decomposition have highest-weight states that are also primary with respect to the full (total) affine algebra. These global characters can now be promoted to total affine characters without leading to an over-counting of states. The resulting partition function -- referred to as the ``generalised supergraviton partition function" -- will contain all of the original supergraviton states, along with the new total affine modes described in section \ref{ssec:totalaffine}.
\end{enumerate}
For the contracted large $\cN=4$ affine algebra of the $T^4$ theory, it is indeed necessary to perform such a subtraction procedure before promotion. The reason for this stems from the single-particle $\frac12$-BPS supergraviton spectrum: the single-strand left-chiral primaries given in \eqref{eq.T4LCP} can be written more explicitly as
\begin{align} \label{eq.T4CPrelations}
    |-\rangle_a\ ,\qquad \psi^{+\Ad}_{-\frac12}|-\rangle_a = |\Ad\rangle_a\ ,\qquad \psi^{+\dot{1}}_{-\frac12} \psi^{+\dot{2}}_{-\frac12} |-\rangle_a = J^+_{-1}|-\rangle_a = |+\rangle_a \ ,
\end{align}
from which we see that the $\psi^{+}$ and $J^+$ affine modes  relate the 1-particle states $|\Ad\rangle_a$ and $|+\rangle_a$ to $|-\rangle_a$. As a consequence of this there will also be relations between states due to total affine modes at the level of multi-particle $\frac12$-BPS and $\frac14$-BPS supergraviton states.

We focus on the case of $N=2$ in this paper and leave the systematic generalisation to higher $N$ for future work \cite{wip}. In $\mathrm{Sym}^2(T^4)$ there are two twist sectors: the untwisted sector $N=\{1,1\}$, consisting of two strands of length 1 and the twisted sector $N=\{2\}$, consisting of one strand of length 2.\footnote{In the notation of \eqref{eq.HtwistSector}, the untwisted sector is from the conjugacy class of the identity ($n_1=2$, $n_2=0$) and the twisted sector is from the conjugacy class of transpositions ($n_1=0$, $n_2=1$).}

The $N=2$ supergraviton partition function\footnote{Throughout this section we will set $\tq=1$ in order to simplify notation. From \eqref{eq.1/4BPSsupergrav}, all of the right-moving parts of states contributing to the supergraviton partition function and index are chiral primaries and therefore have $\tilde{h}=\tj$. The $\tq$ dependence of a given state's contribution is thus fixed in terms of the $\ty$ dependence.} $Z_{T^4}^{\mathrm{graviton}}(q,y,\ty;2)$, from the $p^2$ term of the generating function in \eqref{eq.T4NptPF} can be split into separate untwisted and twisted sector partition functions by picking out the contribution from different values of $n$ in the product. This will allow us to perform the subtraction and promotion steps separately for the two twist sectors.

First we describe the logic of the subtraction procedure in the $T^4$ theory in more detail.

\subsection{Subtraction procedure} \label{ssec:subtract}

As previously mentioned, a subtraction procedure is necessary before promotion in the case of $\mathrm{Sym}^N(T^4)$ due to the relation of certain total global primaries under the action of total affine modes. The underlying cause of this are the relations between the single-strand left-chiral primaries in \eqref{eq.T4CPrelations}. This results in relations between the $\frac14$-BPS multi-supergraviton states appearing in the supergraviton partition function. In the following we argue how knowledge of the single-strand relations is sufficient to correctly implement the subtraction of global characters from the supergraviton partition function.
\begin{figure}[htb]
\begin{center}
\begin{adjustbox}{max totalsize={0.6\textwidth}{\textheight},center}
\begin{tikzpicture}[squarednode/.style={rectangle, draw=black!0, very thick, minimum size=5mm},]
    \node[squarednode] (state+) {$|+\rangle_a$} ;
    \node[squarednode] (state-) [below=2cm of state+] {$|-\rangle_a$};
    \node[squarednode] (stateA) [below right=0.5cm and 2cm of state+] {$|\Ad\rangle_a$};

    \draw[-Stealth] (state+.south) -- node[left=5pt]{$J^{-}_{1}$} (state-.north);
    \draw[-Stealth] (state+.east) -- node[above=2pt]{$\psi^{-\Ad}_{\frac12}$} (stateA.west);
    \draw[-Stealth] (stateA.west) -- node[below right=-8pt]{$\ep_{\Bd\Ad}\psi^{-\Bd}_{\frac12}$} (state-.east);
\end{tikzpicture}
\end{adjustbox}
\caption{\it The relations between single-strand left-chiral primaries for $\mathrm{Sym}^N(T^4)$ under the action of the affine modes $J^{-}_{1}$ and $\psi^{-\Ad}_{\frac12}$.\label{fig:1} The vertical position of each state corresponds to their $j$ value.}
\end{center}
\end{figure}
\begin{enumerate}[start=1,
    label=(\arabic*)]
    
    \item \label{list:1}

    The single-strand left-chiral primaries in \eqref{eq.T4CPrelations} (on a strand of length $a$) are highest-weight states of global characters. However, they are all contained within one affine contracted large $\cN=4$ character due to the relations displayed in figure \ref{fig:1}.\footnote{This has the same content as \eqref{eq.T4CPrelations}, but we are using $\psi^{-\Ad}_{\frac12}$ and $J^-_1$ instead of $\psi^{+\Ad}_{-\frac12}$ and $J^+_{-1}$ because these make the relations between descendants simpler.}

    \item \label{list:2}

    Next consider the short $SU(1,1\,|\,2)_L$ multiplets built upon each of these chiral primaries and the associated short characters. The global descendants in these multiplets are formed by the action of the global modes $L_{-1}$, $J^-_0$ and $G^{-A}_{-\frac12}$. Using the mode algebra in \eqref{eq.commcurrents2} and \eqref{eq.commcurrents3} and the example of the $|+\rangle_a$ multiplets, one can show that the action of the affine modes $\phi^{-\Ad}_{\frac12}$ and $J^-_1$ on such descendant states takes the form
    \begin{align}
        J^-_1\cO|+\rangle_a &= \cO'|+\rangle_a + \cO|-\rangle_a \ ,\\
        \psi^{-\Ad}_{\frac12}\cO|+\rangle_a &= \cO'|+\rangle_a \pm \cO|\Ad\rangle_a \ ,
    \end{align}
    where $\cO$ is any collection of global modes and $\cO'$ comes from the (anti-)commutator of the affine mode with $\cO$. The first terms are states within the same short multiplet as $\cO|+\rangle_a$ and the second terms are in the short multiplets of $|-\rangle_a$ and $|\Ad\rangle_a$ respectively. All relations between global descendants of the single-strand chiral primaries follow similarly and are summarised in figure~\ref{fig:2}. The full global short multiplets are thus related by affine modes, not just the highest-weight states (chiral primaries).

    \begin{figure}[htb]
    \begin{center}
    \begin{adjustbox}{max totalsize={\textwidth}{\textheight},center}
    \begin{tikzpicture}[
    squarednode/.style={rectangle, draw=black!0, very thick, minimum size=5mm},align=center,]
    
        \node[squarednode] (state-) {$(L_{-1})^{n}(J^-_0)^m|-\rangle_a$ \\[-.5ex] + \\[-0.5ex] $n(L_{-1})^{n-1}(J^-_0)^{m+1}|+\rangle_a$};
        \node[squarednode] (state+) [above=3cm of state-] {$(L_{-1})^{n}(J^-_0)^m|+\rangle_a$} ;
        \node[squarednode] (stateA) [below right=1cm and 2cm of state+] {$(L_{-1})^{n}(J^-_0)^m|\Ad\rangle_a$};
        \node[squarednode] (state-2) [below=1.8cm of stateA] {$(L_{-1})^{n}(J^-_0)^m|-\rangle_a$};
    
        \draw[-Stealth] (state+.south) -- node[left=3pt]{$J^{-}_{1}$} (state-.north);
        \draw[-Stealth] (state+.east) -- node[above right=-5pt]{$\psi^{-\Ad}_{\frac12}$} (stateA.west);
        \draw[-Stealth] (stateA.south) -- node[right=3pt]{$\ep_{\Bd\Ad}\psi^{-\Bd}_{\frac12}$} (state-2.north);
        \end{tikzpicture}
        \end{adjustbox}
        \caption{\it As a consequence of the relations between single-strand chiral primaries displayed in figure~\ref{fig:1}, their global descendants are similarly related. The descendant states containing $G^{-A}_{-1/2}$ modes obey identical relations since the affine modes $J^{-}_1$ and $\psi^{-\Ad}_{\frac12}$ (anti-)commute with them.\label{fig:2}}
    \end{center}
    \end{figure}
    
    \item \label{list:3}
    
    The multi-supergraviton partition function, however, does not solely receive contributions from such single-strand states, but also multi-strand states. We now consider how the relations between single-strand left-chiral primaries in step \ref{list:1} induces relations between multi-strand left-chiral primaries, focussing on the case of two strands of length 1 (\textit{i.e.}\ on the untwisted sector of the $N=2$ theory). The modes of the appropriate affine algebra are the total modes \eqref{eq.totalModes} and so the action of $J^{-(\mathrm{T})}_{1}$ and $\psi^{-\Ad(\mathrm{T})}_{\frac12}$ then induce the relations between $N=\{1,1\}$ left-chiral primaries given in figure~\ref{fig:3}.
    
    \begin{figure}[htbp]
    \begin{center}
    \begin{adjustbox}{max totalsize={\textwidth}{\textheight},center}
    \begin{tikzpicture}[
    squarednode/.style={rectangle, draw=black!0, very thick, minimum size=5mm},align=center,]
    
    \node[squarednode] (state++) {$|+\rangle^{(1)}_1|+\rangle^{(2)}_1$};
        
    \node[squarednode] (state+-) [below=3.0cm of state++] {$|+\rangle^{(1)}_1|-\rangle^{(2)}_1 + |-\rangle^{(1)}_1|+\rangle^{(2)}_1$};
    
    \node[squarednode] (state--) [below=3.0cm of state+-] {$|-\rangle^{(1)}_1|-\rangle^{(2)}_1$};
    
    \node[squarednode] (stateA+) [below right=1cm and 2.5cm of state++] {$|\Ad\rangle^{(1)}_1|+\rangle^{(2)}_1 + |+\rangle^{(1)}_1|\Ad\rangle^{(2)}_1$};

    \node[squarednode] (stateA-) [below=3.0cm of stateA+] {$|\Ad\rangle^{(1)}_1|-\rangle^{(2)}_1 + |-\rangle^{(1)}_1|\Ad\rangle^{(2)}_1$};
    
    \node[squarednode] (stateAB) [below right=0.5cm and 2cm of stateA+] {$\ep_{\Ad\Bd}\big(|\Ad\rangle^{(1)}_1|\Bd\rangle_1^{(2)} - |\Bd\rangle^{(1)}_1|\Ad\rangle_1^{(2)}\big)\vspace{1pt}$ \\[-.5ex] + \vspace{1pt}\\[-.5ex] $\big(|+\rangle^{(1)}_1|-\rangle^{(2)}_1 + |-\rangle^{(1)}_1|+\rangle^{(2)}_1\big)$};
    
    \draw[-Stealth] (state++.south) -- node[left=3pt]{$J^{-(\mathrm{T})}_{1}$} (state+-.north);
    \draw[-Stealth] (state+-.south) -- node[left=3pt]{$J^{-(\mathrm{T})}_{1}$} (state--.north);
    \draw[-Stealth] (stateA+.south) -- node[right=3pt]{$J^{-(\mathrm{T})}_{1}$} (stateA-.north);
    \draw[-Stealth] (state++.east) -- node[above right=-5pt]{$\psi^{-\Ad(\mathrm{T})}_{\frac12}$} (stateA+.west);
    \draw[-Stealth] (stateA-.west) -- node[below right=1pt and -10pt]{$\frac12\ep_{\Bd\Ad}\psi^{-\Bd(\mathrm{T})}_{\frac12}$} (state--.east);
    \draw[-Stealth] (stateA+.east) -- node[above right=3pt and -10pt]{$\ep_{\Bd\Ad}\psi^{-\Bd(\mathrm{T})}_{\frac12}$} (stateAB.west);
    \draw[-Stealth] (state+-.east) -- node[above right=-1pt and -7pt]{$\psi^{-\Ad(\mathrm{T})}_{\frac12}$} (stateA-.west);
    \draw[-Stealth] (stateAB.west) -- node[below right=-1pt and -5pt]{$\psi^{-\Cd(\mathrm{T})}_{\frac12}$} (stateA-.east);
    %%%%%%%%%%%%%%%%%%%%%%%%%%%%%%%%%%%%
    \node[squarednode] (stateA+2) [below=2cm of state--] {$|\Ad\rangle^{(1)}_1|+\rangle^{(2)}_1 - |+\rangle^{(1)}_1|\Ad\rangle^{(2)}_1$};
    
    \node (b) [above left=1pt and 10pt of stateA+2] {b)};
    \node (a) [above=8.7cm of b] {a)};
    
    \node[squarednode] (stateA-2) [below=4.5cm of stateA+2] {$|\Ad\rangle^{(1)}_1|-\rangle^{(2)}_1 - |-\rangle^{(1)}_1|\Ad\rangle^{(2)}_1$};
    
    \node[squarednode] (state+-2) [below right=1.9cm and 2cm of stateA+2] {$|+\rangle^{(1)}_1|-\rangle^{(2)}_1 - |-\rangle^{(1)}_1|+\rangle^{(2)}_1$};
    
    \node[squarednode] (stateAB2) [right=1cm of state+-2] {$\sigma_{\Ad\Bd}^{i}\big(|\Ad\rangle^{(1)}_1|\Bd\rangle_1^{(2)} + |\Bd\rangle^{(1)}_1|\Ad\rangle_1^{(2)}\big)\vspace{1pt}$};
    
    \draw[-Stealth] (stateA+2.south) -- node[left=3pt]{$J^{-(\mathrm{T})}_{1}$} (stateA-2.north);
    \draw[-Stealth] (stateA+2.east) -- node[above right=-2pt and -13pt]{$-\sigma_{\Bd\Ad}^{i}\psi^{-\Bd(\mathrm{T})}_{\frac12}\,$} (stateAB2.west);
    \draw[-Stealth] (stateA+2.east) -- node[below left=3pt and -13pt]{$\ep_{\Bd\Ad}\psi^{-\Bd(\mathrm{T})}_{\frac12}\,$} (state+-2.west);
    \draw[-Stealth] (state+-2.west) -- node[right=2pt]{$\psi^{-\Ad(\mathrm{T})}_{\frac12}$} (stateA-2.east);
    \draw[-Stealth] (stateAB2.west) -- node[below=2pt]{$\psi^{-\Cd(\mathrm{T})}_{\frac12}$} (stateA-2.east);

    \end{tikzpicture}
    \end{adjustbox}
    \caption{\it Relations between left-chiral primaries of the $N=\{1,1\}$ sector of $\mathrm{Sym}^2(T^4)$ under the action of the total affine modes $J^{-(\mathrm{T})}_{1}$ and $\psi^{-\Ad(\mathrm{T})}_{\frac12}$. These are induced by the relations between single-strand left-chiral primaries displayed in figure~\ref{fig:1}. The orbits displayed in a) and b) are disconnected. \label{fig:3}}
    \end{center}
    \end{figure}

    \item \label{list:4}

    The final step is then to see that the arguments of step~\ref{list:2} along with the relations between multi-strand left-chiral primaries, discussed in step \ref{list:3} for the $N=\{1,1\}$ sector, lead to relations between their global descendants. The product of short $SU(1,1\,|\,2)_L$ multiplets generically decomposes into one short multiplet and an infinite tower of long multiplets under the diagonal (total) algebra. At the level of characters this can be seen from the product formula in \eqref{eq.SU112CharProd}. Take as an example the chiral primary $|+\rangle_1^{(1)}|+\rangle_1^{(2)}$; this is the highest-weight state of the short total multiplet in the decomposition of the product of two short multiplets built from the single-strand chiral primary $|+\rangle_1$. The product of generic states in these single-strand short multiplets is of the form $\cO_1^{(1)}|+\rangle_1^{(1)}\cO_2^{(2)}|+\rangle_1^{(2)}$ with $\cO_1$ and $\cO_2$ being combinations of global modes. The action of the total affine mode $J^{-(\mathrm{T})}_1$ on a state of this form is then
    \begin{align}
        J^{-(\mathrm{T})}_1\left(\cO_1^{(1)}|+\rangle_1^{(1)}\cO_2^{(2)}|+\rangle_1^{(2)}\right) &= \cO_1^{(1)}|-\rangle_1^{(1)}\cO_2^{(2)}|+\rangle_1^{(2)} + \cO_1^{(1)}|+\rangle_1^{(1)}\cO_2^{(2)}|-\rangle_1^{(2)} \nonumber\\
        &\ + \cO_1^{'(1)}|+\rangle_1^{(1)}\cO_2^{(2)}|+\rangle_1^{(2)} + \cO_1^{(1)}|+\rangle_1^{(1)}\cO_2^{'(2)}|+\rangle_1^{(2)} \ ,
    \end{align}
    with the first two resulting states being in the product of short multiplets built from $|+\rangle_1$ and $|-\rangle_1$ single-strand chiral primaries. The remaining states are within the original product of short multiplets. Similarly, the other relations between multi-strand chiral primaries in figure~\ref{fig:3} result in relations between the products of their associated short multiplets. 
    
\end{enumerate}
A limitation of this argument is that, if a linear combination of a supergraviton state and a singleton state can form an affine primary, then in effect that supergraviton's global multiplet should not be subtracted. In this paper we will neglect this possibility and will point out whenever this effect could modify a result. However, we will show that the conclusions of our paper do not depend on the existence of these additional effects.

\subsection{$T^4$ generalised index for $N=2$}

\subsubsection{Twisted sector: $N=\{2\}$} \label{ssec:N2}

We start by applying the subtraction procedure discussed in \ref{ssec:subtract} to the twisted sector of the $N=2$ theory; from the $n=2$ contribution to the order $p^2$ term of \eqref{eq.T4NptPF} we get the twisted-sector partition function
\begin{align} \label{eq.ZT4N2}
    Z_{T^4}^{\mathrm{graviton}}(q,y,\ty;\{2\}) &\equiv \prod_{m,l,\tilde{m},\tilde{\ell}} \big(1-p^2q^my^l\tq^{\tilde{m}}\ty^{\tilde{\ell}}\big)^{-c_{T^4}^{\mathrm{graviton}}(2,m,l,\tilde{m},\tilde{\ell})} \bigg|_{\substack{p^2\\\tq=1}} \nonumber\\
    &= X_2^{T^4}\!(q,y)\,\omega_2(\ty) \ ,
\end{align}
where $X_2^{T^4}$ is defined in \eqref{eq.XT4def} and we have used the following shorthand for $\omega_k$ (as defined in \eqref{eq.omegakn})
\begin{align} \label{eq.omegakChars}
    \omega_k(\ty) \equiv \omega_k(1,\ty) = \tilde{\chi}_{\frac12;k}(\ty) - 2\tilde{\chi}_{0;k}(\ty) \ .
\end{align}
In this $SU(2)_R$ character decomposition, $\tilde{\chi}_{\frac12;k}$ and $-2\tilde{\chi}_{0;k}$ are the contributions from the right-chiral primaries $|\ald\rangle_k$ and $|\Ad\rangle_k$ respectively.

To identify the underlying single-strand left-chiral primaries that give rise to the various global characters in this partition function it is helpful to temporarily introduce additional fugacities into the single-supergraviton partition function \eqref{eq.T41ptPF}. Introducing $\xi_-$, $\xi_0$ and $\xi_+$ to count $|-\rangle_a$, $|\Ad\rangle_a$ and $|+\rangle_a$ strands respectively, \eqref{eq.ZT4N2} becomes
\begin{align} \label{eq.ZT4N2xi}
    Z_{T^4}^{\mathrm{graviton}}(q,y,\ty,\xi;\{2\}) = \Big(\xi_-\phi^{(s)}_{\frac{1}{2}}(q,y) - 2\xi_0\phi^{(s)}_{1}(q,y) + \xi_+\phi^{(s)}_{\frac{3}{2}}(q,y) \Big)\,\omega_2(\ty) \ .
\end{align}
From this we see that the only twisted-sector total global primaries are the single-strand left-chiral primaries in \eqref{eq.T4CPrelations} with $a=2$, along with any of the right-chiral primaries given in \eqref{eq.T4RCP} (contributing the factor of $\omega_2(\ty)$ to the partition function). We can therefore directly use the arguments of steps~\ref{list:1} and \ref{list:2} of the subtraction procedure to relate the short global characters in \eqref{eq.ZT4N2xi} as shown in figure~\ref{fig:6}.
\begin{figure}[h]
\begin{center}
        \begin{adjustbox}{max totalsize={\textwidth}{\textheight},center}
            \begin{tikzpicture}[squarednode/.style={rectangle, draw=black!0, very thick, minimum size=5mm},align=center,]
                \node[squarednode] (1) {$\xi_+\phi^{(s)}_{\frac{3}{2}}\!(q,y)\,\omega_2(\ty)$};
                \node[squarednode] (2) [below=2cm of 1] {$\xi_-\phi^{(s)}_{\frac{1}{2}}\!(q,y)\,\omega_2(\ty)$};
                \node[squarednode] (3) [below right=0.5cm and 1.5cm of 1] {$-2\xi_0\,\phi^{(s)}_{1}\!(q,y)\,\omega_2(\ty)$};
                \draw[-Stealth] (1.south) -- node[left=3pt]{$J^{-(\mathrm{T})}_1$} (2.north);
                \draw[-Stealth] (1.east) -- node[above right=1pt and -8pt]{$\psi^{-\Ad(\mathrm{T})}_{\frac12}$} (3.west);
                \draw[-Stealth] (3.west) -- node[below right=-1pt and -9pt]{$\ep_{\Ad\Bd}\psi^{-\Bd(\mathrm{T})}_{\frac12}$} (2.east);
            \end{tikzpicture}
        \end{adjustbox}
        \caption{\it Relations between short global characters appearing in the twisted sector supergraviton partition function \eqref{eq.ZT4N2xi} due to total affine modes mapping between their associated short global multiplets. \label{fig:6}}
\end{center}
\end{figure}
The subtraction procedure then yields the ``reduced" twisted-sector supergraviton partition function
\begin{align} \label{eq.T4N2red}
    Z_{T^4}^{\mathrm{red}}(q,y,\ty,\xi;\{2\}) = \xi_-\phi^{(s)}_{\frac{1}{2}}(q,y)\,\omega_2(\ty) \ .
\end{align}
The promotion of total global characters to total affine characters described in \ref{list:iib} then gives the ``generalised" supergraviton partition function in the twisted sector as (we now set $\xi_-=1$)
\begin{align} \label{eq.T4N2gen}
    Z_{T^4}^{\mathrm{gen}}(q,y,\ty;\{2\}) = \mathbf{\Phi}^{(s)}_{\frac{1}{2};2}(q,y)\,\omega_2(\ty) \ ,
\end{align}
where $\mathbf{\Phi}^{(s)}_{j;N}$ are short characters of the total contracted large algebra, detailed in \eqref{eq.clargeN4s}. By decomposing this affine character into global characters
\begin{align} \label{eq.T4N2genPF}
    Z_{T^4}^{\mathrm{gen}}(q,y,\ty;\{2\}) &= \Big[\phi^{(s)}_{\frac12} - 2 \phi^{(s)}_{1} + \phi^{(s)}_{\frac32} - 2 \phi^{(\ell)}_{0,1} - 12\phi^{(\ell)}_{0,2} + 5 \phi^{(\ell)}_{\frac12,\frac32} - 4\phi^{(\ell)}_{1,2} + \cdots\Big]\,\omega_2(\ty) \nonumber\\
    &= Z_{T^4}^{\mathrm{graviton}}(q,y,\ty;\{2\}) + \Big[- 2 \phi^{(\ell)}_{0,1} - 12\phi^{(\ell)}_{0,2} + 5 \phi^{(\ell)}_{\frac12,\frac32} - 4\phi^{(\ell)}_{1,2} + \cdots \Big]\,\omega_2(\ty) \ ,
\end{align}
we see that by following the subtraction and promotion steps we have added to the original twisted-sector supergraviton partition function additional long global multiplets corresponding to total affine descendants. Using the definition \eqref{eq.EMEGNS}, we can find the contribution of the generalised supergraviton partition function \eqref{eq.T4N2gen} to the modified elliptic genus:
\begin{align} \label{eq.T4N2MEG}
    \hat{\cE}^{\mathrm{gen}}_{T^4}(q,y;\{2\}) \equiv \frac12 \big(\ty\pd_{\ty}\big)^2 Z_{T^4}^{\mathrm{gen}}(q,y,\ty;\{2\})\Big|_{\ty=1} = \mathbf{\Phi}^{(s)}_{\frac{1}{2};2}(q,y) \ .
\end{align}

\subsubsection{Untwisted sector: $N=\{1,1\}$} \label{ssec:T4N11}

In contrast to the twisted sector, the untwisted sector contains multi-strand states. From the $n=1$ contribution to the $p^2$ term of \eqref{eq.T4NptPF} we get the untwisted-sector partition function as
\begin{align} \label{eq.ZT4N11}
    Z_{T^4}^{\mathrm{graviton}}(q,y,\ty;\{1,1\}) &\equiv \prod_{m,l,\tilde{m},\tilde{\ell}} \big(1-pq^my^l\tq^{\tilde{m}}\ty^{\tilde{\ell}}\big)^{-c_{T^4}^{\mathrm{graviton}}(1,m,l,\tilde{m},\tilde{\ell})} \bigg|_{\substack{p^2\\\tq=1}} \nonumber\\
    &= \frac{1}{2} \Big(X_{1}^{T^4}\!(q,y)\,\omega_1(\qb,\yb)\Big)^2 + \frac12 X_{1}^{T^4}\!(q^2,y^2)\,\omega_1(\qb^2,\yb^2) \ .
\end{align}
Here the second term is correcting an undercounting in the first term that occurs due to the strand symmetry factor\footnote{This symmetry factor is to account for the indistinguishability of strands of the same length. In the case of the untwisted sector of $N=2$ it is simply $|S_2|^{-1}=\frac12$. The full centraliser subgroup of a twist sector is given in \eqref{eq.Centraliser}.} of $\frac12$ for the case of having identical states on the two strands.

After again introducing the strand-type fugacities $\xi_{\pm,0}$ in order to organise the contributions by the single-strand left-chiral primaries on which states are built, we find the global character expansion (suppressing the functional dependence here so that $Z_{T^4\{1,1\}}^{\mathrm{graviton}} \equiv Z_{T^4}^{\mathrm{graviton}}(q,y,\ty,\xi;\{1,1\})$) reads
\begin{align} \label{eq.ZT4N11xi}
    Z_{T^4\{1,1\}}^{\mathrm{graviton}} &= \frac12 \xi_-^2\bigg[\Big(\phi^{(s)}_{0}\!(q,y)\, \omega_1(\yb)\Big)^2 + \phi^{(s)}_{0}\!(q^2,y^2)\,\omega_1(\yb^2)\bigg] - 2\xi_-\xi_0 \phi^{(s)}_{0}\!(q,y)\phi^{(s)}_{\frac12}\!(q,y)\big(\omega_1(\yb)\big)^2 \nonumber\\
    &\ +  \xi_-\xi_+ \phi^{(s)}_{0}\!(q,y)\phi^{(s)}_{1}\!(q,y)\big(\omega_1(\yb)\big)^2 + \xi_0^2\bigg[2\Big(\phi^{(s)}_{\frac12}\!(q,y)\,\omega_1(\yb)\Big)^2 - \phi^{(s)}_{\frac12}\!(q^2,y^2)\,\omega_1(\yb^2)\bigg] \nonumber\\
    &\ - 2\xi_0\xi_+ \phi^{(s)}_{\frac12}\!(q.y)\phi^{(s)}_{1}\!(q,y)\big(\omega_1(\yb)\big)^2 + \frac12 \xi_+^2 \bigg[\Big(\phi^{(s)}_{1}\!(q,y)\,\omega_1(\yb)\Big)^2 + \phi^{(s)}_{1}\!(q^2,y^2)\,\omega_1(\yb^2)\bigg] \ .
\end{align}
This partition function contains contributions from long global characters coming from the products of short characters, as seen by the formula \eqref{eq.SU112CharProd}. 

To understand the states contributing to these various terms we first note that in the untwisted sector ($N=\{1,1\}$) of the symmetric orbifold Hilbert space, states should be invariant under $S_2\cong\mathbb{Z}_2$ acting on the copy labels. However, the left- and right-moving parts of a state do not have to separately be $\mathbb{Z}_2$ invariant; they can individually transform in a representation of $\mathbb{Z}_2$, \textit{i.e.}\ in the symmetric or anti-symmetric irreducible representations. Overall $\mathbb{Z}_2$ invariance then dictates that a given state should have the same representation on the left and right \cite{Jevicki:2015irq}. This selection rule will help in the identification of states contributing to the partition function \eqref{eq.ZT4N11xi}.

Consider first the order $\xi_-^2$ term in \eqref{eq.ZT4N11xi},
\begin{align} \label{eq.T4N11xim2}
    Z_{T^4\{1,1\}}^{\mathrm{graviton}}\Big|_{\xi_-^2} &= \frac12 \bigg[\Big(\phi^{(s)}_{0}\!(q,y)\, \omega_1(\yb)\Big)^2 + \phi^{(s)}_{0}\!(q^2,y^2)\,\omega_1(\yb^2)\bigg] \nonumber\\
    &= \phi^{(s)}_{0}\!(q,y)\, \frac12\Big[\big(\omega_1(\yb)\big)^2 + \omega_1(\yb^2)\Big] \ ,
\end{align}
where \eqref{eq.SU112CharProd} and \eqref{eq.SU112CharPow} were used to get the second line. There is only a single short $SU(1,1\,|\,2)_L$ character in this term; having a highest-weight state with $h=j=0$ this corresponds to the NS vacuum on both strands, \textit{i.e.}\ $|-\rangle_1^{(1)}|-\rangle_1^{(2)}$. This left-moving state is clearly copy symmetric by itself and therefore to have an overall copy-symmetric state, for any choice of right-chiral primaries \eqref{eq.T4RCP} on the two strands, the right-moving state should also be in the symmetric representation.

This logic agrees with \eqref{eq.T4N11xim2} since, by using the $SU(2)_R$ character decomposition of $\omega_n$ given in \eqref{eq.omegakChars} we find
\begin{align} \label{eq.RSomega}
    \omega_{\mathrm{sym}}(\ty) \equiv \frac12\Big[\big(\omega_1(\yb)\big)^2 + \omega_1(\yb^2)\Big] = \tilde{\chi}_{1;2}(\ty) -2\tilde{\chi}_{\frac12;2}(\ty) + \tilde{\chi}_{0;2}(\ty) \ ,
\end{align}
which is the contribution of all possible right-chiral primaries from \eqref{eq.T4RCP} in copy-symmetric combinations (we will write down their explicit form below in \eqref{eq.zm2states}). Likewise, the analogous contribution from copy-anti-symmetric states is given by
\begin{align} \label{eq.RAomega}
    \omega_{\mathrm{asym}}(\ty) \equiv \frac12\Big[\big(\omega_1(\yb)\big)^2 - \omega_1(\yb^2)\Big] = -2\tilde{\chi}_{\frac12;2}(\ty) + 4\tilde{\chi}_{0;2}(\ty) \ ,
\end{align}
where one of the $\tilde{\chi}_{0;2}$ comes from the $SU(2)_R$ singlet combination of $|\ald\rangle_1$ and $|\bed\rangle_1$, and the remaining three are from the $SU(2)_2$ triplet combination of $|\Ad\rangle_1$ and $|\Bd\rangle_1$. In total, $\big(\omega_1(\ty)\big)^2=\omega_{\mathrm{sym}}(\ty) + \omega_{\mathrm{asym}}(\ty)$ would give the contribution from all possible $N=\{1,1\}$ right-chiral primaries.

We therefore conclude that the states contributing to \eqref{eq.T4N11xim2} are
\begin{subequations} \label{eq.zm2states}
\begin{align}
    \phi^{(s)}_{0}\tilde{\chi}_{1;2}:\qquad &|-\rangle_1^{(1)}|-\rangle_1^{(2)}\Big(|\ald\rangle_1^{(1)}|\bed\rangle_1^{(2)} + |\bed\rangle_1^{(1)}|\ald\rangle_1^{(2)}\Big) \ ,\label{eq.zm2State1}\\
    -2\phi^{(s)}_{0}\tilde{\chi}_{\frac12;2}:\qquad & |-\rangle_1^{(1)}|-\rangle_1^{(2)}\Big(|\ald\rangle_1^{(1)}|\Cd\rangle_1^{(2)} + |\Cd\rangle_1^{(1)}|\ald\rangle_1^{(2)}\Big) \ ,\label{eq.zm2State2}\\
    \phi^{(s)}_{0}\tilde{\chi}_{0;2}:\qquad & |-\rangle_1^{(1)}|-\rangle_1^{(2)}\Big(|\Cd\rangle_1^{(1)}|\Dd\rangle_1^{(2)} - |\Dd\rangle_1^{(1)}|\Cd\rangle_1^{(2)}\Big) \ .\label{eq.zm2State3}
\end{align}
\end{subequations}
We see that the right-moving part is copy symmetric because $\ket{\ald}$ is bosonic while $\ket{\Ad}$ is fermionic.

In a similar manner we can understand the states contributing to the order $\xi_-\xi_0$ term of \eqref{eq.ZT4N11xi}
\begin{align}
    Z_{T^4\{1,1\}}^{\mathrm{graviton}}\Big|_{\xi_-\xi_0} &= - 2\phi^{(s)}_{0}\!(q,y)\,\phi^{(s)}_{\frac12}\!(q,y)\,\big(\omega_1(\yb)\big)^2 \nonumber\\
    &= -2\phi^{(s)}_{\frac12}\!(q,y)\,\Big(\tilde{\chi}_{1;2} -4\tilde{\chi}_{\frac12;2} + 5\tilde{\chi}_{0;2}\Big) \ ,
\end{align}
where once again the $SU(1,1\,|\,2)_L$ and $SU(2)_R$ character product formulae \eqref{eq.SU112CharProd} and \eqref{eq.SU2CharProd} were used. Since this term is contributing to the partition function at order $\xi_-\xi_0$ the left-moving part of the state should contain one $|-\rangle_1$ and one $|\Ad\rangle_1$ chiral primary strand. These can then be in either the symmetric or anti-symmetric representation of $\mathbb{Z}_2$. Matching the copy-symmetry of the right-moving state to that of the left, we can identify the chiral primary states at order $\xi_-\xi_0$ as
\begin{subequations} \label{eq.zmz0states}
\begin{align}
    -2\phi^{(s)}_{\frac12}\big(\tilde{\chi}_{1;2}+\tilde{\chi}_{0;2}\big):\quad &\Big(|-\rangle_1^{(1)}|\Ad\rangle_1^{(2)} \pm |\Ad\rangle_1^{(1)}|-\rangle_1^{(2)}\Big)\Big(|\ald\rangle_1^{(1)}|\bed\rangle_1^{(2)} \pm |\bed\rangle_1^{(1)}|\ald\rangle_1^{(2)}\Big) \ ,\label{eq.zmz0State1}\\
    8\phi^{(s)}_{\frac12}\tilde{\chi}_{\frac12;2}:\quad &\Big(|-\rangle_1^{(1)}|\Ad\rangle_1^{(2)} \pm |\Ad\rangle_1^{(1)}|-\rangle_1^{(2)}\Big)\Big(|\ald\rangle_1^{(1)}|\Cd\rangle_1^{(2)} \pm |\Cd\rangle_1^{(1)}|\ald\rangle_1^{(2)}\Big) \ ,\label{eq.zmz0State3}\\
    -8\phi^{(s)}_{\frac12}\tilde{\chi}_{0;2}:\quad &\Big(|-\rangle_1^{(1)}|\Ad\rangle_1^{(2)} \pm |\Ad\rangle_1^{(1)}|-\rangle_1^{(2)}\Big)\Big(|\Cd\rangle_1^{(1)}|\Dd\rangle_1^{(2)} \mp |\Dd\rangle_1^{(1)}|\Cd\rangle_1^{(2)}\Big) \ .\label{eq.zmz0State5}
\end{align}
\end{subequations}
Of these states, those with left-moving parts in the symmetric representation are related to the states \eqref{eq.zm2states} appearing at order $\xi_-^2$, as seen from figure~\ref{fig:3}. Those with left-moving states in the anti-symmetric representation are new total affine primaries and so should not be subtracted.

As described in step \ref{list:4} of section \ref{ssec:subtract}, as a consequence of these relations between chiral primaries we then have the following relations between order-$\xi_-\xi_0$ terms and order-$\xi_-^2$ terms from the action of the affine mode $\psi^{-\Ad(\mathrm{T})}_{\frac12}$
\begin{subequations}\label{cog10May25}
\begin{align}
    -2\xi_-\xi_0\,\phi^{(s)}_{\frac12} \tilde{\chi}_{1;2} &\,\lto\, \xi_-^2 \phi^{(s)}_{0} \tilde{\chi}_{1;2} \ ,\\
    4\xi_-\xi_0\,\phi^{(s)}_{\frac12} \tilde{\chi}_{\frac12;2} &\,\lto\, -2\xi_-^2 \phi^{(s)}_{0} \tilde{\chi}_{\frac12;2} \ ,\\
    -2\xi_-\xi_0\,\phi^{(s)}_{\frac12}\tilde{\chi}_{0;2} &\,\lto\, \xi_-^2 \phi^{(s)}_{0} \tilde{\chi}_{0;2} \ ,
\end{align}
\end{subequations}
and so these order-$\xi_-\xi_0$ contributions should be subtracted when defining the reduced $N=\{1,1\}$ partition function.  The coefficients appearing on the left of \eqref{cog10May25} correspond to the symmetric representations in \eqref{eq.zmz0states}.

At order $\xi_-\xi_+$ we again have only one short $SU(1,1\,|\,2)_L$ character
\begin{align}
    Z_{T^4\{1,1\}}^{\mathrm{graviton}}\Big|_{\xi_-\xi_+} &= \phi^{(s)}_{0}\!(q,y)\,\phi^{(s)}_{1}\!(q,y)\,\big(\omega_1(\yb)\big)^2 \nonumber\\
    &= \phi^{(s)}_{1}\!(q,y)\,\Big(\tilde{\chi}_{1;2} -4\tilde{\chi}_{\frac12;2} + 5\tilde{\chi}_{0;2}\Big) \ ,
\end{align}
and we find the contributing chiral primaries
\begin{subequations} \label{eq.zmzpstates}
\begin{align}
    \phi^{(s)}_{1}\big(\tilde{\chi}_{1;2}+\tilde{\chi}_{0;2}\big):\quad &\Big(|-\rangle_1^{(1)}|+\rangle_1^{(2)} \pm |+\rangle_1^{(1)}|-\rangle_1^{(2)}\Big)\Big(|\ald\rangle_1^{(1)}|\bed\rangle_1^{(2)} \pm |\bed\rangle_1^{(1)}|\ald\rangle_1^{(2)}\Big) \ ,\label{eq.zmzpState1}\\
    -4\phi^{(s)}_{1}\tilde{\chi}_{\frac12;2}:\quad &\Big(|-\rangle_1^{(1)}|+\rangle_1^{(2)} \pm |+\rangle_1^{(1)}|-\rangle_1^{(2)}\Big)\Big(|\ald\rangle_1^{(1)}|\Cd\rangle_1^{(2)} \pm |\Cd\rangle_1^{(1)}|\ald\rangle_1^{(2)}\Big) \ ,\label{eq.zmzpState2}\\
    4\phi^{(s)}_{1}\tilde{\chi}_{0;2}:\quad &\Big(|-\rangle_1^{(1)}|+\rangle_1^{(2)} \pm |+\rangle_1^{(1)}|-\rangle_1^{(2)}\Big)\Big(|\Cd\rangle_1^{(1)}|\Dd\rangle_1^{(2)} \mp |\Dd\rangle_1^{(1)}|\Cd\rangle_1^{(2)}\Big) \ .\label{eq.zmzpState3}
\end{align}
\end{subequations}
From figure~\ref{fig:3} we see that all of these chiral primaries are related to those in \eqref{eq.zmz0states} by the action of $\psi^{-\Ad(\mathrm{T})}_{\frac12}$ and so by step \ref{list:4} we conclude that all characters at order $\xi_-\xi_+$ should be subtracted.

Whilst the order-$\xi_0^2$ contribution to \eqref{eq.ZT4N11xi} appears complicated, it can be broken down into parts based on their symmetry
\begin{align}
    Z_{T^4\{1,1\}}^{\mathrm{graviton}}\Big|_{\xi_0^2} &= 2\Big(\phi^{(s)}_{\frac12}\!(q,y)\,\omega_1(\yb)\Big)^2 - \phi^{(s)}_{\frac12}\!(q^2,y^2)\,\omega_1(\yb^2) \nonumber\\
    &= \phi^{(s)}_{1}(q,y)\Big[\omega_{\mathrm{sym}}(\ty) +3\,\omega_{\mathrm{asym}}(\ty)\Big] + \sum_{\substack{h\geq1\\\mathrm{odd}}} \phi^{(\ell)}_{0,h}(q,y) \Big[3\,\omega_{\mathrm{sym}}(\ty) +\omega_{\mathrm{asym}}(\ty)\Big] \nonumber\\
    &\ + \sum_{\substack{h\geq2\\\mathrm{even}}} \phi^{(\ell)}_{0,h}(q,y) \Big[\omega_{\mathrm{sym}}(\ty) + 3\,\omega_{\mathrm{asym}}(\ty)\Big] \ ,
\end{align}
where the character product formulae in appendix~\ref{ssec:chars} were used. From this we expect some chiral primaries with copy-symmetric right-moving parts and some with copy-anti-symmetric parts. Using \eqref{eq.RSomega} and \eqref{eq.RAomega} we find the following contribution from chiral primaries at this order
\begin{subequations} \label{eq.xi02states}
    \begin{align}
    \phi^{(s)}_{1}\tilde{\chi}_{1;2}:\quad &\Big(|\Ad\rangle_1^{(1)}|\Bd\rangle_1^{(2)} - |\Bd\rangle_1^{(1)}|\Ad\rangle_1^{(2)}\Big)\Big(|\ald\rangle_1^{(1)}|\bed\rangle_1^{(2)} +|\bed\rangle_1^{(1)}|\ald\rangle_1^{(2)}\Big) \ , \label{eq.xi02states1}\\
    3\phi^{(s)}_{1}\tilde{\chi}_{0;2}:\quad &\Big(|\Ad\rangle_1^{(1)}|\Bd\rangle_1^{(2)} + |\Bd\rangle_1^{(1)}|\Ad\rangle_1^{(2)}\Big)\Big(|\ald\rangle_1^{(1)}|\bed\rangle_1^{(2)} - |\bed\rangle_1^{(1)}|\ald\rangle_1^{(2)}\Big) \ , \label{eq.xi02states2}\\
    -8\phi^{(s)}_{1}\tilde{\chi}_{\frac12;2}:\quad &\Big(|\Ad\ald\rangle_1^{(1)}|\Bd\dot{C}\rangle_1^{(2)} \pm |\Bd\dot{C}\rangle_1^{(1)}|\Ad\ald\rangle_1^{(2)}\Big) \ , \label{eq.xi02states3}\\
    10\phi^{(s)}_{1}\tilde{\chi}_{0;2}:\quad &\Big(|\Ad\rangle_1^{(1)}|\Bd\rangle_1^{(2)} \pm |\Bd\rangle_1^{(1)}|\Ad\rangle_1^{(2)}\Big)\Big(|\Cd\rangle_1^{(1)}|\Dd\rangle_1^{(2)} \pm |\Dd\rangle_1^{(1)}|\Cd\rangle_1^{(2)}\Big) \ ,\label{eq.xi02states4}
\end{align}
\end{subequations}
where in \eqref{eq.xi02states3} we wrote the left and right parts of the state together for notational convenience.
From figure~\ref{fig:3} we see that all of these chiral primaries are related to the chiral primaries in \eqref{eq.zmz0states} by the action of $\psi^{-\Ad(\mathrm{T})}_{\frac12}$ and so by \ref{list:4} we should subtract all characters at order~$\xi_0^2$.

In fact, the relations in figure~\ref{fig:3} and the arguments in step \ref{list:4} indicate that all remaining total global characters appearing in the partition function \eqref{eq.ZT4N11xi} should be subtracted. As noted at the end of section \ref{ssec:subtract}, we cannot a priori exclude the need to retain and promote additional global characters, since affine primaries might be generated through mixing of supergraviton and singleton states. Although such an effect could in principle start at level $h=1$, we will argue in section~\ref{ssec:Fortuity} that our results for $\mathrm{Sym}^{2}(T^4)$ nonetheless give the full result up to at least level $h=2$.

This leaves us with the reduced $N=\{1,1\}$ supergraviton partition function (setting $\xi_-=\xi_0=1$)
\begin{align} \label{eq.T4N11red}
    Z^{\mathrm{red}}_{T^4}(q,y,\ty;\{1,1\}) = \phi^{(s)}_{0}(q,y)\,\omega_{\mathrm{sym}}(\ty) - 2\phi^{(s)}_{\frac12}(q,y)\,\omega_{\mathrm{asym}}(\ty) \ ,
\end{align}
and so by promoting the total global characters to total affine characters we get the untwisted sector contribution to the generalised supergraviton partition function
\begin{align} \label{eq.T4N11gen}
    Z^{\mathrm{gen}}_{T^4}(q,y,\ty;\{1,1\}) = \mathbf{\Phi}^{(s)}_{0;2}(q,y)\,\omega_{\mathrm{sym}}(\ty) - 2\mathbf{\Phi}^{(s)}_{\frac12;2}(q,y)\,\omega_{\mathrm{asym}}(\ty) \ .
\end{align}
Once again we can decompose these into total global characters and $SU(2)_R$ characters (suppressing arguments)
\begin{align} \label{eq.T4PFN11exp}
    Z^{\mathrm{gen}}_{T^4\{1,1\}} &\approx Z_{T^4\{1,1\}}^{\mathrm{graviton}} + 6\phi^{(\ell)}_{0,1}\,\Big(2\tilde{\chi}_{0;2}-\tilde{\chi}_{\frac12;2}\Big) - 2\phi^{(\ell)}_{\frac12,\frac32}\,\Big(17\tilde{\chi}_{0;2}-10\tilde{\chi}_{\frac12;2}+\tilde{\chi}_{1;2}\Big)  + \cdots\ ,
    %&\quad + 2 \phi^{(\ell)}_{0,2} \,\Big(47\tilde{\chi}_{0;2}-31\tilde{\chi}_{\frac12;2}+5\tilde{\chi}_{1;2}\Big) + 2\phi^{(\ell)}_{1,2} \,\Big(16\tilde{\chi}_{0;2}-11\tilde{\chi}_{\frac12;2}+2\tilde{\chi}_{1;2}\Big) + \cdots \ ,
\end{align}
and we see that this process has added long global characters starting at $h=1$, $j=0$ to the supergraviton partition function. The contribution of \eqref{eq.T4N11gen} to the modified elliptic genus is then
\begin{align} \label{eq.T4N11MEG}
    \hat{\cE}^{\mathrm{gen}}_{T^4}(q,y;\{1,1\}) \equiv \frac12 \big(\ty\pd_{\ty}\big)^2 Z_{T^4}^{\mathrm{gen}}(q,y,\ty;\{1,1\})\Big|_{\ty=1} = 2\mathbf{\Phi}^{(s)}_{0;2}(q,y) + 4\mathbf{\Phi}^{(s)}_{\frac{1}{2};2}(q,y) \ .
\end{align}

\subsubsection{The $N=2$ generalised index}

Combining the untwisted sector result \eqref{eq.T4N11MEG} with that of the twisted-sector \eqref{eq.T4N2MEG} we get the $N=2$ generalised supergraviton index
\begin{align} \label{eq.T4genMEG}
    \hat{\cE}^{\mathrm{gen}}_{T^4}(q,y;2) = 2\mathbf{\Phi}^{(s)}_{0;2}(q,y) + 5\mathbf{\Phi}^{(s)}_{\frac{1}{2};2}(q,y) \ .
\end{align}
While this is an extremely compact form in terms of only short affine characters for the case of $N=2$, we expect the contribution also from long affine characters at higher values of $N$. In other words, we expect that long total global characters survive the subtraction procedure in general. We now analyse this new index in more detail.

\subsection{Lowest-lying singleton states} \label{ssec:Fortuity}

As seen from \eqref{eq.T4PFN11exp} and \eqref{eq.T4N2genPF} above, the lowest lying singleton states contributing to the generalised supergraviton modified elliptic genus have the quantum numbers $(h,j)=(1,0)$ in both the twisted and untwisted sectors. Comparing the $q$-expansions for the $N=2$ NS CFT and supergraviton modified elliptic genera, computed from the partition functions \eqref{eq.ZNT4NS} and \eqref{eq.T4NptPF} using the definition \eqref{eq.EMEGNS}, with the generalised supergraviton index \eqref{eq.T4genMEG} we find
\begin{subequations} \label{eq.CFTMEGall}
\begin{align}
    \hat{\cE}^{\mathrm{CFT}}_{\mathrm{NS}}(q,y;2) &= 2 + \sqrt{q}\big(y^{-1} + y\big) + q\big(-12 - 6y^{-2} - 6y^2\big) + q^{\frac32} \big(y^{-3} + 39y^{-1} + 39y + y^3\big) \nonumber\\
    &\hspace{3.05cm}+ q^2\big(-192 + 2y^{-4} -56y^2 -56y^{-2} + 2y^4\big) + \cdots \ ,\label{eq.CFTMEGexp}\\
    \hat{\cE}^{\mathrm{graviton}}_{T^4}(q,y;2) &= 2 + \sqrt{q}\big(y^{-1} + y\big) + q\big(-4 - 6y^{-2} - 6y^2\big) + q^{\frac32}\big(y^{-3} + 6y^{-1} + 6y + y^3\big) \nonumber\\
    &\hspace{3.05cm}+ q^2\big(2 + 2y^{-4} -4y^2 -4y^{-2} + 2y^4\big) + \cdots \ ,\label{eq.gravMEGexp}\\
    \hat{\cE}^{\mathrm{gen}}_{T^4}(q,y;2) &= 2 + \sqrt{q}\big(y^{-1} + y\big) + q\big(-12 - 6y^{-2} - 6y^2\big) + q^{\frac32} \big(y^{-3} + 39y^{-1} + 39y + y^3\big) \nonumber\\
    &\hspace{3.05cm}+ q^2\big(-150 + 2y^{-4} -56y^2 -56y^{-2} + 2y^4\big) + \cdots \ .\label{eq.genMEGexp}
\end{align}
\end{subequations}
From the de Boer bound for the $T^4$ theory \eqref{deBoer_bound_T4} for $N=2$, we only expect agreement of the CFT and supergraviton indices for $h<\frac12$. However, in reality, there is agreement between \eqref{eq.CFTMEGexp} and \eqref{eq.gravMEGexp} also for $h=\frac12$ with the first mismatch occurring only at $(h,j)=(1,0)$.\footnote{This is an accident due to smallness of $N=2$.  For $N=2$, there is no $\frac14$-BPS state at level $h=\frac12$. 
This is clear from figure \ref{fig:MEGdiff}(a), in which all $h=\frac12$ states
are either chiral or anti-chiral primaries on the red unitarity-bound lines and are $\frac12$-BPS\@.  For $\frac12$-BPS states, the supergraviton index always agrees with the CFT index even above the de Boer bound \cite{deBoer:1998us}.} Interestingly, the agreement between the generalised supergraviton index and the CFT index holds for $h<2$: all states contributing to the NS CFT modified elliptic genus for $h\leq\frac32$ are supergraviton and singleton states. We now examine this enhancement of the de Boer bound in more detail.

The difference between the supergraviton and generalised supergraviton modified elliptic genera \eqref{eq.genMEGexp} and \eqref{eq.gravMEGexp} has an expansion
\begin{align} \label{eq.MEGgen-MEGgrav}
    \hat{\cE}^{\mathrm{gen}}_{T^4}(q,y;2) - \hat{\cE}^{\mathrm{graviton}}_{T^4}(q,y;2) = -8q + 33q^{\frac32} \big(y^{-1}+y\big) + \cdots \ .
    %+ q^2\big(-52y^{-2} -152 - 52y^2\big) \ .
\end{align}
We now examine the singleton states that account for these terms in and which cause the enhancement of the matching to the CFT index \eqref{eq.CFTMEGexp}.

\subsubsection*{\underline{$h=1$, $j=0\,$}}

From \eqref{eq.T4N2genPF} and \eqref{eq.T4PFN11exp} we see that the $-8q$ term comes from a mixture of the contributions of twisted and untwisted states to the partition function
\begin{align}
    \left.\left(Z^{\mathrm{gen}}_{T^4\{1,1\}} - Z_{T^4\{1,1\}}^{\mathrm{graviton}}\right)\right|_{q} 
    &= \left. 3\phi^{(\ell)}_{0,1}(q,y)\,\omega_{\mathrm{asym}}(\ty)\right|_{q} = 3 \,\omega_{\mathrm{asym}}(\ty) \overset{\mathrm{MEG}}{\lto} -6 \ ,\\
    \left.\left(Z^{\mathrm{gen}}_{T^4\{2\}} - Z_{T^4\{2\}}^{\mathrm{graviton}}\right)\right|_{q} &= \left. -2\phi^{(\ell)}_{0,1}(q,y)\,\omega_{2}(\ty)\right|_{q} = -2\,\omega_{2}(\ty) \overset{\mathrm{MEG}}{\lto} -2 \ ,
\end{align}
where the final step is obtained by using the definition of the modified elliptic genus in \eqref{eq.EMEGNS}.
Noting that all of these contributions come from short affine characters with $j=\frac12$ in the generalised supergraviton index, we conclude that the states responsible for enhancing the de Boer bound at order $q$ are global primaries with $(h,j)=(1,0)$ that are affine descendants of chiral primaries with $h=j=\frac12$. 

The twisted sector states have the form\footnote{Strictly speaking, the global primary combination in fact has the left-moving form $\big(2\psi^{-\Ad}_{-\frac12} - J^{-}_0\psi^{+\Ad}_{-\frac12}\big)|-\rangle_2$. This second term is equal to $J^-_0|\Ad\rangle_2$ which is a global descendant of a chiral primary. In what follows we suppress all such additional terms that yield a true global primary for clarity. This is sufficient for our counting of states.}
\begin{align} \label{eq.twisted01singletons}
    \psi^{-\Ad}_{-\frac12}|-\rangle_2\otimes\begin{cases}
        \ |\ald\rangle_2\vspace{2pt}\\
        \ |\Bd\rangle_2
    \end{cases}  \ ,
\end{align}
contributing $-2q$ to $\hat{\cE}^{\mathrm{gen}}_{T^4}(q,y;2)$ and the untwisted sector states have the form
\begin{align} \label{eq.untwisted01singletons}
    \sigma^i_{\Ad\Bd} \psi^{-\Ad(\mathrm{T})}_{-\frac12}\Big(|-\rangle_1^{(1)}|\Bd\rangle_1^{(2)} - |\Bd\rangle_1^{(1)}|-\rangle_1^{(2)}\Big) \otimes\begin{cases}
        \ \Big(|\ald\rangle_1^{(1)}|\bed\rangle_1^{(2)} - |\bed\rangle_1^{(1)}|\ald\rangle_1^{(2)}\Big) \vspace{1pt}\\
        \ \Big(|\ald\rangle_1^{(1)}|\Cd\rangle_1^{(2)} - |\Cd\rangle_1^{(1)}|\ald\rangle_1^{(2)}\Big) \vspace{1pt}\\
        \ \Big(|\Cd\rangle_1^{(1)}|\Dd\rangle_1^{(2)} + |\Dd\rangle_1^{(1)}|\Cd\rangle_1^{(2)}\Big)
    \end{cases} \ ,
\end{align}
contributing $-6q$ to $\hat{\cE}^{\mathrm{gen}}_{T^4}(q,y;2)$. Here $\sigma^i_{\Ad\Bd}$ with $i=1,2,3$ give a basis of symmetric matrices that project onto the $SU(2)_2$ triplet representation; see appendix \ref{ssec:alg}.
We note that the $SU(2)_2$ singlet combination of these untwisted sector states that involves the antisymmetric matrix $\epsilon_{\dot{A}\dot{B}}$ is proportional to a combination of total global descendant states of chiral primaries and so have already contributed to the supergraviton index.

\subsubsection*{\underline{$h=\frac32$, $j=\frac12$}}

The order $q^{3/2}$ term of \eqref{eq.MEGgen-MEGgrav} comes from a mixture of twisted and untwisted sector states, as well as a mixture of global descendants and global primaries. In the twisted sector we have the $G^{+\Cd(\mathrm{T})}_{-\frac12}$ total global descendants of \eqref{eq.twisted01singletons} contributing $4q^{\frac32}y$, along with the global primaries\footnote{Once again these states require the addition of terms involving global descendants of chiral primaries in order to have a global primary combination. We neglect these terms here.}
\begin{align}
    \psi^{+\Ad}_{-\frac12} \psi^{-\Bd}_{-\frac12}|-\rangle_2\otimes\begin{cases}
        \ |\ald\rangle_2\vspace{2pt}\\
        \ |\Bd\rangle_2
    \end{cases}  ,\qquad
    J^3_{-1}|-\rangle_2\otimes\begin{cases}
        \ |\ald\rangle_2\vspace{2pt}\\
        \ |\Bd\rangle_2
    \end{cases}  \ ,
\end{align}
which contribute $(4+1)q^{\frac32}y$.

In the untwisted sector we have the $G^{+\Cd(\mathrm{T})}_{-\frac12}$ total global descendants of \eqref{eq.untwisted01singletons} contributing $12q^{\frac32}y$ to the modified elliptic genus, as well as the global primaries
\begin{align}
    G^{+\Ad(T)}_{-\frac32}|-\rangle_1^{(1)}|-\rangle_1^{(2)} \otimes\begin{cases}
        \ \Big(|\ald\rangle_1^{(1)}|\bed\rangle_1^{(2)} + |\bed\rangle_1^{(1)}|\ald\rangle_1^{(2)}\Big) \vspace{1pt}\\
        \ \Big(|\ald\rangle_1^{(1)}|\Cd\rangle_1^{(2)} + |\Cd\rangle_1^{(1)}|\ald\rangle_1^{(2)}\Big) \vspace{1pt}\\
        \ \Big(|\Cd\rangle_1^{(1)}|\Dd\rangle_1^{(2)} - |\Dd\rangle_1^{(1)}|\Cd\rangle_1^{(2)}\Big)
    \end{cases} \ ,
\end{align}
contributing $-4q^{\frac32}y$ and 
\begin{align}
    \psi^{-\Ad(T)}_{-\frac12}\psi^{+\Bd(T)}_{-\frac12}\Big(|-\rangle_1^{(1)}|\Cd\rangle_1^{(2)}-|\Cd\rangle_1^{(1)}|-\rangle_1^{(2)}\Big) \otimes\begin{cases}
        \ \Big(|\ald\rangle_1^{(1)}|\bed\rangle_1^{(2)} - |\bed\rangle_1^{(1)}|\ald\rangle_1^{(2)}\Big) \vspace{1pt}\\
        \ \Big(|\ald\rangle_1^{(1)}|\Cd\rangle_1^{(2)} - |\Cd\rangle_1^{(1)}|\ald\rangle_1^{(2)}\Big) \vspace{1pt}\\
        \ \Big(|\Cd\rangle_1^{(1)}|\Dd\rangle_1^{(2)} + |\Dd\rangle_1^{(1)}|\Cd\rangle_1^{(2)}\Big)
    \end{cases} \ ,
\end{align}
contributing $16q^{\frac32}y$. The states accounting for the $33q^{\frac32}y^{-1}$ term in \eqref{eq.MEGgen-MEGgrav} are then just the $J^{-(\mathrm{T})}_0$ global descendants of the above states.

The fact that the generalised supergraviton and CFT indices agree at orders $q$ and $q^{\frac32}$ can then be explained by decomposing the CFT modified elliptic genus \eqref{eq.CFTMEGexp} into total affine characters (where we suppress the arguments of the characters)
\begin{align} \label{eq.CFTMEGdecomp}
    \hat{\cE}^{\mathrm{CFT}}_{\mathrm{NS}}(q,y;2) = 2\mathbf{\Phi}^{(s)}_{0;2} + 5\mathbf{\Phi}^{(s)}_{\frac12;2} -42\mathbf{\Phi}^{(\ell)}_{0,2;2} - 70\mathbf{\Phi}^{(\ell)}_{0,3;2} - 324\mathbf{\Phi}^{(\ell)}_{0,4;2} + \cdots \ .
\end{align}
The short affine characters here are exactly the generalised supergraviton index that we find and there are no long $\mathbf{\Phi}^{(\ell)}_{0,1;2}$ characters, which would also contribute at orders $q$ and $q^{\frac32}$. The states dual to typical D1-D5-P black hole microstates (accounting for the vast majority of its entropy) should therefore be found in the long character contributions present in \eqref{eq.CFTMEGdecomp}.

The lack of $\mathbf{\Phi}^{(\ell)}_{0,1;2}$ long characters in \eqref{eq.CFTMEGdecomp} was checked explicitly using conformal perturbation theory in \cite{Guo:2019ady,Guo:2020gxm} where it was found that all such states pair up and lift. This can be seen by decomposing the individual twist sector contributions to the modified elliptic genus into affine characters
\begin{align} 
    \hat{\cE}^{\mathrm{CFT}}_{\mathrm{NS}}(q,y;\{1,1\}) &= 2\mathbf{\Phi}^{(s)}_{0;2} + 4\mathbf{\Phi}^{(s)}_{\frac12;2} + 6\mathbf{\Phi}^{(\ell)}_{0,1;2} - 14\mathbf{\Phi}^{(\ell)}_{0,2;2} +28\mathbf{\Phi}^{(\ell)}_{0,3;2} - 42\mathbf{\Phi}^{(\ell)}_{0,4;2} + \cdots \ ,\label{eq.CFTMEGdecomp11}\\
    \hat{\cE}^{\mathrm{CFT}}_{\mathrm{NS}}(q,y;\{2\}) &= 
    \mathbf{\Phi}^{(s)}_{\frac12;2} - 6\mathbf{\Phi}^{(\ell)}_{0,1;2} - 28\mathbf{\Phi}^{(\ell)}_{0,2;2} - 98 \mathbf{\Phi}^{(\ell)}_{0,3;2} - 282\mathbf{\Phi}^{(\ell)}_{0,4;2} + \cdots \ ,\label{eq.CFTMEGdecomp2}
\end{align}
where the contributions of $\mathbf{\Phi}^{(\ell)}_{0,1;2}$ from the two twist sectors cancel, in line with the index being invariant under deformations of the symmetric orbifold theory. Therefore, despite the possibility discussed at the end of section~\ref{ssec:subtract} of additional long characters contributing to the generalised supergraviton index, the validity of our result \eqref{eq.T4N11MEG} is guaranteed at least up to and including $h=\frac32$ since all affine primaries with $h=1$ lift and representation theory prohibits $N=2$ affine primaries with $h=\frac32$ (see appendix~\ref{ssec:chars}).\footnote{Potential corrections at higher levels will be addressed in future work \cite{wip}.}

\section{$K3$ generalised supergraviton index}
\label{sec:K3}

For the case of $\mathrm{Sym}^N(K3)$ a similar story to that discussed in section \ref{sec:T4} for $\mathrm{Sym}^N(T^4)$ plays out. As described in section~\ref{sec:background}, the relevant differences in the context of generalising the supergraviton index are the symmetry algebra and the BPS spectrum. The symmetry algebra is now just the small $\cN=(4,4)$ algebra \eqref{eq.commcurrents2} and the single-strand chiral primary spectrum, as seen in \eqref{eq.K3LCP}, consists of the states $|\alpha\ald\rangle_a$ charged under the $SU(2)_L\times SU(2)_R$ R-symmetry (which were also present in $\mathrm{Sym}^N(T^4)$) and R-charge neutral states $|I\rangle_a$. Of these neutral states, there are the $|\Ad\Bd\rangle_a$ which appear in the $T^4$ theory, and the $|r\rangle_a$ with $r=1,\dots,16$ which come from the $\mathbb{Z}_2$ twisted sector of the $T^4/\mathbb{Z}_2$ description of $K3$ (as argued in section~\ref{sec:background}, this description of $K3$ is sufficient for our purposes).
\begin{figure}[ht]
\begin{center}
    \begin{adjustbox}{max totalsize={0.9\textwidth}{\textheight},center}
    \begin{tikzpicture}[squarednode/.style={rectangle, draw=black!0, very thick, minimum size=5mm},]
        \node[squarednode] (state-) {$|{-}\,\ald\rangle_a$};
        \node[squarednode] (state+) [above=1.25cm of state-] {$|{+}\,\ald\rangle_a$} ;
        \draw[-Stealth] (state+.south) -- node[left=3pt]{$J^{-}_{1}$} (state-.north);
    \end{tikzpicture}
    \end{adjustbox}
    \caption{\it The relations between single-strand chiral primaries of $\mathrm{Sym}^{N}(K3)$ under the action of the affine mode $J^{-}_{1}$. \label{fig:7}}
    \end{center}
\end{figure}
With the aim of finding a generalised supergraviton elliptic genus by repeating the logic laid out in steps~\ref{list:i}--\ref{list:iib} of section~\ref{sec:T4}, we should first understand whether any single-strand chiral primaries are related by the action of modes of the small $\cN=4$ algebra \eqref{eq.commcurrents2}. This turns out to be simpler than in the case of $T^4$, with there being no such relations involving the R-charge neutral states $|I\rangle_a$. The only relations are then between the $|\alpha\ald\rangle_a$ by the action of $J^{\pm}_{\mp1}$, as summarised in figure~\ref{fig:7}.

Since the global subalgebra is also $SU(1,1\,|\,2)_L\times SU(1,1\,|\,2)_R$ here, the arguments given in step~\ref{list:2} of section~\ref{ssec:subtract} can again be used to take the relations between these single-strand chiral primaries to relations between the short $SU(1,1\,|\,2)_L$ multiplets built upon them.

\subsection{Subtraction and promotion: $N=2$}

We will again focus uniquely on the case of $N=2$ here; in the $N=\{2\}$ twisted sector the single-strand relations in figure~\ref{fig:7} are sufficient to understand which global primaries are not affine primaries, whereas in the $N=\{1,1\}$ untwisted sector we will once again need to consider relations between multi-strand states.

Working with the multi-supergraviton elliptic genus refined by the $SU(2)_R$ charge, the twisted and untwisted sector contributions can be obtained from the $n=2$ and $n=1$ contributions to the order $p^2$ terms in \eqref{eq.K3NptEG}. Thus we define
\begin{align}
    Z_{K3}^{\mathrm{graviton}}(q,y,\ty;\{2\}) &\equiv \prod_{m,\ell,\tilde{\ell}} \Big(1-p^2q^my^{\ell}\ty^{\tilde{\ell}}\Big)^{-c^{K3}_{\mathrm{graviton}}(2,m,\ell,\tilde{\ell})}\bigg|_{p^2} = X^{K3}_2(q,y,\ty)\ , \label{eq.ZK3N2}\\
    Z_{K3}^{\mathrm{graviton}}(q,y,\ty;\{1,1\}) &\equiv \prod_{m,\ell,\tilde{\ell}} \Big(1-pq^my^{\ell}\ty^{\tilde{\ell}}\Big)^{-c^{K3}_{\mathrm{graviton}}(1,m,\ell,\tilde{\ell})}\bigg|_{p^2} \nonumber\\
    &= \frac12 \big(X^{K3}_1(q,y,\ty)\big)^2 + \frac12 X^{K3}_1(q^2,y^2,\ty^2) \ ,\label{eq.ZK3N11}
\end{align}
where $X_n^{K3}(q,y,\ty)$ is given in \eqref{eq.XK3def}.

Starting with the twisted sector, we once again introduce additional fugacities $\xi_-$, $\xi_0$ and $\xi_+$ to count $|{-}\,\ald\rangle_a$, $|I\rangle_a$ and $|{+}\,\ald\rangle_a$ strands respectively. Then \eqref{eq.ZK3N2} reads
\begin{align} \label{eq.ZK3N2xi}
    Z_{K3}^{\mathrm{graviton}}(q,y,\ty,\xi;\{2\}) = \xi_-\,\phi^{(s)}_{\frac{1}{2}}\!(q,y)\,\tilde{\chi}_{\frac12;2}(\ty) + 20\xi_0\,\phi^{(s)}_{1}\!(q,y)\,\tilde{\chi}_{0;2}(\ty) + \xi_+\,\phi^{(s)}_{\frac{3}{2}}\!(q,y)\,\tilde{\chi}_{\frac12;2}(\ty) \ ,
\end{align}
and we see that the only total global primaries are the single-strand chiral primaries in \eqref{eq.K3LCP} with $a=2$. From figure~\ref{fig:7} and step \ref{list:2} of section~\ref{ssec:subtract}, the short global multiplet built on the state $|{+}\,\ald\rangle_2$ is related to the multiplet built on $|{-}\,\ald\rangle_2$. The generalised supergraviton index for the twisted sector is thus obtained from \eqref{eq.ZK3N2xi} by subtracting the order $\xi_+$ term and then promoting the remaining global characters to affine characters (of the small $\cN=4$ algebra), giving
\begin{align} \label{eq.ZK3N2gen}
    Z_{K3}^{\mathrm{gen}}(q,y,\ty;\{2\}) = \Phi^{(s)}_{\frac{1}{2};2}(q,y)\,\tilde{\chi}_{\frac12;2}(\ty) + 20\Phi^{(s)}_{1;2}(q,y)\,\tilde{\chi}_{0;2}(\ty) \ ,
\end{align}
where we have set the $\xi$ fugacities to 1.

In terms of global characters, the untwisted-sector supergraviton index \eqref{eq.ZK3N11} with the additional strand-type fugacities is given by (here we use the shorthand $Z_{K3\{1,1\}}^{\mathrm{graviton}}\equiv Z_{K3}^{\mathrm{graviton}}(q,y,\ty,\xi;\{1,1\})$) 
\begin{align} \label{eq.ZK3N11xi}
    Z_{K3\{1,1\}}^{\mathrm{graviton}} &= \xi_-^2 \frac12\Big[ \big(\phi^{(s)}_{0}\!(q,y)\,\tilde{\chi}_{\frac12;1}(\ty)\big)^2 + \phi^{(s)}_{0}\!(q^2,y^2)\,\tilde{\chi}_{\frac12;1}(\ty^2) \Big] + \xi_-\xi_+ \phi^{(s)}_{0}\!(q,y)\,\phi^{(s)}_{1}\!(q,y)\big(\tilde{\chi}_{\frac12;1}(\ty)\big)^2 \nonumber\\
    &\quad+ \xi_+^2 \frac12\Big[\big(\phi^{(s)}_{1}\!(q,y)\,\tilde{\chi}_{\frac12;1}(\ty)\big)^2 + \phi^{(s)}_{1}\!(q^2,y^2)\,\tilde{\chi}_{\frac12;1}(\ty^2) \Big] \nonumber\\
    &\quad + 20\xi_0\xi_-\,\phi^{(s)}_{\frac12}\!(q,y)\,\phi^{(s)}_{0}\!(q,y)\,\tilde{\chi}_{0;1}(\ty)\tilde{\chi}_{\frac12;1}(\ty) + 20\xi_0\xi_+\,\phi^{(s)}_{\frac12}\!(q,y)\, \phi^{(s)}_{1}\!(q,y)\,\tilde{\chi}_{0;1}(\ty)\tilde{\chi}_{\frac12;1}(\ty) \nonumber\\
    &\quad + \xi_0^2\frac12\Big[ \big(20\phi^{(s)}_{\frac12}\!(q,y)\,\tilde{\chi}_{0;1}(\ty)\big)^2 + 20\phi^{(s)}_{\frac12}\!(q^2,y^2)\,\tilde{\chi}_{0;1}(\ty^2)\Big] \ .
\end{align}
To understand which products of global characters we should subtract, first note that states involving different numbers of $|I\rangle_1$ strands cannot be related by affine modes, making them effectively spectator strands. The decomposition \eqref{eq.ZK3N11xi} can then be analysed order by order in $\xi_0$. This proceeds as follows:
\begin{itemize}
    \item At order $\xi_0^2$ \eqref{eq.ZK3N11xi} receives contributions from states constructed from two strands of global descendants of the $|I\rangle_1$ single-strand chiral primaries. These states are not related to any other supergraviton states under the action of total affine modes and so we should keep all global characters at order $\xi_0^2$. Using the character product expansion in \eqref{eq.SU112CharProd} and \eqref{eq.SU112CharPow}, these are
    \begin{align}
        Z_{K3\{1,1\}}^{\mathrm{graviton}}\Big|_{\xi_0^2} &= \bigg[210\Phi^{(s)}_{1;2}(q,y) + 190\sum_{\substack{h\geq1\\\mathrm{odd}}} \Phi^{(\ell)}_{0,h;2}(q,y) + 210\sum_{\substack{h\geq2\\\mathrm{even}}} \Phi^{(\ell)}_{0,h;2}(q,y)\bigg] \tilde{\chi}_{0;2}(\ty) \, .
    \end{align}
    \item At orders $\xi_0\xi_{\pm}$ we have
    \begin{align}
        Z_{K3\{1,1\}}^{\mathrm{graviton}}\Big|_{\xi_0} &=  20\bigg[\xi_-\,\phi^{(s)}_{\frac12}\!(q,y)\,\phi^{(s)}_{0}(q,y) + \xi_+\,\phi^{(s)}_{\frac12}\!(q,y)\, \phi^{(s)}_{1}\!(q,y)\bigg]\,\tilde{\chi}_{0;1}(\ty)\tilde{\chi}_{\frac12;1}(\ty) \nonumber\\
        &= 20\bigg[\xi_-\,\phi^{(s)}_{\frac12}\!(q,y) + \xi_+\bigg(\phi^{(s)}_{\frac32}\!(q,y) + \sum_{h\geq1} \phi^{(\ell)}_{\frac12,h+\frac12}\!(q,y)\bigg)\bigg] \,\tilde{\chi}_{\frac12;2}(\ty) \ ,
    \end{align}    
    where the global character product formula \eqref{eq.SU112CharProd} was used. So we see that there is an $SU(2)_R$ doublet of chiral primaries contributing at order $\xi_-$ and $\xi_+$, containing one $|I\rangle_1$ strand and one $|{\pm}\,\ald\rangle_1$ strand in an overall copy-symmetric combination, \textit{i.e.},
    \begin{equation}
        20\xi_0\xi_{\pm}\phi^{(s)}_{\frac12}\tilde{\chi}_{\frac12;2}:\quad |{\pm}\,\ald\rangle_1^{(1)}|I\rangle_1^{(2)} + |I\rangle_1^{(1)}|{\pm}\,\ald\rangle_1^{(2)} \ .
    \end{equation}
    Due to the one spectator strand, the relation between these states under the action of the affine mode $J^{-(\mathrm{T})}_{1}$ is exactly the same as in the single-strand case -- as given in figure~\ref{fig:9}. The non-chiral primary states appearing at order $\xi_+$ are likewise related to the order $\xi_-$ states in the same way using the arguments of section~\ref{ssec:subtract}. The conclusion is thus that all global characters at order $\xi_0\xi_+$ should be subtracted from the index.
    \item Whilst there are no spectator strands at order $(\xi_0)^0$, the contributing chiral primaries involve strands only of the type $|\alpha\ald\rangle_1$ and so the relations between the multi-strand chiral primaries are a subset of the relations previously discussed in the $T^4$ case (see figure~\ref{fig:3}). \textit{i.e.} by expanding the products of individual global characters in the order $(\xi_0)^0$ terms of \eqref{eq.ZK3N11xi} in terms of total global characters using the formulae in \eqref{eq.SU112CharProd} and \eqref{eq.SU112CharPow}, we find the following short total global characters
    \begin{align} \label{eq.ZK3N11xi0s}
        Z_{K3\{1,1\}}^{\mathrm{graviton}}\Big|_{(\xi_0)^0} &\supset \xi_-^2\phi^{(s)}_0\!(q,y)\tilde{\chi}_{1;2}(\ty) + \xi_-\xi_+ \phi^{(s)}_1\!(q,y) \Big[\tilde{\chi}_{1;2}(\ty) + \tilde{\chi}_{0;2}(\ty)\Big] \nonumber\\
        &\ \ + \xi_+^2 \phi^{(s)}_2\!(q,y)\tilde{\chi}_{1;2}(\ty) \ ,
    \end{align}
    where we have also used the $SU(2)_R$ character product formulae in \eqref{eq.SU2CharProd} and \eqref{eq.SU2CharPow}. The highest-weight states of each of these short global characters are
    \begin{subequations} \label{eq.K3xi0CPs}
    \begin{align} 
        \xi_-^2:\quad &|{-}\,\ald\rangle_1^{(1)}|{-}\bed\rangle_1^{(2)} + |{-}\bed\rangle_1^{(1)}|{-}\,\ald\rangle_1^{(2)} \ , \label{eq.K3xi0CPs1} \\
        \xi_-\xi_+:\quad &\Big(|{+}\rangle_1^{(1)}|{-}\rangle_1^{(2)} \pm |-\rangle_1^{(1)}|+\rangle_1^{(2)}\Big)\Big(|\ald\rangle_1^{(1)}|\bed\rangle_1^{(2)} \pm |\bed\rangle_1^{(1)}|\ald\rangle_1^{(2)}\Big) \ , \label{eq.K3xi0CPs2} \\
        \xi_+^2:\quad &|{+}\,\ald\rangle_1^{(1)}|{+}\bed\rangle_1^{(2)} + |{+}\bed\rangle_1^{(1)}|{+}\,\ald\rangle_1^{(2)} \ , \label{eq.K3xi0CPs3}
    \end{align}
    \end{subequations}
    where the $\pm$ in states \eqref{eq.K3xi0CPs2} corresponds to $SU(2)_R$ triplet and singlets respectively. The relations between the chiral primaries are summarised in figure~\ref{fig:9} and, along with the arguments of section~\ref{ssec:subtract}, tell us that only the short characters corresponding to the state \eqref{eq.K3xi0CPs1} and the singlet state in \eqref{eq.K3xi0CPs2} should survive the subtraction. Furthermore, step~\ref{list:4} in section~\ref{ssec:subtract} tells us that the same subtraction should apply not only to the short total characters in \eqref{eq.ZK3N11xi0s}, but also to all long total characters stemming from the same product of short individual characters in \eqref{eq.ZK3N11xi}.
\end{itemize}
    \begin{figure}[ht]
\begin{center}
        \begin{adjustbox}{max totalsize={0.9\textwidth}{\textheight},center}
            \begin{tikzpicture}[squarednode/.style={rectangle, draw=black!0, very thick, minimum size=5mm},align=center,]
            
                \node[squarednode] (++) {$|{+}\,\ald\rangle_1^{(1)}|{+}\,\bed\rangle_1^{(2)} + |{+}\,\bed\rangle_1^{(1)}|{+}\,\ald\rangle_1^{(2)}$};
                \node[squarednode] (+-) [below=2cm of ++] {$\big(|{+}\rangle_1^{(1)}|-\rangle_1^{(2)} + |{-}\rangle_1^{(1)}|{+}\rangle_1^{(2)}\big)\big(|\ald\rangle_1^{(1)}|\bed\rangle_1^{(2)} + |\bed\rangle_1^{(1)}|\ald\rangle_1^{(2)}\big)$};
                \node[squarednode] (--) [below=2cm of +-] {$|{-}\,\ald\rangle_1^{(1)}|{-}\,\bed\rangle_1^{(2)} + |{-}\,\bed\rangle_1^{(1)}|{-}\,\ald\rangle_1^{(2)}$};
                
                \node[squarednode] (+I) [below right=0.6cm and 2cm of ++] {$|{+}\,\ald\rangle_1^{(1)}|I\rangle_1^{(2)} + |I\rangle_1^{(1)}|{+}\,\ald\rangle_1^{(2)}$};
                \node[squarednode] (-I) [below=2.3cm of +I] {$|{-}\,\ald\rangle_1^{(1)}|I\rangle_1^{(2)} + |I\rangle_1^{(1)}|{-}\,\ald\rangle_1^{(2)}$};

                \draw[-Stealth] (++.south) -- node[left=3pt]{$J^{-(\mathrm{T})}_1$} (+-.north);
                \draw[-Stealth] (+-.south) -- node[left=3pt]{$J^{-(\mathrm{T})}_1$} (--.north);
                \draw[-Stealth] (+I.south) -- node[left=3pt]{$J^{-(\mathrm{T})}_1$} (-I.north);
            \end{tikzpicture}
        \end{adjustbox}
        \caption{\it Relations between multi-strand chiral primaries contributing to the untwisted sector supergraviton index \eqref{eq.ZK3N11xi} for $K3$.\label{fig:9}}
    \end{center}
\end{figure}
As discussed at the end of section \ref{ssec:subtract}, it is possible that additional long characters should be kept and promoted due to the formation of affine primaries from combinations of supergraviton and singleton states, which we have not taken into account here. Since for $\mathrm{Sym}^2(K3)$ the first singleton states are at $h=\frac32$, such an effect can only add to the result of the above discussion starting from that level. In the untwisted sector we are thus left with the reduced supergraviton index up to order $q$
\begin{align}
    Z^{\mathrm{red}}_{K3}(q,y,\ty,\xi;\{1,1\}) &= \xi_-^2 \phi^{(s)}_{0}(q,y)\tilde{\chi}_{1;2}(\ty) + \xi_-\xi_+ \phi^{(s)}_{1}(q,y)\tilde{\chi}_{0;2}(\ty) + 20\xi_0\xi_- \phi^{(s)}_{\frac12}(q,y) \tilde{\chi}_{\frac12;2}(\ty) \nonumber\\
    &\ + \xi_0^2 \Big[210\phi^{(s)}_{1}(q,y) + 190 \phi^{(\ell)}_{0,1}(q,y) \Big] \tilde{\chi}_{0;2}(\ty) + \cdots\ ,
\end{align}
and so, by promoting these total global characters to total affine characters, we find the untwisted sector generalised supergraviton index (setting $\xi_-=\xi_0=\xi_+=1$)
\begin{align} \label{eq.ZK3N11gen}
    Z^{\mathrm{gen}}_{K3}(q,y,\ty;\{1,1\}) &= \Phi^{(s)}_{0;2}(q,y)\tilde{\chi}_{1;2}(\ty) + \Phi^{(s)}_{1;2}(q,y)\tilde{\chi}_{0;2}(\ty) + 20 \Phi^{(s)}_{\frac12;2}(q,y) \tilde{\chi}_{\frac12;2}(\ty) \nonumber\\
    &\ + \Big[210\Phi^{(s)}_{1;2}(q,y) + 190 \Phi^{(\ell)}_{0,1;2}(q,y)\Big] \tilde{\chi}_{0;2}(\ty) \ .
\end{align}

Setting $\ty=1$, the sum of \eqref{eq.ZK3N11gen} and \eqref{eq.ZK3N2gen} then gives the generalised supergraviton elliptic genus for $N=2$ up to and including order $q$
\begin{equation} \label{eq.ZK3gen}
    \cE^{\mathrm{gen}}_{K3}(q,y;2) = 3\Phi^{(s)}_{0;2}(q,y) + 42\Phi^{(s)}_{\frac12;2}(q,y) + 231\Phi^{(s)}_{1;2}(q,y) + 190 \Phi^{(\ell)}_{0,1;2}(q,y) + \cdots\ .
\end{equation}

\subsection{Comparisons}

Since the $Q$-cohomology structure of the BPS spectrum was not considered for $\mathrm{Sym}^N(K3)$ in \cite{Chang:2025rqy}, we cannot make the same comparisons as in section~\ref{ssec:Fortuity} with the lowest-lying singleton states that we have added to the supergraviton index. However, by decomposing the small $\cN=4$ characters in \eqref{eq.ZK3gen} into global characters we find that the lowest-lying singleton states contribute at $(h,j)=(\frac32,\frac12)$, as seen from
\begin{align} \label{eq.K3gen-graviton}
    \cE^{\mathrm{gen}}_{K3}(q,y;2) - \cE^{\mathrm{graviton}}_{K3}(q,y;2) = 2 \phi^{(\ell)}_{\frac12,\frac32} + \cdots \ .
\end{align}
These $(h,j)=(\frac32,\frac12)$ singleton states are only from the twisted sector, as seen from the comparisons
\begin{align}
    \cE^{\mathrm{gen}}_{K3}(q,y;\{1,1\}) - \cE^{\mathrm{graviton}}_{K3}(q,y;\{1,1\}) &=  \mathcal{O}(q^2) \ , \\
    \cE^{\mathrm{gen}}_{K3}(q,y;\{2\}) - \cE^{\mathrm{graviton}}_{K3}(q,y;\{2\}) &= 2 \phi^{(\ell)}_{\frac12,\frac32} + \cdots \ ,
\end{align}
and the explicit form of these states is
\begin{align} \label{eq.twisted1/23/2fortuitous}
    J^3_{-1}|{-}\,\ald\rangle_2 \ .
\end{align}

The de Boer bound for $\mathrm{Sym}^2(K3)$ \eqref{deBoer_bound_K3} is $h\leq\frac34$; namely, the supergraviton and CFT elliptic genera agree for $h=0$ and $h=\tfrac12$, with a mismatch setting in at $h=1$.
This means that fortuitous states start to exist already at $h=1$, below
level $h=\frac32$ where the lowest-lying singleton states \eqref{eq.twisted1/23/2fortuitous} appear.
This is qualitatively different from the case of $T^4$ that was discussed in section~\ref{ssec:Fortuity}.

We can see this explicitly by comparing the $q$-expansions for $N=2$ of the CFT \eqref{eq.cftEGgenNS}, supergraviton \eqref{eq.K3gravEGdef} and generalised supergraviton \eqref{eq.ZK3gen} elliptic genera
\begin{align}
    \cE^{\mathrm{CFT}}_{\mathrm{NS}}(q,y;2) &\approx 3 + \sqrt{q} \Big(42y^{-1} + 42y\Big) + q\Big(234y^{-2} + 360 + 234y^2\Big) \nonumber\\
    &\hspace{4.05cm}+ q^{\frac32} \Big(42y^{-3} -3114y^{-1} - 3114y + 42y^3\Big) + \cdots \ ,\label{eq.K3EGcompare1}\\
    \cE^{\mathrm{graviton}}_{K3}(q,y;2) &\approx 3 + \sqrt{q} \Big(42y^{-1} + 42y\Big) + q\Big(234y^{-2} + 340 + 234y^2\Big) \nonumber\\
    &\hspace{4.05cm}+ q^{\frac32} \Big(42y^{-3} -724y^{-1} - 724y + 42y^3\Big) + \cdots \ ,\label{eq.K3EGcompare2}\\
    \cE^{\mathrm{gen}}_{K3}(q,y;2) &\approx 3 + \sqrt{q} \Big(42y^{-1} + 42y\Big) + q\Big(234y^{-2} + 340 + 234y^2\Big) \nonumber\\
    &\hspace{4.05cm}+ q^{\frac32} \Big(42y^{-3} -722y^{-1} - 722y + 42y^3\Big) + \cdots \ .\label{eq.K3EGcompare3}
\end{align}
We see that although the singletons modify terms starting at $q^\frac32 y^{\pm1}$ this does not change the mismatch of the $q y^0$ term with that of the CFT elliptic genus. We give the explicit form of the CFT states which yield this $20q$ difference (the lowest-lying fortuitous states) in section \ref{sec:App2}.

While the de Boer bound is not improved by the addition of singleton states for $\mathrm{Sym}^2(K3)$, it is still possible that for higher values of $N$ there is improvement. We leave such investigation for future work \cite{wip}.

Since we now have the $K3$ generalised supergraviton index for $N=2$ \eqref{eq.ZK3gen}, under the assumption that supergraviton and singleton states form the complete set of monotone states, we can thus write the first character contributions to the fortuitous index for $\mathrm{Sym}^2(K3)$ as
\begin{align} \label{eq.K3forEG}
    \cE^{\mathrm{for}}_{K3}(q,y;2) \equiv \cE^{\mathrm{CFT}}_{\mathrm{NS}}(q,y;2) - \cE^{\mathrm{gen}}_{K3}(q,y;2) &= 20\Phi^{(\ell)}_{0,1;2} +\cdots \ .
\end{align}
As in the case of $T^4$, we expect that the fact that this fortuitous index decomposes into only long affine characters is a generic feature of the fortuitous index for any $N$.

\subsection{Lowest-lying fortuitous states for $\mathrm{Sym}^2(K3)$} \label{sec:App2}

By comparing the $q$-expansions of the CFT elliptic genus \eqref{eq.K3EGcompare1} and the supergraviton elliptic genus \eqref{eq.K3EGcompare2} we see that the first non-supergraviton states contribute with quantum numbers $(h,j)=(1,0)$. From \eqref{eq.K3gen-graviton} we also see that these are not singletons since those start contributing only at $h=\frac32$.\footnote{This also means that any additional long characters contributing due to affine primary combinations of supergraviton and singleton states do not affect this discussion.}

The CFT states with $(h,j)=(1,0)$ that are not supergraviton states contribute to the elliptic genus at order $qy^0$. Explicitly, in the untwisted sector these are given by (up to copy symmetrisation)
\begin{subequations} \label{kowe21Jun24}
 \begin{align}
  +12q:\qquad&\sigma^i_{\dot{A}\dot{B}}\ket{-\ald}_1\,\psi^{-\dot{A}}_{-{1\over 2}}\ket{\dot{B}\bed}_1
  &\quad&
  (0;{\bf 1,3})
  +2( 1;{\bf 1,3})
  +(2 ;{\bf 1,3}) \ ,\label{eq.BHstate1}\\
  -192q:\qquad&\sigma^i_{\dot{A}\dot{B}}\ket{-\ald}_1\,\alpha^{\dot{A}A}_{-{1\over 2}}\psi^{-\dot{B}}_0\ket{r}_1
  &&(\tfrac{1}{2};{\bf 2,3})_r+(\tfrac{3}{2};{\bf 2,3})_r \ ,\label{eq.BHstate2}
 \end{align}
\end{subequations}
and in the twisted sector
\begin{subequations} \label{kowg21Jun24}
 \begin{align}
  +4q:\qquad&\ep_{\dot{C}\dot{D}}\psi^{-\dot{C}}_0\psi^{-\dot{D}}_0\ket{\dot{A}\dot{B}}_2 + f\, 
  \psi^{-\dot{A}}_{-{1\over 2}}\ket{-\dot{B}}_2&\quad&(1;{\bf 1,1})+(1;{\bf 1,3}) \ , \label{eq.BHstate3}\\
  -12q:\qquad&\sigma^i_{\dot{A}\dot{B}}\alpha^{\dot{A}A}_{-{1\over 2}}\psi^{-\dot{B}}_0\ket{-\ald}_2&&(\tfrac12;{\bf 2,3})+(\tfrac32;{\bf 2,3}) \ ,\label{eq.BHstate4}\\
  +160q:\qquad&\alpha^{\dot{A}A}_{-{1\over 4}}\alpha^{\dot{B}B}_{-{1\over 4}}J^-_{1\over 2}\ket{r}_2&&(1,{\bf 3,3})_r+(1;{\bf 1,1})_r \ ,\label{eq.BHstate5}\\
  +48q:\qquad&\sigma^i_{\dot{A}\dot{B}} \psi^{-\dot{A}}_{-{1\over 4}}\psi^{-\dot{B}}_{{1\over 4}}\ket{r}_2&&(1;{\bf 1,3})_r \ ,\label{eq.BHstate6}
 \end{align}
\end{subequations}
where $r=1,\dots,16$. We have included the combination $(\tj;\bf{d_1},\bf{d_2})$ where $\tj$ is the $\tilde{J}^3_0$ eigenvalue of $SU(2)_R$ in the NS sector and $\bf{d_1}$ and $\bf{d_2}$ are the dimensions of the $SU(2)_1$ and $SU(2)_2$ representations respectively. The coefficient $f$ in \eqref{eq.BHstate3} is chosen so that the states are orthogonal to the supergraviton states $J^-_{0}|\Ad\Bd\rangle_2\sim \ep_{\dot{C}\dot{D}}\psi^{-\dot{C}}_0\psi^{-\dot{D}}_0\ket{\dot{A}\dot{B}}_2 -2\psi^{-\Ad}_{-\frac12}|{-}\Bd\rangle_2$. The single-strand chiral primaries in $\mathrm{Sym}^N(K3)$ are detailed in \eqref{eq.K3LCP}.

The states in \eqref{kowe21Jun24} and \eqref{kowg21Jun24} contribute a total of $+20q$ to the NS CFT elliptic genus and could represent the first fortuitous states in $\mathrm{Sym}^2(K3)$. However, one should first check whether there exist cancellations due to lifting. In order for states to combine and lift they should form a long supersymmetry multiplet when the symmetric orbifold is deformed. Roughly speaking, such states should be related by the action of the deformed right-moving supersymmetry modes $\tilde{G}^{\dot{+}A}_{-\frac12}(\lambda)$ and so form a quartet of states in representations with $(\tj;\bf{d_1},\bf{d_2})$ given by\footnote{As well as these constraints of the various representations of lifted states, they should also be in particular twist sectors since the deformed supersymmetry mode $\tilde{G}^{\dot{+}A}_{-\frac12}(\lambda)$ contains a twist-2 operator. In this case of $N=2$, if the first state with right-moving R charge $\tj$ is in the untwisted sector then the two states with $\tj+\frac12$ must be in the twisted sector and the state with $\tj+1$ must again be in the untwisted sector. Equivalently, the pattern for a twisted sector state with R charge $\tj$ is that the two with $\tj+\frac12$ are in the untwisted sector and the one with $\tj+1$ is in the twisted sector. See \cite{Guo:2019pzk} for further details.}
\begin{align} \label{eq.liftingReps}
    (\tj;{\bf d_1},{\bf d_2}) \quad,\quad (\tj+\tfrac12;{\bf d_1}\otimes{\bf 2},{\bf d_2}) \quad,\quad (\tj+1;{\bf d_1},{\bf d_2}) \ .
\end{align}
Within the set of states \eqref{kowe21Jun24} and \eqref{kowg21Jun24} there are three such combinations of multiplets and assuming that these all lift leads to the reduced set of states
\begin{subequations}\label{eq.BHstate322}
    \begin{align}
    +4q:\qquad&\ep_{\dot{C}\dot{D}}\psi^{-\dot{C}}_0\psi^{-\dot{D}}_0\ket{\dot{A}\dot{B}}_2 + f\, 
  \psi^{-\dot{A}}_{-{1\over 2}}\ket{-\dot{B}}_2&\quad&(1;{\bf 1,1})+(1;{\bf 1,3}) \ , \label{eq.BHstate32}\\
  +16q:\qquad&\ep_{\Ad\Bd}\ep_{AB}\alpha^{\dot{A}A}_{-{1\over 4}}\alpha^{\dot{B}B}_{-{1\over 4}}J^-_{1\over 2}\ket{r}_2&&(1;{\bf 1,1})_r \ ,\label{eq.BHstate52}
    \end{align}
\end{subequations}
These states are therefore the lowest-lying fortuitous states for $\mathrm{Sym}^2(K3)$. We note that the states in \eqref{eq.BHstate32} are also present in $\mathrm{Sym}^2(T^4)$; however, at the level of the modified elliptic genus we did not observe any contributing fortuitous states (see \eqref{eq.CFTMEGall}). There must then be additional states present in $\mathrm{Sym}^2(T^4)$ but not in $\mathrm{Sym}^2(K3)$ that cancel the contribution of the states \eqref{eq.BHstate32}. This means that these states may in fact lift for $T^4$ and so may not be fortuitous in that theory.\footnote{We thank David Turton for bringing our attention to this.}

We note that the states \eqref{eq.BHstate322} are all bosonic and all have vanishing $\tilde{J}^3_0$ eigenvalue in the Ramond sector. These states can be interpreted as microstates of single-centre black holes, based on of the claims of \cite{Sen:2009vz,Dabholkar:2010rm,Chowdhury:2015gbk} that the microstates of single-centre black holes in four dimensions must carry zero angular momentum\footnote{This angular momentum is that of four dimensions \cite{Gaiotto:2005gf} which, when translated to the AdS$_3\times S^3$ setting considered here, corresponds to $\tilde{J}^i_0$.} and are all bosonic. On the other hand, it is also possible that the states in \eqref{kowe21Jun24} and \eqref{kowg21Jun24} that pair up according to \eqref{eq.liftingReps} do not in fact lift (see \cite{Benjamin:2021zkn} for examples of this). If this is the case, these states are also fortuitous but still cancel from the index; they must correspond to non-single-centre microstates, such as multi-centre black hole microstates and the composite fortuitous states described in \cite{Chang:2025rqy}.

\section{Summary and outlook} \label{sec:conc}

In this paper we have constructed and studied extensions of the supersymmetric indices for supergraviton states in the symmetric orbifold theories $\mathrm{Sym}^N(T^4)$ and $\mathrm{Sym}^N(K3)$, relevant to the D1-D5 system, for the case of $N=2$. These generalised supergraviton indices receive contributions from both supergraviton and singleton states, the latter having a bulk description in terms of boundary excitations in AdS$_3\times S^3$. Within the CFT, these excitations are given by total modes (of the form \eqref{eq.totalModes}) of the affine symmetry algebra; namely these are states of the form $\mathcal{O}^{(\mathrm{T})} |\psi\rangle$ where $|\psi\rangle$ is a supergraviton state \eqref{eq.1/4BPSsupergrav} and $\mathcal{O}^{(\mathrm{T})}$ is a non-global mode of the left-moving small or contracted large $\cN=4$ algebra for $T^4$ and $K3$ respectively. Including these singleton states is necessary for precisely counting CFT states that have a realisation in $AdS_3$ supergravity.

We constructed these generalised supergraviton indices, roughly speaking, by promoting the $SU(1,1\,|\,2)$ total global subalgebra characters in the decomposition of the relevant supergraviton indices to characters of the full total affine symmetry algebra. In order to do this without over-counting due to the existence of supergraviton states that are related by the action of total affine modes, a detailed subtraction procedure is first necessary (discussed in detail for $T^4$ in section~\ref{ssec:subtract}). As we discussed at the end of section~\ref{ssec:subtract}, this subtraction procedure may undercount the number of long affine characters in the generalised supergraviton index and because of this we have quoted our results up to level $h=\frac32$ and $h=1$ for $T^4$ and $K3$ respectively. Due to the technical complexity of this subtraction procedure only the case of $N=2$ was considered in this paper, but we hope to report on extensions of these results to higher values of $N$ in the near future~\cite{wip}.

For the $T^4$ theory, already at $N=2$ the generalised supergraviton index (modified elliptic genus) is observed to match the CFT index for conformal dimensions $h\leq\frac32$ -- an enhanced de Boer bound. This is an improvement over the matching of the supergraviton and CFT indices which holds for $h\leq\frac12$ when $N=2$. In other words, the lowest-lying non-supergraviton states contributing to the CFT modified elliptic genus are singleton states. We identify the explicit form of these singleton states at $(h,j)=(1,0)$ and $(\frac32,\frac12)$ which are responsible for the enhancement of the de Boer bound in section~\ref{ssec:Fortuity}. Above $h=\frac32$, it is possible that our method misses additional long affine character contributions to the generalised supergraviton index coming from affine primary combinations of supergraviton and singleton states. At $h=1$ it is known \cite{Guo:2019ady,Guo:2020gxm} that there are no unlifted affine primaries in $\mathrm{Sym}^2(T^4)$ and at $h=\frac32$ affine primaries are forbidden by representation theory.

In the case of the $K3$ theory such an enhancement of the de Boer bound was not observed for the generalised supergraviton elliptic genus; however, it is possible that for higher values of $N$ there is an enhanced matching with the CFT index. The explicit form of the lowest-lying fortuitous states contributing to the CFT elliptic genus are given in section~\ref{sec:App2}. These states likely correspond to typical microstates of the 3-charge D1-D5-P black hole (albeit that these are states at $N=2$). Additional long affine character contributions to the generalised supergraviton index coming from affine primary combinations of supergraviton and singleton states can only start at $q^{\frac32}$ and so do not modify the above results for the first fortuitous states.

These singleton states are particularly interesting in the context of the recent ideas of fortuity -- a classification of BPS states by their behaviour as the value of $N$ is changed. As discussed in section~\ref{sec:intro}, the singleton states considered in this paper are monotone, in line with their similarity to supergraviton states in the bulk. If supergraviton and singleton states give the full set of monotone states (arguments for which we give below) then this allows one to define the fortuitous index as $I_N^{\mathrm{for}} = I^{\mathrm{CFT}}_{N} - I_N^{\mathrm{gen}}$, with $I_N^{\mathrm{gen}}$ given for $N=2$ and (up to $h=\frac32$) in \eqref{eq.T4genMEG} for $T^4$ and in \eqref{eq.ZK3gen} (up to $h=1$) for $K3$. Being an index, $I_N^{\mathrm{for}}$ does not necessarily capture all fortuitous states; however, above the black hole parabola those that are captured are expected to be microstates of single-centre black holes \cite{Sen:2009vz,Dabholkar:2010rm,Chowdhury:2015gbk}
and our findings in section~\ref{sec:App2} are consistent with that expectation.

As suggested by \cite{Bena:2025pcy}, a more general concept of fortuity may also exist in terms of the behaviour of BPS states or supergravity solutions as a control parameter, that is not necessarily $N$, is changed. For example, multi-centred microstate geometries (smooth horizonless supersymmetric solutions of five-dimensional supergravity) appear to exhibit fortuitous-like behaviour under changes of quantised magnetic fluxes, as well as $N=n_1n_5$. On top of this, some microstate geometries -- termed scaling solutions \cite{Denef:2002ru,Bena:2006kb}-- contain a parameter which dictates the depth of the AdS$_2$ throat region of the solution. This means that, on the classical level at least, these microstate geometries (of which superstrata are a subset) can arbitrarily closely approximate a classical black hole (see~\cite{Bena:2022rna,Bena:2022ldq} for reviews). How such a picture relates to the fortuity classification is yet to be understood, but this all suggests that supergravity also exhibits a range of monotonous and fortuitous behaviour.

Possible further evidence that these ideas of fortuity could be more general are the observations of \cite{deMelloKoch:2025cec} that the decomposition of gauge-invariant operators in $U(N)$ vector-matrix models into primary and secondary invariants and their behaviour under varying $N$ is in some sense a bosonic analogue to fortuity.

We conclude with some observations about the results of this paper and give some possible future directions. At a technical level, we note that while the supergraviton indices admitted decompositions into characters of the $SU(1,1\,|\,2)$ global algebra, (effectively by construction) the generalised supergraviton indices also admit a decomposition into characters of the full affine algebra of the CFTs. These new indices are in some sense the completion of the supergraviton spectrum under even integer spectral flow (which is an automorphism of the spectrum within the NS sector).  From the CFT viewpoint, they are thus more natural quantities than the supergraviton indices and it is for this reason that we expect that they are the full monotone indices.
Note that spectral flow invariance is not enough for an index to be a weak Jacobi form, which the CFT elliptic genus is.  One might imagine constructing a weak Jacobi form by summing over $SL(2,\bbZ)$ images of the supergraviton index \cite{Dijkgraaf:2000fq,Manschot:2007ha}.  For that, one needs the polar part (the part below the black hole parabola, $h<j^2/N+N/4$) of the index as input, but the supergraviton index fails to provide that data above the de Boer bound but still below the black-hole parabola, where it is different from the polar part of the CFT elliptic genus \cite{deBoer:1998us}.  Whether or not the generalised supergraviton index provides that data is an interesting open question but, to answer it, we need to go to larger $N$ than $N=2$ studied in this paper in order to get higher resolution on the $(j,h)$ plane.  We leave such investigation for future work.  
Another comment is that, although it is clear what singleton states are being added in our promotion procedure, summing over $SL(2,\bbZ)$ images would not give us a similar explicit picture of what individual states are being added.  This is a well-known feature of such sums; Rademacher/Fareytail expansions \cite{Dijkgraaf:2000fq,Manschot:2007ha} give the count of states as a result of summation over $SL(2,\bbZ)$ images but do not give a picture of individual states that are being counted.

\begin{figure}[t]
\begin{adjustbox}{center}
    \begin{subfigure}[h]{0.5\textwidth}
        \includegraphics[width=\linewidth]{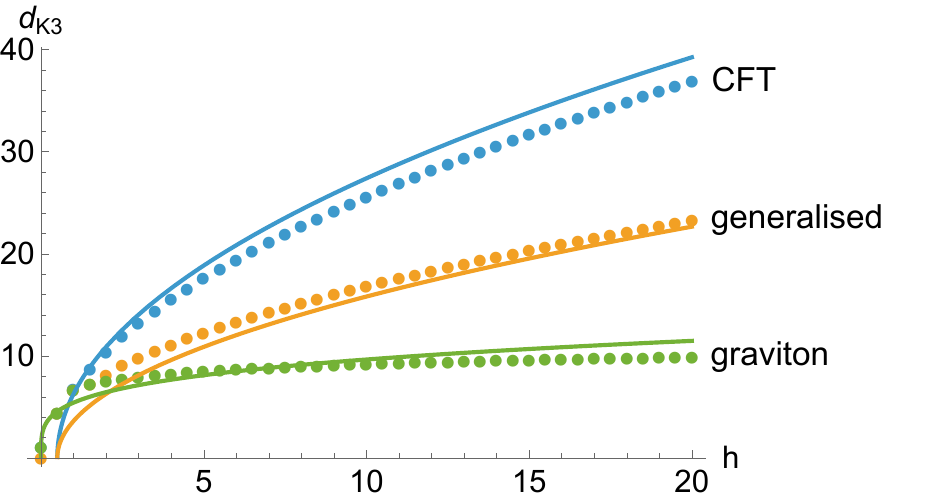}
        \caption{$K3$}
    \end{subfigure}
    \begin{subfigure}[h]{0.5\textwidth}
        \includegraphics[width=\linewidth]{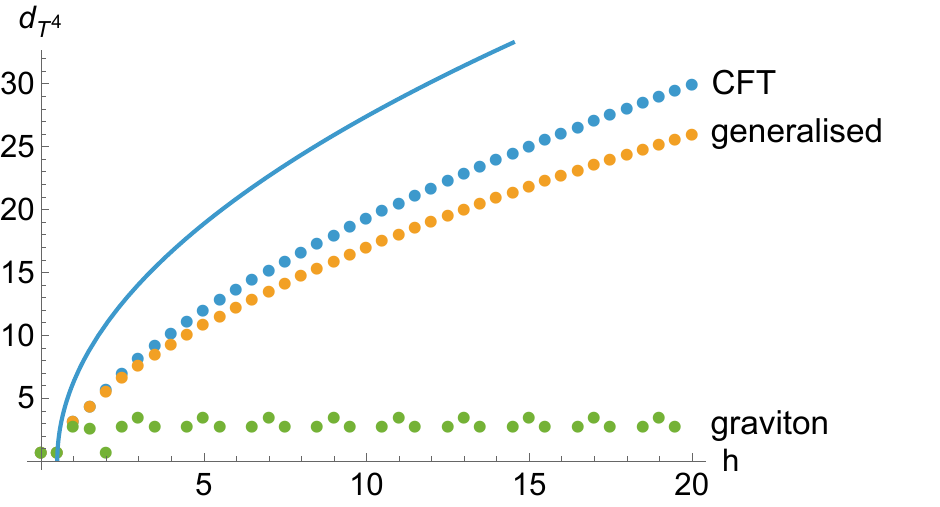}
        \caption{$T^4$}
    \end{subfigure}
\end{adjustbox}\caption{\it The log of absolute values of signed degeneracies \eqref{eq.degens} coming from the CFT, supergraviton and generalised supergraviton indices are plotted against conformal dimensions for $N=2$. (a): For $K3$, the degeneracies from the CFT elliptic genus (blue dots) scales like $\sim h^{1/2}$, similar at large $h$ to the log of the Cardy growth $2\pi\sqrt{c h/6}$ for $c=6N=12$ (solid blue curve). The generalised supergraviton elliptic genus (orange dots) grows as $\sim h^{1/2}$ for large $h$ (orange curve is Cardy growth with $c=4$), parametrically faster than the degeneracies from the supergraviton index (green dots) which scale as $\sim \# h^{1/4}$ for large $h$ (green curve). (b): For $T^4$ there is a similar pattern with the CFT and generalised supergraviton index degeneracies scaling as $\sim h^{1/2}$, parametrically faster than the supergravitons. The blue curve is the same $c=12$ Cardy growth as in (a). The signed degeneracies in (b) grow slower than those in (a) due to the modified elliptic genus being a less good measure of the (unsigned) degeneracies of the spectra than the elliptic genus of $K3$.
\label{fig:EntropyPlots2}}
\end{figure}

The generalised supergraviton index is somewhere between the supergraviton index and the CFT elliptic genus.
Due to this property it is interesting to consider the contribution of the states captured by the generalised supergraviton indices to the black hole entropy. While it is known that in the large $N$ limit, the growth of the elliptic genus for large operator dimensions captures the D1-D5-P black hole entropy~\cite{Strominger:1996sh}, our case of $N=2$ is far from this regime. We can, however, for now simply compare the growth of the CFT and supergraviton indices to that of our generalised supergraviton indices for $N=2$. We define the logarithm of signed degeneracies from the various (modified) elliptic genera for ($T^4$) $K3$ as follows
\begin{equation}
\begin{aligned} \label{eq.degens}
    d^{\mathrm{CFT}}_{K3}(h) &\equiv \log \abs{\cE^{\mathrm{CFT}}_{\mathrm{NS}}(q,1;2)\big|_{q^h}} \quad,& d^{\mathrm{CFT}}_{T^4}(h) &\equiv \log \abs{\hat{\cE}^{\mathrm{CFT}}_{\mathrm{NS}}(q,1;2)\big|_{q^h}} \ ,\vspace{1pt}\\
    d^{\mathrm{graviton}}_{K3}(h) &\equiv \log \abs{\cE^{\mathrm{graviton}}_{K3}(q,1;2)\big|_{q^h}} \quad,& d^{\mathrm{graviton}}_{T^4}(h) &\equiv \log \abs{\hat{\cE}^{\mathrm{graviton}}_{T^4}(q,1;2)\big|_{q^h}}\ ,\vspace{1pt}\\
    d^{\mathrm{gen}}_{K3}(h) &\equiv \log \abs{\cE^{\mathrm{gen}}_{K3}(q,1;2)\big|_{q^h}} \quad,& d^{\mathrm{gen}}_{T^4}(h) &\equiv \log \abs{\hat{\cE}^{\mathrm{gen}}_{T^4}(q,1;2)\big|_{q^h}} \ ,
\end{aligned}
\end{equation}
and plot them up to $h=20$ in figure \ref{fig:EntropyPlots2}.
For the K3, the CFT elliptic genus grows like $d^{\mathrm{CFT}}_{K3}(h)\sim h^{1/2}$ for large $h$ as expected from a Cardy-like behavior\footnote{This $\sim e^{\# h^{1/2}}$ growth of the index also holds in the regime $1\ll h\ll N$ and is much slower than the $\sim e^{\# h}$ Hagedorn growth of the BPS partition function of the symmetric orbifold theory in that regime~\cite{Benjamin:2016pil,Keller:2011xi}. Mechanically this is primarily due to the cancellations of lifted states from the index, but it is also in line with the expectation that the theory is dual to a large radius supergravity somewhere in its moduli space and that the index is protected. The KK spectrum partition function scales as $\sim e^{\# h^{3/4}}$ for $1\ll h\ll N$, consistent with the index growth~\cite{Benjamin:2016pil,Shigemori:2019orj}.}~\cite{Cardy:1986ie}. The growth of supergraviton states is parametrically slower~\cite{Shigemori:2019orj,Mayerson:2020acj}, with $d^{\mathrm{graviton}}_{K3}(h)\sim h^{1/4}$. The observed growth of states of our generalised supergraviton index is slower than that of the full CFT spectrum, but still with $d^{\mathrm{gen}}_{K3}(h)\sim h^{1/2}$. It would be interesting to see how the growth of the generalised supergraviton indices depends on $N$.

While the growth of spectra for $T^4$ is the same, the modified elliptic genus is a less refined measure of degeneracies and so the observed growth is slower, as seen in figure~\ref{fig:EntropyPlots2}. Nevertheless, we still observe the generalised supergraviton index growth to be more similar to the CFT than the supergraviton spectrum.

In the future it would be interesting to see if these generalised supergraviton indices exhibit any of the large $N$ stabilisation and discrete symmetry properties that were observed for the full CFT index (Hodge elliptic genus) in~\cite{Benjamin:2017rnd}.

Since the total mode states considered in this paper require the existence of an affine symmetry algebra of the CFT, generalised supergraviton indices for other AdS$_3/$CFT$_2$ holographic setups, such as those of~\cite{Eberhardt:2018sce,Eberhardt:2017pty,Datta:2017ert}, could also be interesting to study.

Lastly, since these singleton states can be thought of as describing degrees of freedom at the AdS boundary they are involved in connecting the IR ``near-horizon" AdS$_3$ to the UV spacetime, much as the Schwarzian mode for AdS$_2$ \cite{Maldacena:2016upp}.   It could be interesting to consider ``partial singletons'' -- affine generators like $L_{-n}$ acting diagonally on a subset of strands -- and whether they can be interpreted as degrees of freedom connecting local AdS regions to the ambient spacetime \cite{Mathur:2011gz}. While pure speculation, such states may be related to the supermaze constructions of~\cite{Bena:2022wpl,Bena:2023rzm} which live entirely in the internal space.

%%%%%%%%%%%%%%%%%%%%%%%%%%%%%%%%%%%%%%%%%%%%%%%%%%%%%%%%%%%%%%%%%%%%
\section*{Acknowledgements}

We would like to thank Chi-Ming Chang, Stefano Giusto, Rodolfo Russo, David Turton, and Haoyu Zhang for fruitful discussions.
We thank CEA Saclay where this work was partially done during the workshops
``Black-Hole Microstructure VI''
and ``Black-Hole Microstructure VII''.
We also thank Yukawa Institute for Theoretical Physics at Kyoto University, where this work was partially done during the workshop YITP-I-25-01 on ``Black Hole, Quantum Chaos and Quantum Information.''
This work was supported in part by MEXT
KAKENHI Grant Numbers 21H05184 and 24K00626.

%%%%%%%%%%%%%%%%%%%%%%%%%%%%%%%%%%%%%%%%%%%%%%%%%%%%%%%%%%%%%%%%%%%%
\appendix

\section{Notation and conventions}
\label{sec:App1}

\subsection{Symmetry algebra} \label{ssec:alg}

In the free symmetric orbifold theory $\mathrm{Sym}^N(T^4)$ the fundamental bosons and fermions $\pd X$ and $\psi$ have dimensions $h=1$ and $h=\tfrac12$ respectively and the small $\mathcal{N}=4$ symmetry currents are then made up of them via (our conventions mostly follow those of \cite{Avery:2010qw})
\begin{equation}
\begin{aligned} \label{eq.currentsdef}
    T(z) &= \frac12 \ep_{AB}\ep_{\dot{A}\dot{B}}\, \pd X^{\dot{A}A}(z) \pd X^{\dot{B}B}(z) +\frac12 \ep_{\alpha\beta}\ep_{\dot{A}\dot{B}}\,\psi^{\alpha\dot{A}}(z)\psi^{\beta\dot{B}}(z) \ ,\\
    J^a(z) &= \frac14 \ep_{\alpha\beta}\ep_{\dot{A}\dot{B}}(\sigma^{aT})^{\beta}{}_{\gamma}\,\psi^{\alpha \dot{A}}(z)\psi^{\gamma\dot{B}}(z) \ ,\\
    G^{\alpha A}(z) &= \ep_{\dot{A}\dot{B}}\, \psi^{\alpha\dot{A}}(z) \pd X^{\dot{B}A}(z) \ ,
\end{aligned}
\end{equation}
and similarly for the anti-holomorphic currents $\tilde{T}(\zt)$, $\tilde{J}^a(\zt)$ and $\tilde{G}^{\dot{\alpha}A}(\zt)$. The $J^a$ and $\tilde{J}^{\dot{a}}$ are $SU(2)$ currents coming from the ``external" $SO(4)_E\cong SU(2)_L\times SU(2)_R$ R-symmetry of the theory with the indices $a$ and $\dot{a}$ being vector representation indices of $SU(2)_L$ and $SU(2)_R$ respectively.

The remaining indices are for doublet representations as follows: $\alpha,\beta=+,-$ for $SU(2)_L$, $\dot{\alpha},\dot{\beta}=\dot{+},\dot{-}$ for $SU(2)_R$, $A,B=1,2$ for $SU(2)_1$, and $\dot{A},\dot{B}=\dot{1},\dot{2}$ for $SU(2)_2$. These latter two $SU(2)$ are from the ``internal" $SO(4)_I=SU(2)_1\times SU(2)_2$. All $SU(2)$ doublet indices are raised and lowered using the anti-symmetric tensor, with $\ep_{+-}=\ep^{-+}=\ep_{12}=\ep^{21}=1$ (and analogously for the dotted indices) and contractions are given by $\ep_{AB}\ep^{BC}=\ep^{CB}\ep_{BA}=\delta^C_A$.

In \eqref{eq.currentsdef}, $\sigma^{aT}$ are the transpose of the usual Pauli sigma matrices (with $a=1,2,3$), which are given by
\begin{equation}
    (\sigma^1)^{\alpha}{}_{\beta} = \begin{pmatrix}
        0&1\\1&0
    \end{pmatrix} \ ,\qquad (\sigma^2)^{\alpha}{}_{\beta} = \begin{pmatrix}
        0&-i\\i&0
    \end{pmatrix} \ ,\qquad (\sigma^3)^{\alpha}{}_{\beta} = \begin{pmatrix}
        1&0\\0&-1
    \end{pmatrix}\ .
    \label{def_sigma^i}
\end{equation}
We define $(\sigma^i)^{\dot{A}}{}_{\dot{B}}$ to be given by the expression \eqref{def_sigma^i}, and raise and lower their indices using the $\epsilon$ tensor; {\it e.g.}, $\sigma^i_{\dot{A}\dot{B}}=\epsilon_{\dot{A}\dot{C}}(\sigma^i)^{\dot{C}}{}_{\dot{B}}$.

It is often more useful to use the mode decomposition of the fundamental bosons and fermions (and so also the currents) on the complex plane using
\begin{equation} \label{eq.modes}
    \cO(z) = \sum_{n} \cO_n\, z^{-h-n} \ ,
\end{equation}
where $h$ is the holomorphic conformal dimension of $\cO$ and the domain of $n$ can be the integers or half integers.

In the symmetric orbifold theories $\mathrm{Sym}^N(T^4)$ and $\mathrm{Sym}^N(K3)$ the total current modes generate a small $\mathcal{N}=4$ superconformal algebra with central charge $c=6N$ (we drop the total mode label here for ease of notation)
\begin{subequations} \label{eq.commcurrents2}
    \begin{align}
        \big[L_m,L_n\big] &= (m-n)L_{m+n} + \frac{c}{12}m(m^2-1)\delta_{m+n,0} \ , \label{LLcomm2}\\
        \big[J^a_{m},J^b_{n}\big] &= \frac{c}{12}m\,\delta^{ab}\delta_{m+n,0} +  i\epsilon^{ab}_{\ \ c}\,J^c_{m+n} \ ,\label{JJcomm2}\\
        \big\{ G^{\alpha A}_{r} , G^{\beta B}_{s} \big\} &= \epsilon^{AB} \bigg[\epsilon^{\alpha\beta}\frac{c}{6}\Big(\frac14-r^2\Big)\delta_{r+s,0} + \big(\sigma^{aT}\big)^{\!\alpha}_{\,\gamma}\:\epsilon^{\beta\gamma}(r-s)J^a_{r+s} - \epsilon^{\alpha\beta}L_{r+s} \bigg] \ ,\label{GGcomm2}\\
        \big[J^a_{m},G^{\alpha A}_{r}\big] &= \frac12 \big(\sigma^{aT}\big)^{\!\alpha}{}_{\beta}\, G^{\beta A}_{m+r} \ ,\label{JGcomm2}\\
        \big[L_{m},G^{\alpha A}_{r}\big] &= \Big(\frac{m}{2}  -r\Big)G^{\alpha A}_{m+r} \ ,\label{LGcomm2}\\
        \big[L_{m},J^a_n\big] &= -nJ^a_{m+n} \ . \label{LJcomm2}
    \end{align}
\end{subequations}
The current mode algebra of $\mathrm{Sym}^N(T^4)$ is in fact enhanced to a contracted large $\cN=4$ superconformal algebra, again with central charge $c=6N$, given by \eqref{eq.commcurrents2} along with the additional commutation relations
\begin{subequations} \label{eq.commcurrents3}
    \begin{align}
        \big[ \alpha^{\dot{A}A}_{n} ,  \alpha^{\dot{B}B}_{m}\big] &= n \ep^{AB}\ep^{\dot{A}\dot{B}} \delta_{n+m,0} \ , \label{eq.alalcomm2}\\
        \big\{ \psi^{\alpha \dot{A}}_{r}, \psi^{\beta \dot{B}}_{s}\big\} &= -\ep^{\alpha\beta}\ep^{\dot{A}\dot{B}} \delta_{r+s,0} \ , \label{eq.psipsicomm2}\\
        \big[ L_n, \alpha^{\dot{A}A}_{m} \big] &= -m\, \alpha^{\dot{A}A}_{n+m} \ ,\label{eq.Lalcomm2} \\
        \big[ L_n, \psi^{\alpha\dot{A}}_{r} \big] &= -\big(\tfrac{n}{2} + r\big) \psi^{\alpha\dot{A}}_{n+r} \ , \label{eq.Lpsicomm2}\\
        \big[ J^a_n, \alpha^{\dot{A}A}_{m} \big] &= 0 \ , \label{eq.Jalcomm2}\\
        \big[ J^a_n, \psi^{\alpha\dot{A}}_{r} \big] &= \tfrac12 \big(\sigma^{a T}\big)^{\alpha}{}_{\beta}\, \psi^{\beta\dot{A}}_{n+r} \ , \label{eq.Jpsicomm2}\\
        \big[G^{\alpha A}_{r} , \alpha^{\dot{B}B}_{m}\big] &= -m \ep^{AB}\psi^{\alpha\dot{B}}_{r+m} \ , \label{eq.Galcomm2}\\
        \big\{G^{\alpha A}_{r} , \psi^{\beta\dot{A}}_{s} \big\} &= \ep^{\alpha\beta} \alpha^{\dot{A}A}_{r+s} \ .\label{eq.Gpsicomm2}
    \end{align}
\end{subequations}
A different basis for the $SU(2)_L$ currents in \eqref{eq.commcurrents2} is also often used; instead of $J^a_n$ we have $J^{\pm,3}_n$ where
\begin{equation}
    J^{\pm}_n = J^1_n \pm i J^2_n \ .
\end{equation}
The non-zero mode algebra relations involving the $J^{\pm}_{n}$ are then
\begin{subequations} \label{eq.commcurrentsJpm}
    \begin{align}
        \big[J^+_n, J^-_m \big] &= n\frac{c}{6}\delta_{n+m,0} + 2J^3_{n+m} \ , \label{eq.JpJmcomm}\\
        \big[J^3_n,J^{\pm}_m\big] &= \pm J^{\pm}_{n+m} \ , \label{eq.J3Jpmcomm}\\
        \big[L_n,J^{\pm}_m\big] &= -m J^{\pm}_{n+m} \ , \label{eq.LJpmcomm}\\
        \big[J^{\pm}_{n}, G^{\mp A}_{r}] &= G^{\pm A}_{n+r} \ , \label{eq.JpmGcomm}\\
        \big[J^{\pm}_{n}, \psi^{\mp \Ad}_{r}] &= \psi^{\pm \Ad}_{n+r} \ .\label{eq.Jpmpsicomm}
    \end{align}
\end{subequations}

\subsection{Special functions}

The various CFT elliptic genera used in the main body require the Jacobi theta functions $\vartheta_i(\nu,\tau)$. These can be defined starting from the classical Jacobi theta function \cite{DHoker:2022dxx}
\begin{equation}
    \vartheta(\nu,\tau) = \sum_{n=-\infty}^{\infty} e^{2\pi in\nu} e^{\pi i n^2\tau} \ ,
\end{equation}
where $q=e^{2\pi i\tau}$ and $y=e^{2\pi i \nu}$ connects with notation used in \ref{ssec:index}. This function can then be generalised by introducing the characteristics $\alpha,\beta\in\mathbb{C}$ to 
\begin{equation}
    \vartheta_{\alpha,\beta}(\nu,\tau) \equiv e^{\pi i\alpha^2\tau} e^{2\pi i\alpha(\nu+\beta)}\, \vartheta(\nu+\alpha\tau+\beta,\tau) \ ,
\end{equation}
where $\vartheta(\nu,\tau)=\vartheta_{0,0}(\nu,\tau)$. Restricting to theta functions with a well-defined parity under $\nu\to-\nu$ yields the four functions

\begin{subequations} \label{eq.Jtheta}
\begin{align}
    \vartheta_{1}(\nu,\tau) &\equiv -\vartheta_{\frac12,\frac12}(\nu,\tau) \ ,\\
    \vartheta_{2}(\nu,\tau) &\equiv \vartheta_{\frac12,0}(\nu,\tau) \ ,\\
    \vartheta_{3}(\nu,\tau) &\equiv \vartheta_{0,0}(\nu,\tau) \ ,\\
    \vartheta_{4}(\nu,\tau) &\equiv \vartheta_{0,\frac12}(\nu,\tau) \ .
\end{align}
\end{subequations}
In the context of the partition function of the seed sigma model on $T^4$ and various characters given in \ref{ssec:chars} the Dedekind eta function is also useful; this is defined as
\begin{equation} \label{eq.Dedeta}
    \eta(q) = q^{\frac{1}{24}} \prod_{n=1}^{\infty} (1-q^n) \ .
\end{equation}

\subsection{Character formulae}
\label{ssec:chars}

Here we detail the forms of the various characters used throughout this paper. Since we work in this paper with partition functions and indices with a factor of $(-1)^F$ inside the trace, giving a relative sign for bosons and fermions, we use conventions for characters which also have this relative sign for states within the representation. However, we do not include in the characters an overall sign for the lowest-weight state of the representation. We include these explicitly as relative signs between characters in the partition function or index.

\subsubsection*{$SU(2)_R$}

The characters of the right-moving R-symmetry group $SU(2)_R$ labelled by eigenvalues $\tj$ of the $\tilde{J}^3_0$ generator are given by
\begin{align} \label{eq.SU2Char}
    \tilde{\chi}_{\tj;N}(\ty) = \ty^{N} \frac{\ty^{2\tj+1} - \ty^{-2\tj-1}}{\ty-\ty^{-1}} \ ,
\end{align}
where we have included a factor of $\ty^{N}$ coming from the spectral flow of the elliptic genus from the Ramond sector in which it is naturally defined to the NS sector in which we can compare with the supergraviton spectrum.

\subsubsection*{Small $\cN=4$}

For the small $\cN=4$ algebra there exists short characters $\Phi^{(s)}_{j;N}$ for representations built on a chiral primary with $h=j$ for values $j=0,{1\over 2},1,\dots, {N\over 2}$ and long characters $\Phi^{(\ell)}_{j,h;N}$ for representations built on a primary with $h>j$ for values $j=0,{1\over 2},1,\dots, {N-1\over 2}$ \cite{Eguchi:1987sm}. In the NS sector these take the form
\begin{align}
    \Phi^{(s)}_{j;N}(q,y) &= F(q,y)\,\varphi^{(s)}_{j;N}(q,y)\ ,\vspace{2pt}\label{eq.smallN4s}
    \\
    \Phi^{(\ell)}_{j,h;N}(q,y) &= F(q,y)\,\varphi^{(\ell)}_{j,h;N}(q,y) \ ,\label{eq.smallN4l}
\end{align}
where the Verma module is accounted for by the factor \cite{Eguchi:1987wf}
\begin{align}
    F(q,y) &= \prod_{k=1}^\infty
    {(1-yq^{k-1/2})^2(1-y^{-1}q^{k-1/2})^2\over (1-q^k)^2(1-y^2q^k)(1-y^{-2}q^{k-1})}
    ={-iq^{1/4}\vartheta_4(\nu,\tau)^2\over \vartheta_1(2\nu,\tau)\,\eta(q)^3} \ ,
\end{align}
where $q=e^{2\pi i\tau}$ and $y=e^{2\pi i\nu}$ and the null states are accounted for by the factor
\begin{align}
    \varphi^{(s)}_{j;N}(q,y)
    &= q^j \sum_{k\in\bbZ} \bigg[{y^{2(N+1)k+2j}\over (1-yq^{k+1/2})^2}
    - {y^{-2(N+1)k-2(j+1)}\over (1-y^{-1}q^{k+1/2})^2}
    \bigg]\,
    q^{(N+1)k^2+(2j+1)k} \ ,\vspace{2pt}\\
    \varphi^{(\ell)}_{j,h;N}(q,y)
    &= q^h\sum_{k\in\bbZ}\Big(y^{2(N+1)k+2j}-y^{-2(N+1)k-2(j+1)}\Big) \,q^{(N+1)k^2+(2j+1)k} \ .
\end{align}

\subsubsection*{Contracted large $\cN=4$}

There exist short characters of the contracted large $\mathcal{N}=4$ algebra for $h=j=0,\frac12,1,\dots,\frac{N-1}{2}$ which in the NS sector are given by \cite{Petersen:1989zz,Petersen:1989pp}
\begin{equation} \label{eq.clargeN4s}
    \mathbf{\Phi}^{(s)}_{j;N}(q,y) = \frac{q^{\frac18}}{\eta(q)^3}\, \Phi^{(\ell)}_{0,0;1}(q,y)\,\Phi^{(s)}_{j;N-1}(q,y) \ ,
\end{equation}
where $\Phi^{(s)}_{j;N}(q,y)$ and $\Phi^{(\ell)}_{j,h;N}(q,y)$ are the NS sector characters of the small $\mathcal{N}=4$ algebra \eqref{eq.smallN4s} and \eqref{eq.smallN4l}. Long characters of the contracted large algebra exist for $h>j=0,\frac12,1,\dots,\frac{N-2}{2}$ and in the NS sector are given by
\begin{equation} \label{eq.clargeN4l}
    \mathbf{\Phi}^{(\ell)}_{j,h;N}(q,y) = \frac{q^{\frac18}}{\eta(q)^3}\, \Phi^{(\ell)}_{0,0;1}(q,y)\,\Phi^{(\ell)}_{j,h;N-1}(q,y) \ .
\end{equation}

\subsubsection*{$SU(1,1\,|\,2)$}

The global subalgebra of both the small and contracted large  $\cN=4$ superconformal algebras is given by the supergroup $SU(1,1\,|\,2)$ which has short representations with characters
\begin{align} \label{eq.globalN4s}
    \phi_{j}^{(s)}(q,y)
    = {q^j\over 1-q}{
    y^{2j}(y-2q^{\frac12}+y^{-1}q)
    -y^{-2j}(y^{-1}-2q^{\frac12}+yq)
    \over  y-y^{-1}} \ ,
\end{align}
and long representations with characters
\begin{align} \label{eq.globalN4l}
    \phi_{j,h}^{(\ell)}(q,y)
    &=
    \phi_{j,h}^{(s)}(q,y)
    -2\phi_{j+\frac12,h+\frac12}^{(s)}(q,y)
    +\phi_{j+1,h+1}^{(s)}(q,y)
    \notag\\
    &= {q^h\over 1-q}
    {(y^{2j+1}-y^{-2j-1})(y^{-1}-2q^{\frac12}+qy)(y-2q^{\frac12}+qy^{-1})\over y-y^{-1}} \ .
\end{align}

\subsubsection*{Product formulae}

In sections \ref{sec:T4} and \ref{sec:K3} it will be necessary to decompose products of short $SU(1,1\,|\,2)_L$ characters and characters with squared fugacities into sums of short and long characters. These decompositions are given by
\begin{equation} \label{eq.SU112CharProd}
    \phi^{(s)}_{j_1}(q,y)\times \phi^{(s)}_{j_2}(q,y) = \phi^{(s)}_{j_1+j_2}(q,y) + \sum_{h\geq j_1+j_2} \sum_{j=|j_1-j_2|}^{j_1+j_2-1} \phi^{(\ell)}_{j,h}(q,y) \ .
\end{equation}
and
\begin{equation} \label{eq.SU112CharPow}
    \phi^{(s)}_{j}(q^2,y^2) = \phi^{(s)}_{2j}(q,y) + \sum_{h\geq 2j} \sum_{j'=0}^{2j-1} (-1)^{h+j'} \phi^{(\ell)}_{j',h}(q,y) \ .
\end{equation}
The series $\sum_{h\geq j_1+j_2}$ in \eqref{eq.SU112CharProd} runs over the domain $h\in\{j_1+j_2, j_1+j_2+1, j_1+j_2+2, \dots\}$ where these values of $h$ can be integer if $j_1+j_2\in\mathbb{Z}$ and half-integer if $j_1+j_2\in\mathbb{Z}+\frac12$. The equivalent character product formulae for the $SU(2)_R$ characters \eqref{eq.SU2Char} are
\begin{equation} \label{eq.SU2CharProd}
    \tilde{\chi}_{j_1;N_1}(\bar{y}) \times \tilde{\chi}_{j_2;N_2}(\bar{y}) = \sum_{j=|j_1-j_2|}^{j_1+j_2} \tilde{\chi}_{j;N_1+N_2}(\bar{y}) \ ,
\end{equation}
and
\begin{equation} \label{eq.SU2CharPow}
    \tilde{\chi}_{j_1;N}(\bar{y}^2) = \sum_{j=0}^{2j_1} (-1)^{2j_1-j} \tilde{\chi}_{j;2N}(\bar{y}) \ .
\end{equation}

%%%%%%%%%%%%%%%%%%%%%%%%%%%%%%%%%%%%%%%%%%%%%%%%%%%%%%%%%%%%%%%%%%%%%%%%%%%%%%
\bibliographystyle{utphys}
\bibliography{IndexPaper.bib}

\providecommand{\href}[2]{#2}\begingroup\raggedright\begin{thebibliography}{10}

\bibitem{Strominger:1996sh}
A.~Strominger and C.~Vafa, ``{Microscopic origin of the Bekenstein-Hawking entropy},'' \href{http://dx.doi.org/10.1016/0370-2693(96)00345-0}{{\em Phys.Lett.} {\bfseries B379} (1996) 99--104}, \href{http://arxiv.org/abs/hep-th/9601029}{{\ttfamily arXiv:hep-th/9601029 [hep-th]}}.

\bibitem{deBoer:1998us}
J.~de~Boer, ``{Large N elliptic genus and AdS / CFT correspondence},'' \href{http://dx.doi.org/10.1088/1126-6708/1999/05/017}{{\em JHEP} {\bfseries 9905} (1999) 017},
\href{http://arxiv.org/abs/hep-th/9812240}{{\ttfamily arXiv:hep-th/9812240 [hep-th]}}.
%%CITATION = HEP-TH/9812240;%%.

\bibitem{Maldacena:1998bw}
J.~M. Maldacena and A.~Strominger, ``{AdS(3) black holes and a stringy exclusion principle},'' {\em JHEP} {\bfseries 12} (1998) 005,
\href{http://arxiv.org/abs/hep-th/9804085}{{\ttfamily arXiv:hep-th/9804085}}.
%%CITATION = HEP-TH/9804085;%%.

\bibitem{Maldacena:1999bp}
J.~M. Maldacena, G.~W. Moore, and A.~Strominger, ``{Counting BPS black holes in toroidal Type II string theory},'' \href{http://arxiv.org/abs/hep-th/9903163}{{\ttfamily arXiv:hep-th/9903163}}.

\bibitem{Chang:2022mjp}
C.-M. Chang and Y.-H. Lin, ``{Words to describe a black hole},'' \href{http://dx.doi.org/10.1007/JHEP02(2023)109}{{\em JHEP} {\bfseries 02} (2023) 109}, \href{http://arxiv.org/abs/2209.06728}{{\ttfamily arXiv:2209.06728 [hep-th]}}.

\bibitem{Chang:2024zqi}
C.-M. Chang and Y.-H. Lin, ``{Holographic covering and the fortuity of black holes},'' \href{http://arxiv.org/abs/2402.10129}{{\ttfamily arXiv:2402.10129 [hep-th]}}.

\bibitem{Kinney:2005ej}
J.~Kinney, J.~M. Maldacena, S.~Minwalla, and S.~Raju, ``{An Index for 4 dimensional super conformal theories},'' \href{http://dx.doi.org/10.1007/s00220-007-0258-7}{{\em Commun. Math. Phys.} {\bfseries 275} (2007) 209--254}, \href{http://arxiv.org/abs/hep-th/0510251}{{\ttfamily arXiv:hep-th/0510251}}.

\bibitem{Grant:2008sk}
L.~Grant, P.~A. Grassi, S.~Kim, and S.~Minwalla, ``{Comments on 1/16 BPS Quantum States and Classical Configurations},'' \href{http://dx.doi.org/10.1088/1126-6708/2008/05/049}{{\em JHEP} {\bfseries 05} (2008) 049}, \href{http://arxiv.org/abs/0803.4183}{{\ttfamily arXiv:0803.4183 [hep-th]}}.

\bibitem{Chang:2013fba}
C.-M. Chang and X.~Yin, ``{1/16 BPS states in $\mathcal N=$ 4 super-Yang-Mills theory},'' \href{http://dx.doi.org/10.1103/PhysRevD.88.106005}{{\em Phys. Rev. D} {\bfseries 88} no.~10, (2013) 106005}, \href{http://arxiv.org/abs/1305.6314}{{\ttfamily arXiv:1305.6314 [hep-th]}}.

\bibitem{Choi:2022caq}
S.~Choi, S.~Kim, E.~Lee, and J.~Park, ``{The shape of non-graviton operators for SU(2)},'' \href{http://dx.doi.org/10.1007/JHEP09(2024)029}{{\em JHEP} {\bfseries 09} (2024) 029}, \href{http://arxiv.org/abs/2209.12696}{{\ttfamily arXiv:2209.12696 [hep-th]}}.

\bibitem{Choi:2023vdm}
J.~Choi, S.~Choi, S.~Kim, J.~Lee, and S.~Lee, ``{Finite N black hole cohomologies},'' \href{http://dx.doi.org/10.1007/JHEP12(2024)029}{{\em JHEP} {\bfseries 12} (2024) 029}, \href{http://arxiv.org/abs/2312.16443}{{\ttfamily arXiv:2312.16443 [hep-th]}}.

\bibitem{Chang:2023zqk}
C.-M. Chang, L.~Feng, Y.-H. Lin, and Y.-X. Tao, ``{Decoding stringy near-supersymmetric black holes},'' \href{http://dx.doi.org/10.21468/SciPostPhys.16.4.109}{{\em SciPost Phys.} {\bfseries 16} no.~4, (2024) 109}, \href{http://arxiv.org/abs/2306.04673}{{\ttfamily arXiv:2306.04673 [hep-th]}}.

\bibitem{Choi:2023znd}
S.~Choi, S.~Kim, E.~Lee, S.~Lee, and J.~Park, ``{Towards quantum black hole microstates},'' \href{http://dx.doi.org/10.1007/JHEP11(2023)175}{{\em JHEP} {\bfseries 11} (2023) 175}, \href{http://arxiv.org/abs/2304.10155}{{\ttfamily arXiv:2304.10155 [hep-th]}}. [Erratum: JHEP 03, 091 (2025)].

\bibitem{Budzik:2023vtr}
K.~Budzik, H.~Murali, and P.~Vieira, ``{Following Black Hole States},'' \href{http://arxiv.org/abs/2306.04693}{{\ttfamily arXiv:2306.04693 [hep-th]}}.

\bibitem{deMelloKoch:2024pcs}
R.~de~Mello~Koch, M.~Kim, S.~Kim, J.~Lee, and S.~Lee, ``{Brane-fused black hole operators},'' \href{http://arxiv.org/abs/2412.08695}{{\ttfamily arXiv:2412.08695 [hep-th]}}.

\bibitem{Bena:2022ldq}
I.~Bena, E.~J. Martinec, S.~D. Mathur, and N.~P. Warner, ``{Snowmass White Paper: Micro- and Macro-Structure of Black Holes},'' \href{http://arxiv.org/abs/2203.04981}{{\ttfamily arXiv:2203.04981 [hep-th]}}.

\bibitem{Bena:2022rna}
I.~Bena, E.~J. Martinec, S.~D. Mathur, and N.~P. Warner, ``{Fuzzballs and Microstate Geometries: Black-Hole Structure in String Theory},'' \href{http://arxiv.org/abs/2204.13113}{{\ttfamily arXiv:2204.13113 [hep-th]}}.

\bibitem{Shigemori:2020yuo}
M.~Shigemori, ``{Superstrata},'' \href{http://dx.doi.org/10.1007/s10714-020-02698-8}{{\em Gen. Rel. Grav.} {\bfseries 52} no.~5, (2020) 51}, \href{http://arxiv.org/abs/2002.01592}{{\ttfamily arXiv:2002.01592 [hep-th]}}.

\bibitem{Chang:2024lxt}
C.-M. Chang, Y.~Chen, B.~S. Sia, and Z.~Yang, ``{Fortuity in SYK Models},'' \href{http://arxiv.org/abs/2412.06902}{{\ttfamily arXiv:2412.06902 [hep-th]}}.

\bibitem{Chang:2025rqy}
C.-M. Chang, Y.-H. Lin, and H.~Zhang, ``{Fortuity in the D1-D5 system},'' \href{http://arxiv.org/abs/2501.05448}{{\ttfamily arXiv:2501.05448 [hep-th]}}.

\bibitem{Brown:1986nw}
J.~D. Brown and M.~Henneaux, ``{Central Charges in the Canonical Realization of Asymptotic Symmetries: An Example from Three-Dimensional Gravity},'' \href{http://dx.doi.org/10.1007/BF01211590}{{\em Commun. Math. Phys.} {\bfseries 104} (1986) 207--226}.

\bibitem{Kanitscheider:2007wq}
I.~Kanitscheider, K.~Skenderis, and M.~Taylor, ``{Fuzzballs with internal excitations},'' {\em JHEP} {\bfseries 06} (2007) 056,
\href{http://arxiv.org/abs/0704.0690}{{\ttfamily arXiv:0704.0690 [hep-th]}}.
%%CITATION = 0704.0690;%%.

\bibitem{Bena:2017xbt}
I.~Bena, S.~Giusto, E.~J. Martinec, R.~Russo, M.~Shigemori, D.~Turton, and N.~P. Warner, ``{Asymptotically-flat supergravity solutions deep inside the black-hole regime},'' \href{http://dx.doi.org/10.1007/JHEP02(2018)014}{{\em JHEP} {\bfseries 02} (2018) 014}, \href{http://arxiv.org/abs/1711.10474}{{\ttfamily arXiv:1711.10474 [hep-th]}}.

\bibitem{Giusto:2015dfa}
S.~Giusto, E.~Moscato, and R.~Russo, ``{AdS$_{3}$ holography for 1/4 and 1/8 BPS geometries},'' \href{http://dx.doi.org/10.1007/JHEP11(2015)004}{{\em JHEP} {\bfseries 11} (2015) 004},
\href{http://arxiv.org/abs/1507.00945}{{\ttfamily arXiv:1507.00945 [hep-th]}}.
%%CITATION = ARXIV:1507.00945;%%.

\bibitem{deBoer:1998ip}
J.~de~Boer, ``{Six-dimensional supergravity on S**3 x AdS(3) and 2-D conformal field theory},'' \href{http://dx.doi.org/10.1016/S0550-3213(99)00160-1}{{\em Nucl.Phys.} {\bfseries B548} (1999) 139--166},
\href{http://arxiv.org/abs/hep-th/9806104}{{\ttfamily arXiv:hep-th/9806104 [hep-th]}}.
%%CITATION = HEP-TH/9806104;%%.

\bibitem{Breckenridge:1996is}
J.~C. Breckenridge, R.~C. Myers, A.~W. Peet, and C.~Vafa, ``{D-branes and spinning black holes},'' \href{http://dx.doi.org/10.1016/S0370-2693(96)01460-8}{{\em Phys. Lett. B} {\bfseries 391} (1997) 93--98}, \href{http://arxiv.org/abs/hep-th/9602065}{{\ttfamily arXiv:hep-th/9602065}}.

\bibitem{Mathur:2011gz}
S.~D. Mathur and D.~Turton, ``{Microstates at the boundary of AdS},'' \href{http://dx.doi.org/10.1007/JHEP05(2012)014}{{\em JHEP} {\bfseries 1205} (2012) 014},
\href{http://arxiv.org/abs/1112.6413}{{\ttfamily arXiv:1112.6413 [hep-th]}}.
%%CITATION = ARXIV:1112.6413;%%.

\bibitem{Mathur:2012tj}
S.~D. Mathur and D.~Turton, ``{Momentum-carrying waves on D1-D5 microstate geometries},'' \href{http://dx.doi.org/10.1016/j.nuclphysb.2012.05.014}{{\em Nucl.Phys.} {\bfseries B862} (2012) 764--780},
\href{http://arxiv.org/abs/1202.6421}{{\ttfamily arXiv:1202.6421 [hep-th]}}.
%%CITATION = ARXIV:1202.6421;%%.

\bibitem{Lunin:2012gp}
O.~Lunin, S.~D. Mathur, and D.~Turton, ``{Adding momentum to supersymmetric geometries},'' \href{http://dx.doi.org/10.1016/j.nuclphysb.2012.11.017}{{\em Nucl.Phys.} {\bfseries B868} (2013) 383--415},
\href{http://arxiv.org/abs/1208.1770}{{\ttfamily arXiv:1208.1770 [hep-th]}}.
%%CITATION = ARXIV:1208.1770;%%.

\bibitem{Giusto:2013bda}
S.~Giusto and R.~Russo, ``{Superdescendants of the D1D5 CFT and their dual 3-charge geometries},'' \href{http://dx.doi.org/10.1007/JHEP03(2014)007}{{\em JHEP} {\bfseries 1403} (2014) 007},
\href{http://arxiv.org/abs/1311.5536}{{\ttfamily arXiv:1311.5536 [hep-th]}}.
%%CITATION = ARXIV:1311.5536;%%.

\bibitem{Chang:2025}
Chi-Ming Chang, Ying-Hsuan Lin and Haoyu Zhang, Work in progress.

\bibitem{HaoyuTalk}
See talk given by Haoyu Zhang, ``Fortuity in AdS$_3$/CFT$_2$'' at {\it Black-Hole Microstructure VII},~~\url{https://www.youtube.com/watch?v=VpWCM0UaqCg}.

\bibitem{Gava:2002xb}
E.~Gava and K.~S. Narain, ``{Proving the PP wave / CFT(2) duality},'' \href{http://dx.doi.org/10.1088/1126-6708/2002/12/023}{{\em JHEP} {\bfseries 12} (2002) 023}, \href{http://arxiv.org/abs/hep-th/0208081}{{\ttfamily arXiv:hep-th/0208081}}.

\bibitem{Maldacena:1997re}
J.~M. Maldacena, ``{The large N limit of superconformal field theories and supergravity},'' {\em Adv. Theor. Math. Phys.} {\bfseries 2} (1998) 231--252,
\href{http://arxiv.org/abs/hep-th/9711200}{{\ttfamily arXiv:hep-th/9711200}}.
%%CITATION = HEP-TH/9711200;%%.

\bibitem{David:2002wn}
J.~R. David, G.~Mandal, and S.~R. Wadia, ``{Microscopic formulation of black holes in string theory},'' \href{http://dx.doi.org/10.1016/S0370-1573(02)00271-5}{{\em Phys. Rept.} {\bfseries 369} (2002) 549--686},
\href{http://arxiv.org/abs/hep-th/0203048}{{\ttfamily arXiv:hep-th/0203048}}.
%%CITATION = HEP-TH/0203048;%%.

\bibitem{Avery:2009tu}
S.~G. Avery, B.~D. Chowdhury, and S.~D. Mathur, ``{Emission from the D1D5 CFT},'' \href{http://dx.doi.org/10.1088/1126-6708/2009/10/065}{{\em JHEP} {\bfseries 10} (2009) 065},
\href{http://arxiv.org/abs/0906.2015}{{\ttfamily arXiv:0906.2015 [hep-th]}}.
%%CITATION = 0906.2015;%%.

\bibitem{Seiberg:1999xz}
N.~Seiberg and E.~Witten, ``{The D1 / D5 system and singular CFT},'' \href{http://dx.doi.org/10.1088/1126-6708/1999/04/017}{{\em JHEP} {\bfseries 04} (1999) 017}, \href{http://arxiv.org/abs/hep-th/9903224}{{\ttfamily arXiv:hep-th/9903224}}.

\bibitem{Gaberdiel:2018rqv}
M.~R. Gaberdiel and R.~Gopakumar, ``{Tensionless string spectra on AdS$_{3}$},'' \href{http://dx.doi.org/10.1007/JHEP05(2018)085}{{\em JHEP} {\bfseries 05} (2018) 085}, \href{http://arxiv.org/abs/1803.04423}{{\ttfamily arXiv:1803.04423 [hep-th]}}.

\bibitem{Eberhardt:2018ouy}
L.~Eberhardt, M.~R. Gaberdiel, and R.~Gopakumar, ``{The Worldsheet Dual of the Symmetric Product CFT},'' \href{http://dx.doi.org/10.1007/JHEP04(2019)103}{{\em JHEP} {\bfseries 04} (2019) 103}, \href{http://arxiv.org/abs/1812.01007}{{\ttfamily arXiv:1812.01007 [hep-th]}}.

\bibitem{Eberhardt:2019ywk}
L.~Eberhardt, M.~R. Gaberdiel, and R.~Gopakumar, ``{Deriving the AdS$_{3}$/CFT$_{2}$ correspondence},'' \href{http://dx.doi.org/10.1007/JHEP02(2020)136}{{\em JHEP} {\bfseries 02} (2020) 136}, \href{http://arxiv.org/abs/1911.00378}{{\ttfamily arXiv:1911.00378 [hep-th]}}.

\bibitem{Eberhardt:2020akk}
L.~Eberhardt, ``{AdS$_{3}$/CFT$_{2}$ at higher genus},'' \href{http://dx.doi.org/10.1007/JHEP05(2020)150}{{\em JHEP} {\bfseries 05} (2020) 150}, \href{http://arxiv.org/abs/2002.11729}{{\ttfamily arXiv:2002.11729 [hep-th]}}.

\bibitem{Eberhardt:2019qcl}
L.~Eberhardt and M.~R. Gaberdiel, ``{String theory on AdS$_3$ and the symmetric orbifold of Liouville theory},'' \href{http://dx.doi.org/10.1016/j.nuclphysb.2019.114774}{{\em Nucl. Phys. B} {\bfseries 948} (2019) 114774}, \href{http://arxiv.org/abs/1903.00421}{{\ttfamily arXiv:1903.00421 [hep-th]}}.

\bibitem{Dei:2019osr}
A.~Dei, L.~Eberhardt, and M.~R. Gaberdiel, ``{Three-point functions in AdS$_{3}$/CFT$_{2}$ holography},'' \href{http://dx.doi.org/10.1007/JHEP12(2019)012}{{\em JHEP} {\bfseries 12} (2019) 012}, \href{http://arxiv.org/abs/1907.13144}{{\ttfamily arXiv:1907.13144 [hep-th]}}.

\bibitem{Schwimmer:1986mf}
A.~Schwimmer and N.~Seiberg, ``{Comments on the N=2, N=3, N=4 Superconformal Algebras in Two-Dimensions},''
\href{http://dx.doi.org/10.1016/0370-2693(87)90566-1}{{\em Phys.Lett.} {\bfseries B184} (1987) 191}.
%%CITATION = PHLTA,B184,191;%%.

\bibitem{Gaberdiel:2015uca}
M.~R. Gaberdiel, C.~Peng, and I.~G. Zadeh, ``{Higgsing the stringy higher spin symmetry},'' \href{http://dx.doi.org/10.1007/JHEP10(2015)101}{{\em JHEP} {\bfseries 10} (2015) 101},
\href{http://arxiv.org/abs/1506.02045}{{\ttfamily arXiv:1506.02045 [hep-th]}}.
%%CITATION = ARXIV:1506.02045;%%.

\bibitem{Hampton:2018ygz}
S.~Hampton, S.~D. Mathur, and I.~G. Zadeh, ``{Lifting of D1-D5-P states},'' \href{http://dx.doi.org/10.1007/JHEP01(2019)075}{{\em JHEP} {\bfseries 01} (2019) 075}, \href{http://arxiv.org/abs/1804.10097}{{\ttfamily arXiv:1804.10097 [hep-th]}}.

\bibitem{Benjamin:2021zkn}
N.~Benjamin, C.~A. Keller, and I.~G. Zadeh, ``{Lifting 1/4-BPS states in $AdS_{3}\times S^{3}\times T^{4}$},'' \href{http://dx.doi.org/10.1007/JHEP10(2021)089}{{\em JHEP} {\bfseries 10} (2021) 089}, \href{http://arxiv.org/abs/2107.00655}{{\ttfamily arXiv:2107.00655 [hep-th]}}.

\bibitem{Guo:2019ady}
B.~Guo and S.~D. Mathur, ``{Lifting of level-1 states in the D1D5 CFT},'' \href{http://dx.doi.org/10.1007/JHEP03(2020)028}{{\em JHEP} {\bfseries 03} (2020) 028}, \href{http://arxiv.org/abs/1912.05567}{{\ttfamily arXiv:1912.05567 [hep-th]}}.

\bibitem{Guo:2020gxm}
B.~Guo and S.~D. Mathur, ``{Lifting at higher levels in the D1D5 CFT},'' \href{http://dx.doi.org/10.1007/JHEP11(2020)145}{{\em JHEP} {\bfseries 11} (2020) 145}, \href{http://arxiv.org/abs/2008.01274}{{\ttfamily arXiv:2008.01274 [hep-th]}}.

\bibitem{Guo:2022ifr}
B.~Guo, M.~R.~R. Hughes, S.~D. Mathur, and M.~Mehta, ``{Universal lifting in the D1-D5 CFT},'' \href{http://dx.doi.org/10.1007/JHEP10(2022)148}{{\em JHEP} {\bfseries 10} (2022) 148}, \href{http://arxiv.org/abs/2208.07409}{{\ttfamily arXiv:2208.07409 [hep-th]}}.

\bibitem{Hughes:2023apl}
M.~R.~R. Hughes, S.~D. Mathur, and M.~Mehta, ``{Lifting of two-mode states in the D1-D5 CFT},'' \href{http://dx.doi.org/10.1007/JHEP01(2024)183}{{\em JHEP} {\bfseries 01} (2024) 183}, \href{http://arxiv.org/abs/2309.03321}{{\ttfamily arXiv:2309.03321 [hep-th]}}.

\bibitem{Hughes:2023fot}
M.~R.~R. Hughes, S.~D. Mathur, and M.~Mehta, ``{Lifting of superconformal descendants in the D1-D5 CFT},'' \href{http://dx.doi.org/10.1007/JHEP04(2024)129}{{\em JHEP} {\bfseries 04} (2024) 129}, \href{http://arxiv.org/abs/2311.00052}{{\ttfamily arXiv:2311.00052 [hep-th]}}.

\bibitem{Fabri:2025rok}
M.~Fabri, A.~Sfondrini, and T.~Skrzypek, ``{Perturbed symmetric-product orbifold: first-order mixing and puzzles for integrability},'' \href{http://arxiv.org/abs/2504.13091}{{\ttfamily arXiv:2504.13091 [hep-th]}}.

\bibitem{Benjamin:2016pil}
N.~Benjamin, ``{A Refined Count of BPS States in the D1/D5 System},'' \href{http://dx.doi.org/10.1007/JHEP06(2017)028}{{\em JHEP} {\bfseries 06} (2017) 028}, \href{http://arxiv.org/abs/1610.07607}{{\ttfamily arXiv:1610.07607 [hep-th]}}.

\bibitem{Deger:1998nm}
S.~Deger, A.~Kaya, E.~Sezgin, and P.~Sundell, ``{Spectrum of D = 6, N=4b supergravity on AdS in three-dimensions x S**3},'' \href{http://dx.doi.org/10.1016/S0550-3213(98)00555-0}{{\em Nucl. Phys.} {\bfseries B536} (1998) 110--140},
\href{http://arxiv.org/abs/hep-th/9804166}{{\ttfamily arXiv:hep-th/9804166 [hep-th]}}.
%%CITATION = HEP-TH/9804166;%%.

\bibitem{Larsen:1998xm}
F.~Larsen, ``{The Perturbation spectrum of black holes in N=8 supergravity},'' \href{http://dx.doi.org/10.1016/S0550-3213(98)00564-1}{{\em Nucl. Phys. B} {\bfseries 536} (1998) 258--278}, \href{http://arxiv.org/abs/hep-th/9805208}{{\ttfamily arXiv:hep-th/9805208}}.

\bibitem{DHoker:2022dxx}
E.~D'Hoker and J.~Kaidi, ``{Lectures on modular forms and strings},'' \href{http://arxiv.org/abs/2208.07242}{{\ttfamily arXiv:2208.07242 [hep-th]}}.

\bibitem{Dijkgraaf:1996xw}
R.~Dijkgraaf, G.~W. Moore, E.~P. Verlinde, and H.~L. Verlinde, ``{Elliptic genera of symmetric products and second quantized strings},'' \href{http://dx.doi.org/10.1007/s002200050087}{{\em Commun. Math. Phys.} {\bfseries 185} (1997) 197--209}, \href{http://arxiv.org/abs/hep-th/9608096}{{\ttfamily arXiv:hep-th/9608096}}.

\bibitem{Kachru:2016igs}
S.~Kachru and A.~Tripathy, ``{The Hodge-elliptic genus, spinning BPS states, and black holes},'' \href{http://dx.doi.org/10.1007/s00220-017-2910-1}{{\em Commun. Math. Phys.} {\bfseries 355} no.~1, (2017) 245--259}, \href{http://arxiv.org/abs/1609.02158}{{\ttfamily arXiv:1609.02158 [hep-th]}}.

\bibitem{Benjamin:2017rnd}
N.~Benjamin and S.~M. Harrison, ``{Symmetries of the refined D1/D5 BPS spectrum},'' \href{http://dx.doi.org/10.1007/JHEP11(2017)091}{{\em JHEP} {\bfseries 11} (2017) 091}, \href{http://arxiv.org/abs/1708.02244}{{\ttfamily arXiv:1708.02244 [hep-th]}}.

\bibitem{ArabiArdehali:2024lyz}
A.~Arabi~Ardehali and H.~Krishna, ``{Topologically charged BPS microstates in AdS$_{3}$/CFT$_{2}$},'' \href{http://dx.doi.org/10.1007/JHEP03(2025)069}{{\em JHEP} {\bfseries 03} (2025) 069}, \href{http://arxiv.org/abs/2411.13824}{{\ttfamily arXiv:2411.13824 [hep-th]}}.

\bibitem{Maldacena:2016upp}
J.~Maldacena, D.~Stanford, and Z.~Yang, ``{Conformal symmetry and its breaking in two dimensional Nearly Anti-de-Sitter space},'' \href{http://dx.doi.org/10.1093/ptep/ptw124}{{\em PTEP} {\bfseries 2016} no.~12, (2016) 12C104}, \href{http://arxiv.org/abs/1606.01857}{{\ttfamily arXiv:1606.01857 [hep-th]}}.

\bibitem{vafa:1995bm}
C.~Vafa, ``{Instantons on D-branes},'' \href{http://dx.doi.org/10.1016/0550-3213(96)00075-2}{{\em Nucl. Phys. B} {\bfseries 463} (1996) 435--442}, \href{http://arxiv.org/abs/hep-th/9512078}{{\ttfamily arXiv:hep-th/9512078}}.

\bibitem{wip}
M. R. R. Hughes and M. Shigemori, work in progress.

\bibitem{Jevicki:2015irq}
A.~Jevicki and J.~Yoon, ``{$S_N$ Orbifolds and String Interactions},'' \href{http://dx.doi.org/10.1088/1751-8113/49/20/205401}{{\em J. Phys. A} {\bfseries 49} no.~20, (2016) 205401}, \href{http://arxiv.org/abs/1511.07878}{{\ttfamily arXiv:1511.07878 [hep-th]}}.

\bibitem{Guo:2019pzk}
B.~Guo and S.~D. Mathur, ``{Lifting of states in 2-dimensional $N = 4$ supersymmetric CFTs},'' \href{http://dx.doi.org/10.1007/JHEP10(2019)155}{{\em JHEP} {\bfseries 10} (2019) 155}, \href{http://arxiv.org/abs/1905.11923}{{\ttfamily arXiv:1905.11923 [hep-th]}}.

\bibitem{Sen:2009vz}
A.~Sen, ``{Arithmetic of Quantum Entropy Function},'' \href{http://dx.doi.org/10.1088/1126-6708/2009/08/068}{{\em JHEP} {\bfseries 08} (2009) 068}, \href{http://arxiv.org/abs/0903.1477}{{\ttfamily arXiv:0903.1477 [hep-th]}}.

\bibitem{Dabholkar:2010rm}
A.~Dabholkar, J.~Gomes, S.~Murthy, and A.~Sen, ``{Supersymmetric Index from Black Hole Entropy},'' \href{http://dx.doi.org/10.1007/JHEP04(2011)034}{{\em JHEP} {\bfseries 04} (2011) 034}, \href{http://arxiv.org/abs/1009.3226}{{\ttfamily arXiv:1009.3226 [hep-th]}}.

\bibitem{Chowdhury:2015gbk}
A.~Chowdhury, R.~S. Garavuso, S.~Mondal, and A.~Sen, ``{Do All BPS Black Hole Microstates Carry Zero Angular Momentum?},'' \href{http://dx.doi.org/10.1007/JHEP04(2016)082}{{\em JHEP} {\bfseries 04} (2016) 082},
\href{http://arxiv.org/abs/1511.06978}{{\ttfamily arXiv:1511.06978 [hep-th]}}.
%%CITATION = ARXIV:1511.06978;%%.

\bibitem{Gaiotto:2005gf}
D.~Gaiotto, A.~Strominger, and X.~Yin, ``{New connections between 4-D and 5-D black holes},'' \href{http://dx.doi.org/10.1088/1126-6708/2006/02/024}{{\em JHEP} {\bfseries 0602} (2006) 024},
\href{http://arxiv.org/abs/hep-th/0503217}{{\ttfamily arXiv:hep-th/0503217 [hep-th]}}.
%%CITATION = HEP-TH/0503217;%%.

\bibitem{Bena:2025pcy}
I.~Bena and N.~P. Warner, ``{Microstate Geometries},'' \href{http://arxiv.org/abs/2503.17310}{{\ttfamily arXiv:2503.17310 [hep-th]}}.

\bibitem{Denef:2002ru}
F.~Denef, ``{Quantum quivers and Hall / hole halos},'' \href{http://dx.doi.org/10.1088/1126-6708/2002/10/023}{{\em JHEP} {\bfseries 10} (2002) 023}, \href{http://arxiv.org/abs/hep-th/0206072}{{\ttfamily arXiv:hep-th/0206072}}.

\bibitem{Bena:2006kb}
I.~Bena, C.-W. Wang, and N.~P. Warner, ``{Mergers and Typical Black Hole Microstates},'' \href{http://dx.doi.org/10.1088/1126-6708/2006/11/042}{{\em JHEP} {\bfseries 11} (2006) 042},
\href{http://arxiv.org/abs/hep-th/0608217}{{\ttfamily arXiv:hep-th/0608217}}.
%%CITATION = HEP-TH/0608217;%%.

\bibitem{deMelloKoch:2025cec}
R.~de~Mello~Koch, A.~Ghosh, and H.~J.~R. Van~Zyl, ``{Bosonic Fortuity in Vector Models},'' \href{http://arxiv.org/abs/2504.14181}{{\ttfamily arXiv:2504.14181 [hep-th]}}.

\bibitem{Dijkgraaf:2000fq}
R.~Dijkgraaf, J.~M. Maldacena, G.~W. Moore, and E.~P. Verlinde, ``{A Black hole Farey tail},'' \href{http://arxiv.org/abs/hep-th/0005003}{{\ttfamily arXiv:hep-th/0005003}}.

\bibitem{Manschot:2007ha}
J.~Manschot and G.~W. Moore, ``{A Modern Farey Tail},'' \href{http://dx.doi.org/10.4310/CNTP.2010.v4.n1.a3}{{\em Commun. Num. Theor. Phys.} {\bfseries 4} (2010) 103--159}, \href{http://arxiv.org/abs/0712.0573}{{\ttfamily arXiv:0712.0573 [hep-th]}}.

\bibitem{Keller:2011xi}
C.~A. Keller, ``{Phase transitions in symmetric orbifold CFTs and universality},'' \href{http://dx.doi.org/10.1007/JHEP03(2011)114}{{\em JHEP} {\bfseries 03} (2011) 114}, \href{http://arxiv.org/abs/1101.4937}{{\ttfamily arXiv:1101.4937 [hep-th]}}.

\bibitem{Shigemori:2019orj}
M.~Shigemori, ``{Counting Superstrata},'' \href{http://dx.doi.org/10.1007/JHEP10(2019)017}{{\em JHEP} {\bfseries 10} (2019) 017}, \href{http://arxiv.org/abs/1907.03878}{{\ttfamily arXiv:1907.03878 [hep-th]}}.

\bibitem{Cardy:1986ie}
J.~L. Cardy, ``{Operator Content of Two-Dimensional Conformally Invariant Theories},'' \href{http://dx.doi.org/10.1016/0550-3213(86)90552-3}{{\em Nucl. Phys. B} {\bfseries 270} (1986) 186--204}.

\bibitem{Mayerson:2020acj}
D.~R. Mayerson and M.~Shigemori, ``{Counting D1-D5-P microstates in supergravity},'' \href{http://dx.doi.org/10.21468/SciPostPhys.10.1.018}{{\em SciPost Phys.} {\bfseries 10} no.~1, (2021) 018}, \href{http://arxiv.org/abs/2010.04172}{{\ttfamily arXiv:2010.04172 [hep-th]}}.

\bibitem{Eberhardt:2018sce}
L.~Eberhardt and I.~G. Zadeh, ``{$\mathcal{N}=(3,3)$ holography on ${\rm AdS}_3 \times ({\rm S}^3 \times {\rm S}^3 \times {\rm S}^1)/\mathbb Z_2$},'' \href{http://dx.doi.org/10.1007/JHEP07(2018)143}{{\em JHEP} {\bfseries 07} (2018) 143}, \href{http://arxiv.org/abs/1805.09832}{{\ttfamily arXiv:1805.09832 [hep-th]}}.

\bibitem{Eberhardt:2017pty}
L.~Eberhardt, M.~R. Gaberdiel, and W.~Li, ``{A holographic dual for string theory on AdS$_{3}$\texttimes{}S$^{3}$\texttimes{}S$^{3}$\texttimes{}S$^{1}$},'' \href{http://dx.doi.org/10.1007/JHEP08(2017)111}{{\em JHEP} {\bfseries 08} (2017) 111}, \href{http://arxiv.org/abs/1707.02705}{{\ttfamily arXiv:1707.02705 [hep-th]}}.

\bibitem{Datta:2017ert}
S.~Datta, L.~Eberhardt, and M.~R. Gaberdiel, ``{Stringy $\mathcal{N}=(2,2)$ holography for AdS${_3}$},'' \href{http://dx.doi.org/10.1007/JHEP01(2018)146}{{\em JHEP} {\bfseries 01} (2018) 146}, \href{http://arxiv.org/abs/1709.06393}{{\ttfamily arXiv:1709.06393 [hep-th]}}.

\bibitem{Bena:2022wpl}
I.~Bena, S.~D. Hampton, A.~Houppe, Y.~Li, and D.~Toulikas, ``{The (amazing) super-maze},'' \href{http://dx.doi.org/10.1007/JHEP03(2023)237}{{\em JHEP} {\bfseries 03} (2023) 237}, \href{http://arxiv.org/abs/2211.14326}{{\ttfamily arXiv:2211.14326 [hep-th]}}.

\bibitem{Bena:2023rzm}
I.~Bena, A.~Houppe, D.~Toulikas, and N.~P. Warner, ``{Maze topiary in supergravity},'' \href{http://dx.doi.org/10.1007/JHEP03(2025)120}{{\em JHEP} {\bfseries 03} (2025) 120}, \href{http://arxiv.org/abs/2312.02286}{{\ttfamily arXiv:2312.02286 [hep-th]}}.

\bibitem{Avery:2010qw}
S.~G. Avery, ``{Using the D1D5 CFT to Understand Black Holes},''
\href{http://arxiv.org/abs/1012.0072}{{\ttfamily arXiv:1012.0072 [hep-th]}}.
%%CITATION = ARXIV:1012.0072;%%.

\bibitem{Eguchi:1987sm}
T.~Eguchi and A.~Taormina, ``{Unitary Representations of $N=4$ Superconformal Algebra},'' \href{http://dx.doi.org/10.1016/0370-2693(87)91679-0}{{\em Phys. Lett. B} {\bfseries 196} (1987) 75}.

\bibitem{Eguchi:1987wf}
T.~Eguchi and A.~Taormina, ``{Character Formulas for the $N=4$ Superconformal Algebra},'' \href{http://dx.doi.org/10.1016/0370-2693(88)90778-2}{{\em Phys. Lett. B} {\bfseries 200} (1988) 315}.

\bibitem{Petersen:1989zz}
J.~L. Petersen and A.~Taormina, ``{Characters of the $N=4$ Superconformal Algebra With Two Central Extensions},'' \href{http://dx.doi.org/10.1016/0550-3213(90)90084-Q}{{\em Nucl. Phys. B} {\bfseries 331} (1990) 556--572}.

\bibitem{Petersen:1989pp}
J.~L. Petersen and A.~Taormina, ``{Characters of the $N=4$ Superconformal Algebra With Two Central Extensions: 2. Massless Representations},'' \href{http://dx.doi.org/10.1016/0550-3213(90)90141-Y}{{\em Nucl. Phys. B} {\bfseries 333} (1990) 833--854}.

\end{thebibliography}\endgroup
\end{document}